\documentclass[12pt]{thesis}
\usepackage[pctex32]{lscape}
\usepackage[pctex32]{graphicx}
\usepackage{amssymb,amsmath}
\usepackage{truetype}

\newcommand{\et}{{\it et~al. }}
\renewcommand{\baselinestretch}{2}

\newcommand{\fn}[1]{\renewcommand{\baselinestretch}{.5}\footnote{#1}}
\newcommand{\cpn}[2]{\renewcommand{\baselinestretch}{.5}{\large\caption[#1]{#2}}}
\begin{document}
\bibliographystyle{unsrt}
\pagenumbering{roman}
\begin{titlepage} 
\vspace*{0.5in} 
\centering{GENERALIZED CRITICAL POINTS ANALYSIS OF ACETYLENE VIBRATIONAL DYNAMICS}\\ 
\vspace*{2.0in} \centering{by\\XINLI DING}\\
\vspace*{2.0in} \centering{A DISSERTATION\\Presented to the Department of
Chemistry\\ \vspace{-0.2in}and the Graduate School of the University of Oregon\\
\vspace{-0.2in}in partial fulfillment of the requirements\\ \vspace{-0.2in}for
the degree of\\ \vspace{-0.2in}Doctor of Philosophy\\March 2004}
\end{titlepage} 
\vspace*{0.0in} 
\setcounter{page}{2} 
\noindent ``Generalized Critical Points Analysis of Acetylene Vibrational Dynamics'', a dissertation prepared by Xinli Ding in partial fulfillment of the requirements for the Doctor of Philosophy degree in the Department of Chemistry.  This dissertation has been approved and accepted by: \\ \vspace{0in}

\hbox to 4.75in{\hrulefill} \vspace{-0.2in}%
\noindent Dr. Jeffrey A. Cina, Chair of the Examining Committee\\
\vspace{0in}


\hbox to 3.25in{\hrulefill} \vspace{-0.2in} \noindent Date\\ \vspace{0.5in}

\noindent Committee in charge: \hbox to 0.27in{\hfill} Dr. Jeffrey A. Cina, Chair \vspace{-0.2in}\\ 
\hbox to 2in{\hfill}Dr. Michael E. Kellman, Advisor \vspace{-0.2in}\\ 
\hbox to 2in{\hfill}Dr. David R. Herrick \vspace{-0.2in}\\ 
\hbox to 2in{\hfill}Dr. Michael M. Haley \vspace{-0.2in}\\
\hbox to 2in{\hfill}Dr. Jens U. N\"ockel \vspace{0.in}\\ 
\noindent Accepted by:\\
\vspace{0in}

\hbox to 4.75in{\hrulefill} \vspace{-0.2in} \noindent Dean of the Graduate School
\pagebreak

\vspace*{0.0in}
\centerline{An Abstract of the Dissertation of}
\noindent Xinli Ding {\hfill} for the degree of {\hfill} Doctor of Philosophy\\ 
\noindent in the Department of Chemistry {\hfill} to be taken {\hfill} March 2004\vspace*{0.2in} \\
\noindent{Title: GENERALIZED CRITICAL POINTS ANALYSIS  OF ACETYLENE \\ 
\hspace*{0.47in} VIBRATIONAL DYNAMICS}
\vspace{-0.1in}\\ 

\noindent Approved: \hbox to 3.5in{\hrulefill} \vspace{-0.2in} \\
\noindent \hspace*{2.0in}Dr. Jeffrey A. Cina \\ \vspace{0in}

Classical tools of nonlinear dynamics are used to study the highly excited vibrations of small molecules.  For effective Hamiltonians with one polyad number (approximate constant of motion), previously developed methods locate new anharmonic modes using the critical points in the reduced classical phase space.  Theoretical arguments are given for generalizing the method to more than one polyad number.  As the simplest classical invariant structure, critical points of the reduced phase space are solved analytically without relying on either integrating trajectories or visual inspection.  These critical points, especially those that are linearly stable, are expected to indicate regions with the same type of classical dynamics as well as quantum modes of vibration.   

The pure bending Hamiltonian of acetylene (C$_2$H$_2$) is analyzed to demonstrate the effectiveness of critical points analysis.  Four families of critical points are born in distinct bifurcations, each corresponding to a novel anharmonic mode.  These modes are visualized with custom computer-generated animations.  Their origin and nature are qualitatively explained through separate consideration of DD-I and $\ell$ resonance alone.  Quantum survival probability verifies that the Local and Counter Rotator modes are the stable modes of vibration at high excitation.   

The same analysis is extended for the first time to the acetylene stretch-bend system, which has never been analyzed classically with all the resonance couplings.  Preliminary results are obtained for the polyad series containing the C-H stretch overtones.  The local C-H stretch critical points family, induced by the stretch-stretch ($K_{11/33}$) resonance, is located and shown to bifurcate into at least 4 new families when the stretch-bend resonances are included.    The new families indicate that the mixing between the stretch and bend may result in novel vibrational modes.

This dissertation includes my previously co-authored materials.


\vspace*{0.25in} \begin{center}{CURRICULUM VITA}\end{center}\vspace*{0.2in} 
NAME OF AUTHOR:  Xinli Ding\vspace{0.1in}\\ 
GRADUATE AND UNDERGRADUATE SCHOOLS ATTENDED:\vspace{-6pt}\\ 
\hspace*{0.2in}University of Oregon\vspace{-0.2in}\\ 
\hspace*{0.2in}Peking University, Beijing, P.R.China\vspace{0.1in}\\ 
DEGREES AWARDED:\vspace{-6pt}\\ 
\hspace*{0.2in}Doctor of Philosophy in Chemistry, 2004, University of Oregon\vspace{-0.2in}\\
\hspace*{0.2in}Bachelor of Science in Chemistry, 1997, Peking University\vspace{0.1in}\\ 
AREAS OF SPECIAL INTEREST:\vspace{-6pt}\\ 
\hspace*{0.2in} Chemical Physics \vspace{-0.2in}\\
\hspace*{0.2in} Nonlinear Dynamics \vspace{-0.2in}\\
\hspace*{0.2in} Computer Aided Visualization \vspace{0.1in}\\
PROFESSIONAL EXPERIENCE:\vspace{-6pt}\\ 
\hspace*{0.2in}Research Assistant, Department of Chemistry, University of Oregon, Eugene,\vspace{-0.2in}\\ 
\hspace*{0.4in}1997-2004\vspace{-6pt}\\ 
\hspace*{0.2in}Teaching Assistant, Department of Chemistry, University of Oregon, Eugene,\vspace{-0.2in}\\ 
\hspace*{0.4in}1997-1998\vspace{-1.2in}\\ 
\newpage 
\vspace*{0.4in} 
\noindent PUBLICATIONS:\vspace{0.2in}\\
\hspace*{0.2in} * Under the name V.~Tyng \vspace{0.2in} \\
\hspace*{0.2in}[1] Y.~Shuangbo, V.~Tyng, and M.E. Kellman.  Spectral patterns of isomerizing \vspace{-0.2in}\\ 
\hspace*{0.4in} systems. {\em J. Phys. Chem. A.}, 107:8345, 2003.\vspace{2pt}\\
\hspace*{0.2in}[2] M.E. Kellman, M.W. Dow, and V.~Tyng.   Dressed basis for highly excited \vspace{-0.2in}\\ 
\hspace*{0.4in} molecular vibrations.  {\em J. Chem. Phys.}, 118:9519, 2003.\vspace{2pt}\\
\hspace*{0.2in}[3] J.F. Svitak, V.~Tyng, and M.E. Kellman.   Bifurcation analysis of higher \vspace{-0.2in}\\ 
\hspace*{0.4in} m:n resonance spectroscopic  {Hamiltonian}.  {\em J. Chem. Phys.}, 106:10797, 2002.\vspace{2pt}\\
\hspace*{0.2in}[4] M.E. Kellman and V.~Tyng.   Bifurcation effects in coupled {Bose-Einstein} \vspace{-0.2in}\\
\hspace*{0.4in} condensates.  {\em Phys. Rev. A}, 66:013602/1, 2002.\vspace{2pt}\\
\hspace*{0.2in}[5] C.~Zhou, D.~Xie, R.~Chen, G.~Yan, H.~Guo, V.~Tyng, and M.E. Kellman. \vspace{-0.2in}\\
\hspace*{0.4in} Quantum calculation of highly excited vibrational energy levels of  {CS}$_2$ \vspace{-0.2in}\\
\hspace*{0.4in} ($\tilde{{X}}$) on a new empirical potential energy surface and  semiclassical \vspace{-0.2in}\\
\hspace*{0.4in} analysis of 1:2 {Fermi} resonance. {\em Spectro. Acta. A}, 58A:727, 2002.\vspace{2pt}\\
\hspace*{0.2in}[6] M.E. Kellman, J.P. Rose, and V.~Tyng.  Spectral patterns and ultrafast dyn- \vspace{-0.2in}\\
\hspace*{0.4in} amics in planar acetylene. {\em European Physical Journal D}, 14:225, 2001.\vspace{2pt}\\
\hspace*{0.2in}[7] M.~Joyeux, D.~Sugny, V.~Tyng, M.E. Kellman, H.~Ishikawa, R.W. Field, \vspace{-0.2in}\\
\hspace*{0.4in} C.~Beck, and R.Schinke.  Semiclassical study of the isomerization \vspace{-0.2in}\\
\hspace*{0.4in} states of {HCP}. {\em J. Chem.Phys.}, 112:4162, 2000.\vspace{-6pt}\\
\newpage

\vspace*{0.25in} \begin{center}{ACKNOWLEDGMENTS}\end{center}\vspace{-6pt}

In finishing this thesis the author is indebted to the following people:

\begin{itemize}
\item Foremost, my advisor, Michael Kellman, for research opportunity, academic guidance and financial support over the years.   
\item Dr. John Svitak for valuable help early in my research, and Dr. Shuangbo Yang who continues to be an inspiring co-worker and collaborator.  
\item Jeffrey Cina, Travis Humble and Mary Rohrdanz, for intellectually stimulating discussions on {\it Electrodynamics of Continuous Media} and beyond.
\item Fellow students who provided support and editorial feedback during the writing of this thesis: Polly Berseth, Kerry Breno, Carrie Daniels-Hafer, Fred Harris and Michelle Knowles.
\item Finally and most importantly, Erich Wolf, without whom this thesis would not be possible.  
\end{itemize}

\vspace*{0.3in}

\hspace*{0.2in} This thesis is also available online at the following URL: 

\hspace*{0.3in}{http://darkwing.uoregon.edu/{$\sim$}meklab/DingThesis/}



\pagebreak

\setcounter{page}{8} 
\tableofcontents \listoffigures \listoftables\pagebreak

\setcounter{page}{1} \pagenumbering{arabic} 
\renewcommand{\baselinestretch}{2}{\large\normalsize}%
\addtocontents{toc}{\protect\vspace*{12pt}}%
\setcounter{page}{6} \pagenumbering{arabic}
\addtocontents{toc}{\protect\vspace*{0.5in}}
\chapter[\protect\uppercase{Introduction}]{Introduction}\label{ch.ch1}
\addtocontents{toc}{\protect\vspace{0.25in}}

\section[Molecular Dynamics Encoded In Spectra]{\underline{Molecular Dynamics Encoded In Spectra}}
\addtocontents{toc}{\protect\vspace*{7pt}}

In its earlier years, spectroscopy was an important tool in determining the structure of molecules.  Today, the equilibrium structure, as well as the related spectroscopic constants, continues to be actively pursued.  However, the accumulated theory, techniques and data have also enabled researchers to use spectroscopy as a probe of dynamical processes, such as collisions, energy transfer, and chemical reactions.  

The question is how to decode dynamical information from the experimental or calculated spectra.  Spectra recorded in either the frequency or time domain should reflect the same physical behavior.  These domains are formally connected by a Fourier transform \cite{Fourier}.  Electromagnetic radiation affects molecules on at least four levels in the order of increasing energy: the nuclear spin, rotational, vibrational and electronic {\it Degrees Of Freedom} (DOF).   The interactions among even these four levels within a molecule make it nontrivial to analyze the total dynamics.  For chemists, the vibrational DOF are of special interest since they are closely related to chemical reaction processes \cite{VDOF1}.   The focus of this thesis is the vibrational dynamics of small polyatomic molecules (3-4 atoms).   On the one hand, they are much simpler than most organic or biological molecule, where explicit treatment of all DOF is usually difficult if not impossible.  On the other hand, these small molecules already exhibit a rich range of dynamical behavior, especially among the highly excited states, that is yet to be fully understood \cite{NesbittField}.  

\section[Modes of Vibration]{\underline{Modes of Vibration}}
\addtocontents{toc}{\protect\vspace*{7pt}}

We start with the frequency-domain spectra.  Each resolved level in the spectrum corresponds to one quantum eigenstate of the molecular Hamiltonian.  The decoding of dynamics involves assignment of these levels with {\it quantum numbers} corresponding to their modes of vibration.  The word ``mode" has been rather liberally used to designate patterns in vibrations.  The conventional meaning is that there are certain coordinates in which the vibration appears particularly simple.  Especially, it refers to the case when the vibration can be separated into {\it independent periodic oscillations} in these coordinates.  Such a separation is crucial for comprehending molecular vibrations that may be too complex for direct visualization.  Therefore, mode designation is more than mere labeling because it reflects the pattern of underlying dynamics.

A well-known case is the {\it normal modes}.  When the amplitudes of vibration remain small, the force between any two atoms can be approximated as harmonic, i.e. proportional to the displacement from equilibrium configuration.  Diagonalization of the resulting force constant matrix yields $3N-6$ (or $3N-5$ for linear molecules) normal mode coordinates, $N$ being the number of atoms in the molecule \cite{Wilson}.  The classical motion of the atoms (determined by Newtonian mechanics) is decomposable into oscillations along these normal mode coordinates with characteristic frequencies.  In the quantum system, the nodes of the eigenstates are distributed along the normal coordinates (each intersecting the coordinate perpendicularly).  A set of normal mode quantum numbers can be assigned from counting the nodes.
 
At higher energy, as the vibrational amplitudes increase, inter-atomic forces deviate from the harmonic approximation.  This requires the inclusion of additional terms, such as anharmonic coefficients and resonance couplings, in describing the vibrational dynamics.  The former distorts the normal modes, while the latter not only distort but also mix them.  Eventually, the mixing would become so extensive that the normal mode picture ceases to be a valid representation of the dynamics.  

\begin{figure}[hbtp]  
\begin{center} \includegraphics[width=4.13in]{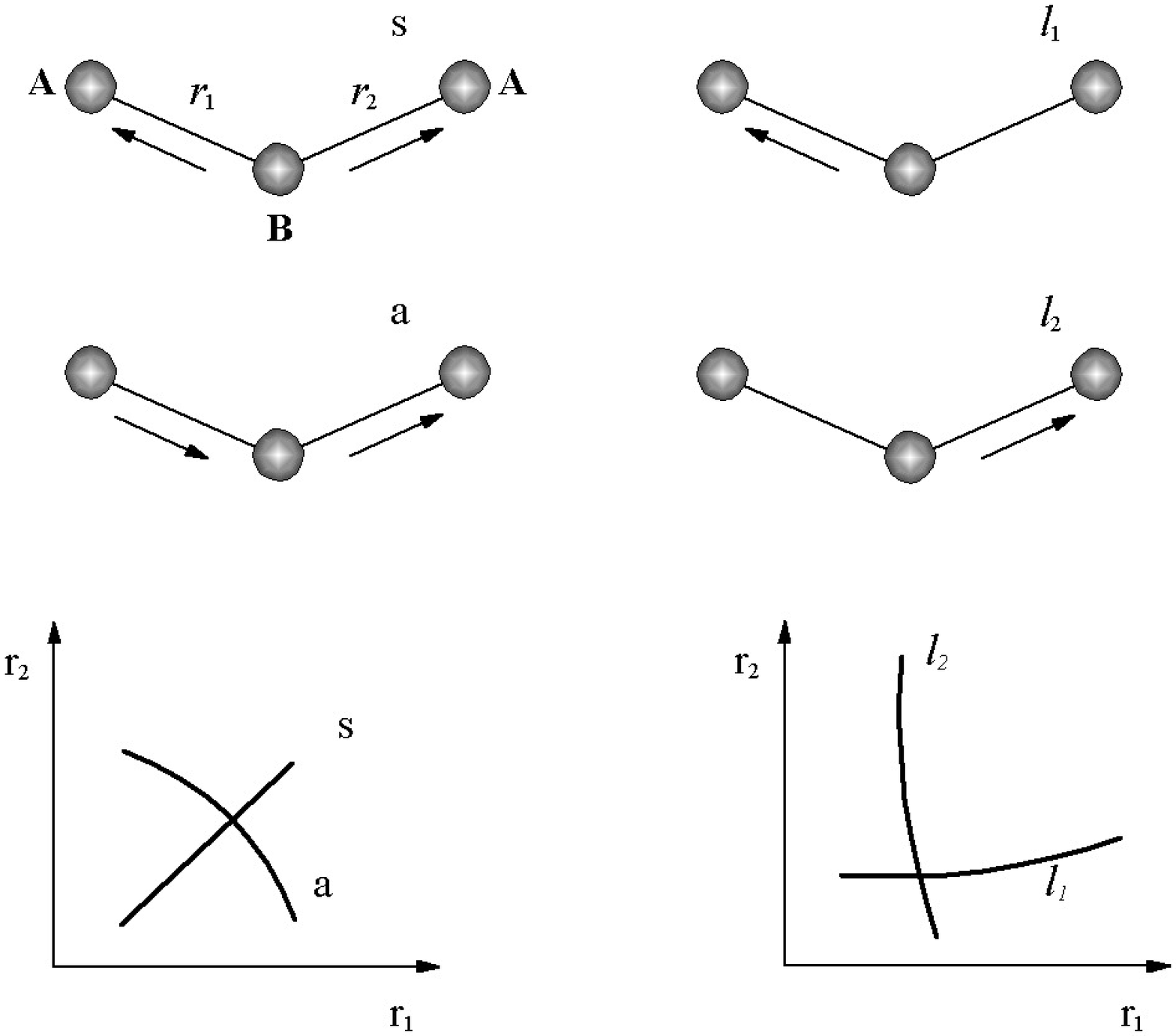}\end{center}
\cpn{Normal and local stretching modes in an ABA molecule} {Normal and local stretching modes in an ABA molecule.  In the normal symmetric (s) and antisymmetric (a) modes, the two A-B bonds vibrate in concert.  In contrast, the local modes 1 ($\ell_1$) and 2 ($\ell_2$) have most vibrational amplitude in only one bond. \label{fig1.1}}
\end{figure}

It is believed that at {\it sufficiently} high energy, the vibrational dynamics enters the ``bag of atoms" regime \cite{NesbittField}, which is characterized by a general lack of periodicity.  In practice, however, this regime may not be easily attained.  In at least some and perhaps most molecules, even when the normal modes break down dynamics could acquire new types of periodicity corresponding to new modes of vibration.  For example, in H$_2$O \cite{LocalWater} and O$_3$ \cite{KellmanOzone}, the stretching states exhibit a transition from normal to {\it local mode} behavior as the energy of excitation is increased.  In Fig.~\ref{fig1.1}, the normal symmetric and antisymmetric modes have two A-B bonds stretch in phase and out of phase by $\pi$, respectively.  In the local modes, only one of the A-B bonds is vibrating.   The more excited states can be assigned in terms of local mode quantum numbers, but not the normal mode ones.    Here the local mode is not just another way to describe the molecule, but a special one that directly reflects the underlying dynamics.  More exotic modes, such as the ``precessional" mode \cite{XiaoPre} and more complicated collective motions \cite{TaylorCHBrClF}, are also known to exist.  

\newpage    
\newpage 

These vibrational modes strongly influence the molecules' chemical behavior.  When a molecule is between collisions and free of other external interactions, its intrinsic reaction rate, if one exists, is determined by the flow of vibrational energy within the molecule, a process termed {\it Intramolecular Vibrational Relaxation} (IVR) \cite{NesbittField,ZewailIVR}.  The complete characterization of IVR requires a ``map" of all possible paths of energy flow.  Yet, in actual modeling of reactions, such a detailed description is often simplified into a statistical treatment.  A most popular assumption is the {\it Rice-Ramsperger-Kassel-Marcus} (RRKM) model, which assumes unrestricted energy flow among all vibrational DOF \cite{RRKMReview}.  This model is assumed increasingly valid at high internal energy and coupling, especially when the corresponding classical dynamics is chaotic \cite{Berne}.  

In contrast, energy in a stable vibrational mode remains trapped in a small fraction of all the possible energy distributions.   IVR and reaction probability would be very different from the RRKM limit when the molecule is in such a state.  Then, one has to be cautious about using a statistical model.   These stable modes may be of great use in coherently controlling chemical reactions \cite{LaserControl}, because excitation energy in these modes may remain long enough for optical manipulation.

Since the famous Fermi-Ulam-Pasta simulation in 1953 \cite{FermiUlamPasta}, molecular systems have served as an application for the mathematical theory of nonlinear dynamics, and a motivation for its continued development.  The past 50 years have seen an increasing interest in applying classical nonlinear mechanics in studying dynamics of microscopic systems.  

\section[Goal and Structure of Thesis]{\underline{Goal and Structure of Thesis}}
\addtocontents{toc}{\protect\vspace*{7pt}}

The identification and assignment of vibrational modes are crucial for analyzing the dynamics of highly excited molecules.  For over a decade, the Kellman group has been using the mathematical tools of bifurcation \cite{Kuznetsov} and catastrophe theory \cite{catastrophe} to study classical and quantum behavior in the vibration of small molecules.  The group developed the method of critical points analysis in order to uncover the nonlinear, anharmonic modes at high excitation via analytical detection (as opposed to numerical search).  Its earliest formulation focused on the integrable two-oscillator-single-resonance systems,  according to the critical point(s) in the reduced classical phase space \cite{Xiao-Kellman}.  This was later extended to coupled 3-oscillator systems, which are nonintegrable \cite{Zi-MinH2O1}.  These two methods are reviewed in references \cite{KellmanReview,KellmanAnnRev}.  This thesis is a further generalization that takes into account Hamiltonians with multiple polyad numbers.  The result is a DOF-independent method of generalized critical points analysis based on a more rigorous foundation.  The vibrational dynamics, especially the pure bending subsystem of acetylene (C$_2$H$_2$) is studied using the generalized critical points analysis.

Chapter 2 reviews the relevant background information: the effective quantum Hamiltonian and polyads, basic tools in classical mechanics, and topics on the quantum-classical correspondence.

Chapter 3 first describes existing procedures of critical points analysis for treating integrable and nonintegrable systems with one polyad number.  Then general considerations are carried out with regard to the behavior of critical points of an arbitrary effective Hamiltonian with polyad number(s).

Chapter 4 applies this analysis to the C$_2$H$_2$ pure bending system.  The result reveals the existence of new families of critical points, born in bifurcations at increasing energy.  The result is compared to those from the separate consideration of single resonances, giving a qualitative description of the nature of the new critical points.  The stable families of critical points correspond to stable quantum modes of vibration.   

Chapter 5 extends the analysis to the stretch-bend acetylene system including all the resonances.  The fate of the C-H normal stretch mode overtones is considered under increasing excitation. The preliminary result suggests that a chain of bifurcations first create a local C-H stretch mode, which then bifurcates into more complex stretch-bend modes.

\newpage Chapter 6 summarizes contributions made in this thesis, and discusses possible future directions of research.

\vspace*{0.25in} 

\begin{picture}(370,5) \thicklines \put(0,0){\line(1,0){370}} \end{picture}

\begin{center}{GLOSSARY}\end{center}\vspace{-6pt}

\begin{picture}(370,5) \thicklines \put(0,0){\line(1,0){370}} \end{picture}

\begin{align}
&\mbox{DOF} & \cdots \cdots \cdots \cdots \cdots &\mbox{\,\,\,\,\,\,\,\,Degrees Of Freedom} \nonumber\\ 
&\mbox{IVR} & \cdots \cdots \cdots \cdots \cdots &\mbox{\,\,\,\,\,\,\,\,Intramolecular Vibrational Redistribution} \nonumber\\ 
&\mbox{PES} &  \cdots \cdots \cdots \cdots \cdots &\mbox{\,\,\,\,\,\,\,\,Potential Energy Surface} \nonumber\\ 
&\mbox{PO} &  \cdots \cdots \cdots \cdots \cdots &\mbox{\,\,\,\,\,\,\,\,Periodic Orbit(s)} \nonumber\\ 
&\mbox{PPS} &  \cdots \cdots \cdots \cdots \cdots &\mbox{\,\,\,\,\,\,\,\,Polyad Phase Sphere(s)} \nonumber\\ 
&\mbox{RRKM} &  \cdots \cdots \cdots \cdots \cdots &\mbox{\,\,\,\,\,\,\,\,Rice-Ramsperger-Kassel-Marcus model} \nonumber\\ 
&\mbox{SOS} & \cdots \cdots \cdots \cdots \cdots &\mbox{\,\,\,\,\,\,\,\,Surface(s) Of Section} \nonumber\\ 
&\mbox{ZOS} & \cdots \cdots \cdots \cdots \cdots &\mbox{\,\,\,\,\,\,\,\,Zero-Order State(s)} \nonumber
\end{align}

\begin{picture}(370,5) \thicklines \put(0,0){\line(1,0){370}} \end{picture}%
\addtocontents{toc}{\protect\vspace*{12pt}}
\chapter[\protect\uppercase{Background Information}]{Background Information}\label{ch.ch2}%
\addtocontents{toc}{\protect\vspace{0.25in}}

\section[Effective Quantum Hamiltonian and Polyads]{\underline{Effective Quantum Hamiltonian and Polyads}}
\addtocontents{toc}{\protect\vspace*{7pt}}

The molecular vibrational spectrum is often modeled by an effective Hamiltonian obtained from fitting the spectral levels \fn{The fit is to either the resolved experimental spectra or the spectra from theoretical calculations \cite{HCP}.}.  Taking $\hbar=1$, the general form of an effective Hamiltonian is: 
\begin{eqnarray}
\hat{H}_{eff} = \hat{H}_0+\sum \hat{V}_{2}^{ij}+\sum \hat{V}_{3}^{ijk}+...  \label{generalHeff} 
\end{eqnarray}
\noindent with
\begin{align}
\hat{H}_0 =& \sum_i \omega_i \left( \hat{n}_i + \frac{d_i}{2} \right) + \sum_{i,j;i \leq j} x_{ij} \left( \hat{n}_i + \frac{d_i}{2} \right) \left( \hat{n}_j + \frac{d_j}{2} \right) + \cdots \label{QH0} \\
\hat{V}_{2}^{ij}= & V_{ij}[(\hat{a}_i^\dagger)^m(\hat{a}_j)^n+(\hat{a}_j^\dagger)^n(\hat{a}_i)^m]  \\
\hat{V}_{3}^{ijk}=&V_{ijk}[(\hat{a}_i^\dagger)^m (\hat{a}_j)^n (\hat{a}_k)^p + (\hat{a}_j^\dagger)^n (\hat{a}_k^\dagger)^p (\hat{a}_i)^m]  \label{QHeff}
\end{align}

The $\hat{H}_0$ term is in the form of a Dunham expansion \cite{Herzberg}.  $\hat{n}_i$ is a zero-order mode (e.g. normal or local modes) number operator, whose eigenvalue is $n_i$.  $d_i$ is the degeneracy of mode $i$: $1$ for non-degenerate modes and $2$ for doubly degenerate modes (such as the bending of a linear molecule).  Each $\hat{V}_{2}^{ij}$ term in eqn. (\ref{generalHeff}) represents a resonance that couples the modes $i$ and $j$.  It exchanges $m$ quanta in mode $i$ and $n$ quanta in mode $j$.  $\hat{V}_{3}^{ijk}$ acts in a similar manner among three modes $i$, $j$ and $k$.  The operators $\hat{a}_i^\dagger$, $\hat{a}_i$ have matrix elements identical to those of harmonic raising and lowering operators, i.e.
\begin{align}
\hat{a}_i^\dagger | n_i \rangle &= \sqrt{n_i+1} \, | n_i+1 \rangle  \label{raising} \\
\hat{a}_i | n_i \rangle  &= \sqrt{n_i} \, | n_i-1 \rangle        \label{lowering}  \\
\hat{a}_i^\dagger \hat{a}_i  |n_i \rangle & =\hat{n}_i|n_i \rangle = n_i \, | n_i \rangle
\end{align}

\noindent $|n_1, n_2, \cdots, n_N \rangle$ comprise a set of eigenstates of $\hat{H}_0$.  They are referred to as the {\it Zero Order States} (ZOS).  In the basis spanned by the ZOS, the matrix form of $\hat{H}_{eff}$ is obtained from eqns. (\ref{raising},\ref{lowering}).  Diagonalization of this matrix yields quantum eigenfunctions in terms of the ZOS.  In order to compare these eigenfunctions to the molecular coordinate space (such as bond length and angle), one needs to assume for each $n_i$ an oscillator model, e.g. of harmonic \cite{MartensEzra} or Morse \cite{RankinMiller} type.  


The resonance coupling terms $\hat{V}$ cause the ZOS to mix in the eigenfunctions.  The quantum numbers $n_i$ then are no longer good quantum numbers.  However, certain linear combinations of them, known as the {\it polyad numbers}, may remain conserved in the fitting Hamiltonian (and approximately conserved in the exact molecular Hamiltonian).  In triatomic molecules like H$_2$O, there is often an approximate 2:1 frequency ratio between one normal stretching mode $n_1$ and the bending mode $n_2$ \cite{JaffeBrumer} \fn{Meanwhile, the other non-interacting stretching normal mode $n_3$ can be regarded as a ``spectator".  Because the number of quanta in it is constant, it is absorbed into the other parameters when we only consider a specific $n_3$ manifold.}.  This ratio leads to the inclusion of a {\it Fermi resonance} term in $\hat{H}_{eff}$.  The Fermi resonance (1) takes  one quantum out of $n_1$ and adds two quanta to $n_2$, and (2) takes two quanta out of $n_2$ and adds one quantum to $n_1$.  When the equilibrium configuration of the molecule is non-linear, the bending $n_2$ mode is singly degenerate ($d_2=1$).  The effective two-mode Hamiltonian with Fermi resonance is
\begin{align}
\hat{H}_{Fermi}=& \omega_1 (\hat{n}_1+\frac{1}{2}) +\omega_2 (\hat{n}_2+\frac{1}{2}) +x_{11} (\hat{n}_1+\frac{1}{2})^2 + x_{12}(\hat{n}_1+\frac{1}{2}) (\hat{n}_2+\frac{1}{2}) \nonumber\\
& \quad + x_{22}(\hat{n}_2+\frac{1}{2})^2 +V_{Fermi} [ \hat{a}_1^\dagger(\hat{a}_2)^2 + (\hat{a}_2^\dagger )^2 \hat{a}_1 ] \label{FermiExample}
\end{align}
\noindent The polyad number  
\begin{eqnarray}
\hat{P}=2\hat{n}_1 + \hat{n}_2 
\end{eqnarray}
\noindent remains conserved since it commutes with the Hamiltonian
\begin{eqnarray}
\mbox{[} \hat{P}, \hat{H}_{Fermi} \mbox{]} = \hat{P} \hat{H}_{Fermi}-\hat{H}_{Fermi} \hat{P} =0   \label{commu}
\end{eqnarray}

\begin{figure}[hbtp]
\begin{center} \includegraphics[width=3.46in]{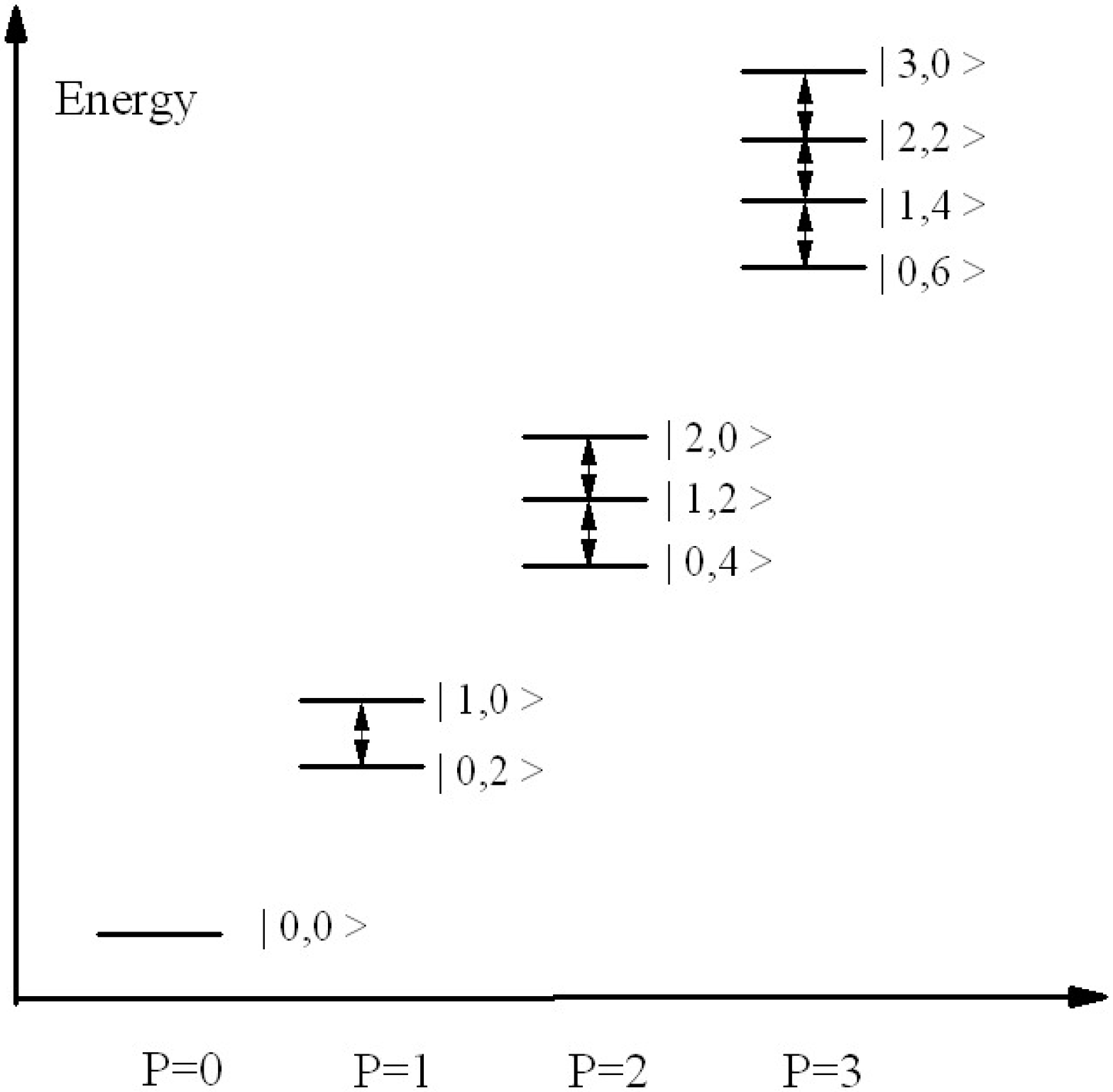}\end{center}
\cpn{Polyad structures in spectra} {Polyad structures in spectra, a schematic illustration.  The states with the same polyad number $P$ may strongly interact with each other due to their closeness.  Interpolyad couplings, on the other hand, are relatively weak due to the large energy spacing between polyads. \label{polyadclusters}}
\end{figure}

In the quantum Hamiltonian, the presence of $\hat{P}$ means that the resonance coupling only couples ZOS with the same polyad number.  For example, there are four ZOS $\vert n_1, n_2 \rangle$ interacting within $P=3$:
\begin{eqnarray}
\vert 3,0 \rangle \leftrightarrow \vert 2,2 \rangle \leftrightarrow \vert 1,4 \rangle \leftrightarrow \vert 0,6 \rangle \nonumber 
\end{eqnarray}
\noindent States belong to the same polyad appear in clusters in the spectra, as illustrated in Fig.~\ref{polyadclusters}.  Experimentally, it is often the observation of such clustering that leads to the adoption of a polyad model \cite{FieldPolyadRecog}.

\newpage   
\newpage 

Under the polyad model there is no coupling between one polyad and another.  The Hamiltonian matrix therefore can be separated into blocks, each containing the intra-polyad couplings.  The blocks can be individually diagonalized, which substantially reduces the amount of computation involved. In the time domain, the existence of polyad(s) imposes an approximate restriction on the energy flow: IVR can only occur within the same polyad \cite{SmithWinn1}.  

\begin{figure}[hbtp]
\begin{center}\includegraphics[width=2.65in]{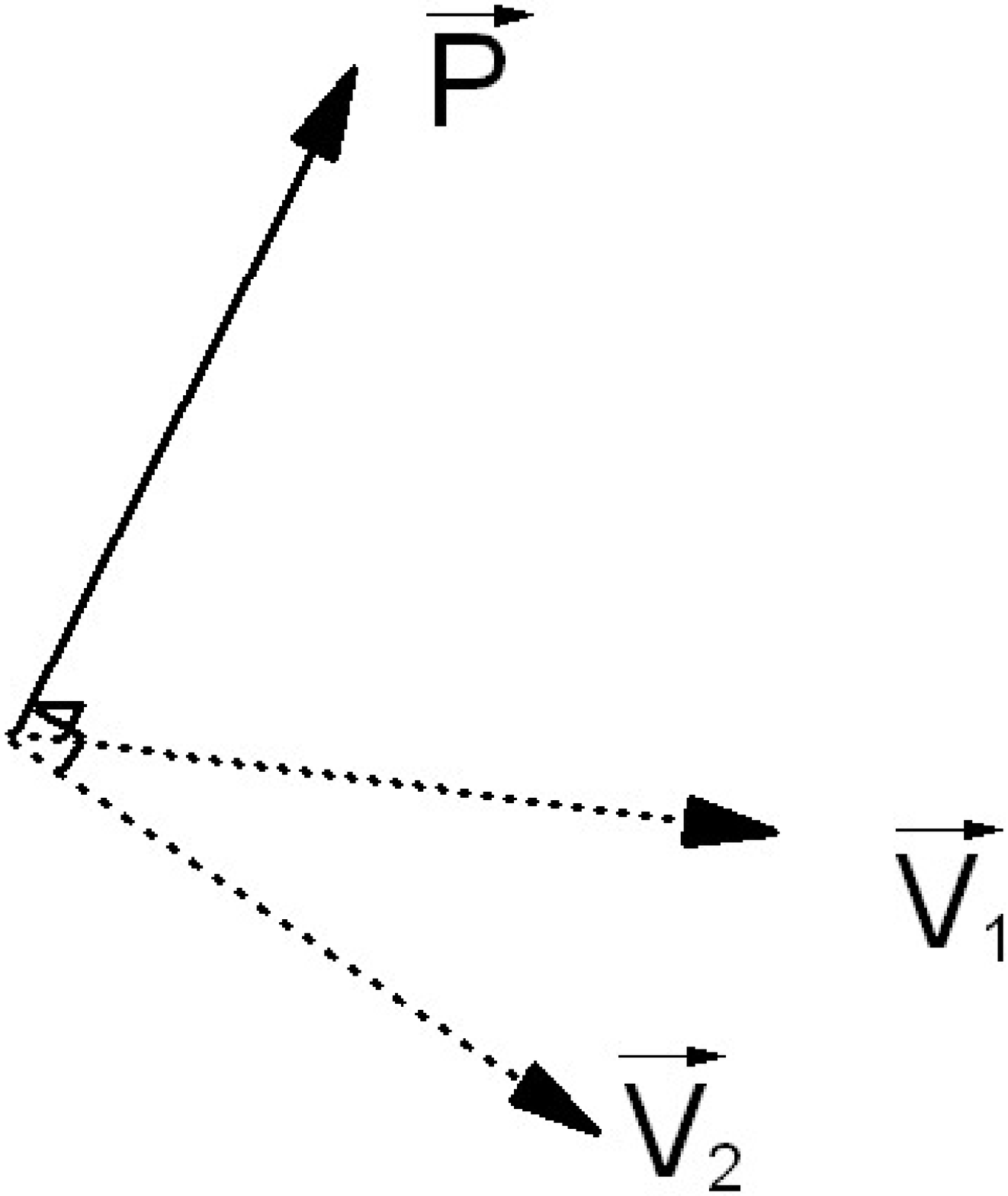}\end{center}
\cpn{Locating a polyad number from resonance vectors} {Locating a polyad number from resonance vectors, a schematic illustration adapted from Fig.~1 of \cite{KellmanAnnRev}. \label{polyads}}
\end{figure}

A systematic method to locate polyad numbers was devised by Kellman \cite{KellmanVector}.  The method is closely related to an earlier van Vleck perturbation study by Fried and Ezra \cite{FriedEzra}.  Let there be $N$ zero-order vibrational modes, excluding the spectator ones.  Each resonance term $\hat{V}$ can be represented by a {\it resonance vector} in the $N$-dimensional linear space $\vec{V}_i = \{ n_1, n_2, \cdots n_N \}$.  All the resonance vectors $\vec{V}_i$ in $\hat{H}_{eff}$, taken together, form a linear subspace with $M$ dimensions ($M \leq N$).  Orthogonal to this subspace is another ($N-M$) dimensional subspace from which the ($N-M$) polyad numbers are found.  This is graphically illustrated in Fig.~\ref{polyads} for $N=3, M=2$.  The coefficients in the polyad  number can be taken to be any set of linearly independent vectors $\vec{P}_j$, which span the ($N-M$) dimensional subspace.  In the Fermi resonance system, for example, the resonance vector $\vec{V}_{Fermi}=\{1, -2 \}$ gives the polyad number $P=2n_1+n_2$, in accordance with the vector $\vec{P}=\{2, 1 \}$ orthogonal to $\vec{V}_{Fermi}$. 

\newpage   
\newpage 

Although the total number of polyad numbers is fixed, each of them is not uniquely defined.  There is the liberty of multiplying $\vec{P}_j$ by an  arbitrary factor or linearly combining any number of $\vec{P}_j$.  The usual choice is to match their coefficients to the approximate integer ratios among the zero-order mode frequencies, as these ratios lead to the inclusion of the respective resonance terms in $\hat{H}_{eff}$ in the first place \cite{ReinhardtHynes}. 

Even when the more comprehensive Potential Energy Surface (PES) is available, effective Hamiltonians are often constructed from the PES with perturbative methods \cite{SibertC2H22,JoueuxSibertIso} in order to simplify the subsequent analysis.  The effective Hamiltonian is not only a more reliable model (compared to PES) for the highly excited states in triatomic or larger molecules, but also easily gives the useful insight of the polyad structure.   

The conservation of polyad numbers is never exact \fn{Physical laws of rigorous conservation are based on fundamental symmetry, known as Noether's Theorem (see Chapter 12.7 of \cite{Goldstein}). As an example, the conservation of energy and linear, angular momentum result from the homogeneity of time, space and the isotropy of space, respectively.  This is not the case with polyad numbers.}.  The degree of their conservation can be estimated by the uncertainty relationship $\Delta E \cdot \Delta t \geq \hbar$.  Spectral data recorded at low frequency resolution (larger $\Delta E$) decodes dynamics at shorter timescale (smaller $\Delta t$), and vice versa.  In molecules, spectral peaks well described by a polyad Hamiltonian may break into finer structures when scrutinized at finer frequency resolution.  This is caused by the small coupling terms not included in the effective Hamiltonian, which exercise their effects (including the breaking down of the polyad number) at longer time scales \cite{SEPReview}.  As an example, the acetylene pure bending Hamiltonian of Chapter 4 is fitted to spectra recorded with a resolution of 2 cm$^{-1}$ or finer.  The corresponding uncertainty $\Delta t=2.6$ picosecond is much longer than the bending vibration period (50 femtoseconds).  Since the polyad structure is still present at $2.6 ps$ time scale, the polyad numbers predicted from the effective Hamiltonian can be assumed valid at the same time scale or longer.

\section[Basic Concepts in Classical Mechanics]{\underline{Basic Concepts in Classical Mechanics}}
\addtocontents{toc}{\protect\vspace*{7pt}}

Although microscopic systems are governed by quantum mechanics, classical mechanics continues to be an important tool in understanding molecular processes due to the following fundamental and empirical reasons. (1) Quantum mechanics is built upon classical mechanics, as opposed to being a self-consistent theory.  Various semiclassical methods serve as a bridge between the quantum and classical worlds.  (2) Even when the classical description is not exact, it provides an intuitive tool for the human researcher, whose perceptions are unfortunately macroscopic and therefore classical, to understand the microscopic phenomena.  (3) In large and/or highly excited systems, treating the whole system quantum mechanically can become challenging, making classical and semiclassical methods useful supplements.

Below we discuss some basic concepts in classical mechanics (of the Hamiltonian formulation) that are pertinent to the topic of this thesis.  The reader is referred to Tabor \cite{Tabor} for a general introduction to classical mechanics with emphasis on the nonlinear dynamics.  The textbook by Goldstein \cite{Goldstein} may serve as a more comprehensive reference.
 
\subsection{2.2.1 Heisenberg's Correspondence Principle}
\addtocontents{toc}{\protect\vspace*{5pt}}

Heisenberg's Correspondence Principle provides an important connection between the quantum and classical worlds \cite{Heisenberg}.  It relates raising and lowering operators in quantum mechanics to Fourier components of action-angle variables in classical mechanics \cite{Child}:
\begin{align}
\hat{a}_i^\dagger & \rightarrow  \sqrt{n_i+\frac{d_i}{2}} e^{i\phi_i}= \sqrt{I_i} e^{i\phi_i}  \nonumber\\
\hat{a}_i & \rightarrow  \sqrt{n_i+\frac{d_i}{2}} e^{-i\phi_i} = \sqrt{I_i} e^{-i\phi_i}  \label{Heisenberg}
\end{align}

The $N$-mode quantum Hamiltonian $\hat{H}_{eff}$ is mapped to an $N$ DOF classical Hamiltonian $H_{eff}$ in canonical variables $(I_i, \phi_i)$. The $\hat{n}_i$ terms in $\hat{H}_0$ transform as
\begin{eqnarray}
I_i = n_i + \frac{d_i}{2}  \label{nitrans}
\end{eqnarray}
\noindent Hence, substitution of eqn. (\ref{Heisenberg}) into (\ref{FermiExample}) yields  
\begin{eqnarray}
H_{Fermi}=\omega_1 I_1+\omega_2 I_2 +x_{11} I_1^2+x_{12} I_1 I_2+x_{22} I_2^2 +2V_{Fermi}\sqrt{I_1^2 I_2} \cos[\phi_1-2\phi_2]   \label{Classical1}
\end{eqnarray}

Classically, the action-angle variables are defined as the conserved action $I_i$ and conjugate angle $\phi_i$ of the zero-order oscillator of mode $i$.  In particular, when the oscillator is 1-dimensional and harmonic, they can be related to the Cartesian coordinate and momentum by
\begin{align}  q_i  &= \sqrt{I_i} \cos \phi_i , & p_i & = \sqrt{I_i} \sin \phi_i  \end{align}

\subsection{2.2.2 Hamiltonian Classical Dynamics}
\addtocontents{toc}{\protect\vspace*{5pt}}

The $(I_i, \phi_i)$ variables form a set of canonically conjugate coordinates in Hamiltonian mechanics.  Their time evolution (also known as equations of motion) has the elegant form 
\begin{align}
\dot{I_i} =\frac{d I_i}{d t}  & = -\frac{\partial H}{\partial \phi_i} \label{EOM1} \\
\dot{\phi_i} = \frac{d \phi_i}{d t} &= \frac{\partial H}{\partial I_i}  \label{EOM2}
\end{align}
\noindent Once the initial condition $\{I_{i}(0), \phi_{i}(0) \}$ is given, the subsequent solution $\{I_i (t), \phi_i(t) \}$ is determined by integrating eqns. (\ref{EOM1},\ref{EOM2}).  $\{I_i (t), \phi_i(t) \}$ is known as a {\it phase space trajectory} or trajectory.  When the Hamiltonian $H$ is independent of time, a trajectory cannot intersect with itself in phase space, although it could retrace the same closed orbit repeatedly.

The equations of motion remain formally invariant.  After a transformation between two sets of canonical coordinates (called a {\it canonical transformation}), e.g. $ (I_i, \phi_i) \rightarrow (J_i, \Phi_i)$, we have:
\begin{align}
\dot{J}_i = & -\frac{\partial H}{\partial \Phi_i} \\
\dot{\Phi}_i = & \frac{\partial H}{\partial J_i}  
\end{align}

\subsection{2.2.3 Constants of Motion}
\addtocontents{toc}{\protect\vspace*{5pt}}

In a time-independent Hamiltonian, the energy is conserved, i.e. is a {\it constant of motion}.  Additional constants of motion may be present due to the polyad numbers.  In the Fermi resonance system, the polyad number $\hat{P}= 2\hat{n}_1+\hat{n}_2$ corresponds to a constant of motion through eqn. (\ref{nitrans}): 
\begin{eqnarray}
I = 2I_1+I_2 = P+ \frac{3}{2}
\end{eqnarray}
\noindent Like the quantum commutator in eqn. (\ref{commu}), the Poisson bracket between  $I$ and $H_{Fermi}$ also vanishes.  In order for the angle $\theta$ conjugate to $I$ to satisfy
\begin{eqnarray}
\dot{I} = -\frac{\partial H_{Fermi}}{\partial \theta} = 0
\end{eqnarray}
\noindent $\theta$ does not appear explicitly in the Hamiltonian.  Such a variable is known as a {\it cyclic angle} \cite{Tabor}.  This property leads  naturally to a canonical transformation $$(I_1, \phi_1, I_2, \phi_2) \rightarrow (I, \theta, I_z, \Psi)$$
\noindent with 
\begin{align}
I = & \frac{2I_1 +I_2}{2}, & \theta = & \phi_1+ 2\phi_2 \\
I_z = & \frac{2I_1-I_2}{2},  & \Psi = & \phi_1-2\phi_2
\end{align}
\noindent In the new coordinates, the classical Hamiltonian in eqn. (\ref{Classical1}) is expressed  as
\begin{align}
H_{Fermi}= & \omega_1(I+I_z)+\omega_2(I-I_z) +x_{11}(I+I_z)^2+x_{12}(I^2 - I_z^2)+x_{22}(I - I_z)^2 \nonumber\\
&+2V_{Fermi}\sqrt{(I+I_z)^2(I-I_z)} \cos \Psi   \label{FermiClassical}
\end{align}

\noindent Since its value does not change with time, $I$ can be regarded as an external parameter.  $\theta$ is absent from the Hamiltonian, and has limited physical significance.  The non-trivial part of the dynamics is captured in a 2-dimensional phase space $(I_z, \Psi)$, called the {\it reduced phase space}.  In general, in an $N$ DOF Hamiltonian with $(N-M)$ constants of motion, the phase space can be reduced from $2N$ to $2M$ dimensions by a similar transformation.  The details of such transformations are discussed in Appendix A.

A system is called {\it integrable} is there are as many constants of motion as the number of DOF.  In an integrable system, it is possible to express the Hamiltonian in $N$ constants of motion only (without their cyclic angles).  Then the equations of motion can be solved analytically without recourse to numerical integration.  A Hamiltonian with 1 DOF is always integrable when the energy is conserved.    

\subsection{2.2.4 Invariant Phase Space Structures}
\addtocontents{toc}{\protect\vspace*{5pt}}

An {\it invariant phase space structure} is defined as any lower-dimensional subset of the phase space that is mapped onto itself by the equations of motion.  These structures are the ``landmarks" that delineate regions in phase space with different kinds of dynamics.  The qualitative description of all these regions is called the {\it phase portrait} \cite{Kuznetsov}.

\begin{figure}[hbtp]
\begin{center}\includegraphics[width=3.13in]{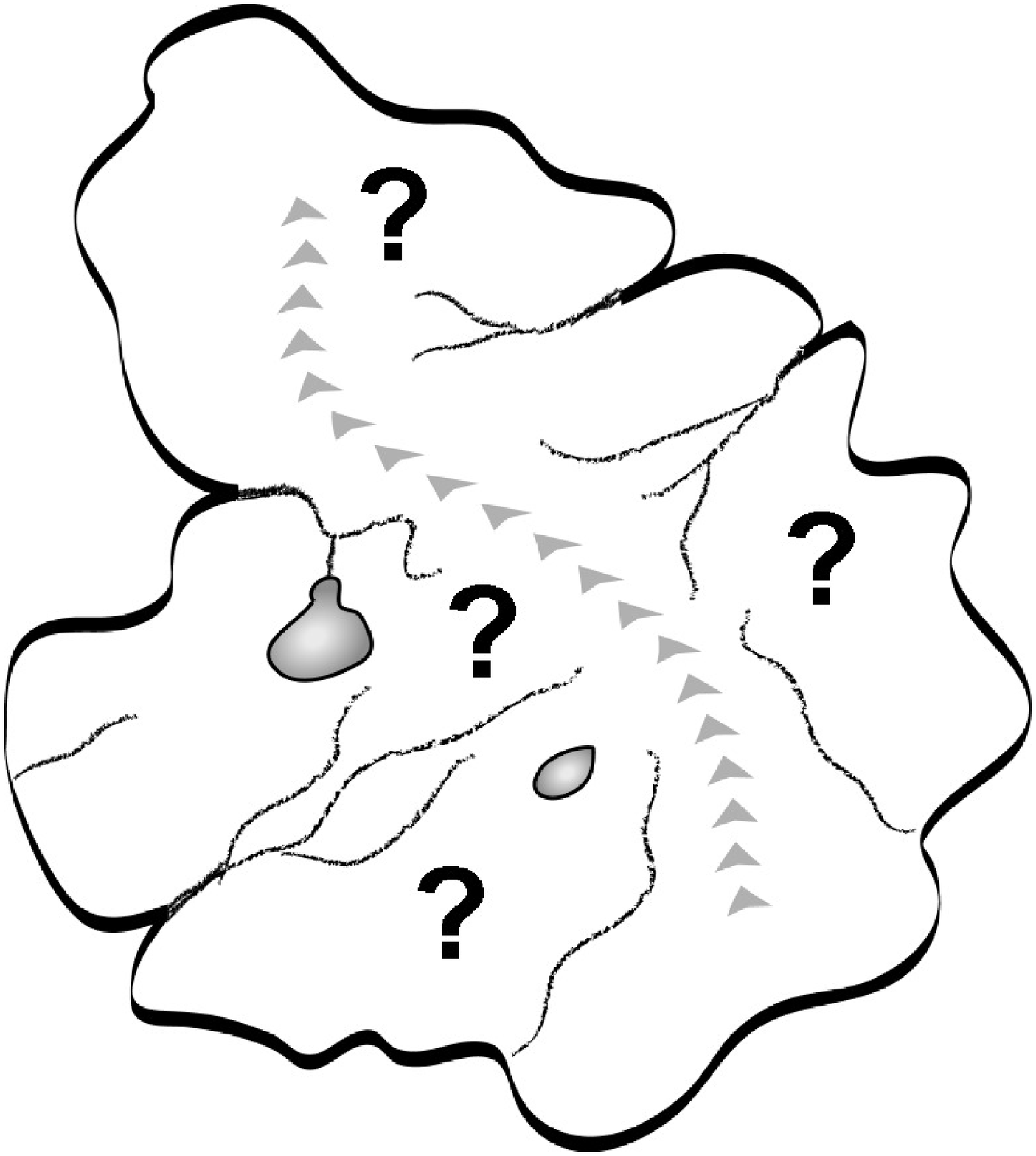}\end{center}
\cpn{Hydrologic landmarks on an unknown continent} {Hydrologic landmarks on an unknown continent, a schematic illustration.  Regions with different hydrologic dynamics are summarized by a map with the watershed (peaks in the center), rivers and lakes.     \label{landmark}}
\end{figure}

We are primarily interested in non-integrable systems, whose phase spaces are at least 4-dimensional and not easy to visualize directly.  In order to illustrate the role played by these landmarks, the following analogy seems appropriate.  Construction of the phase portrait of a new dynamical system can be compared to mapping out the hydrologic flow on an unknown continent \fn{This comparison certainly should not be taken literally.  For example, the autonomous flow in the phase space has neither ``sources" nor ``sinks".}.   Invariant phase space structures then serve purposes similar to the watersheds, rivers and lakes et cetera in the exploration of the continental water system.  In the cartoon map of Fig.~\ref{landmark}, the general flow trends on the surface is well characterized by the landmarks in the map, even though they do not account for the fate of every single raindrop, and whole regions may be left out of the picture (the areas marked with ``?").   When there is no prior knowledge about the system, finding these landmarks is the first step of a systematic exploration.

\newpage    
\newpage 

\noindent \underline {Critical Points} \,\,\,  The simplest invariant phase space structure is a {\it critical point} \cite{WigginsBook}.  These are defined as points where the equations of motion (\ref{EOM1}, \ref{EOM2}) vanish:
\begin{align}
\dot{I}_i = \dot{\phi}_i = - \frac{\partial H}{\partial \phi_i} = \frac{\partial H}{\partial I_i}=0   \label{CP}
\end{align}
\noindent Note that the term ``critical points" has also been used referring to where the gradient of a given function vanishes \cite{Weissen}, and the function may not be related to any dynamical property.  However, in Hamiltonian systems this definition coincides with the one defined above in eqn. (\ref{CP}).

Critical points are the simplest invariant structure because they have the lowest dimensionality 0, and because they can be exactly solved for as roots of simultaneous equations.

\begin{figure}[hbtp]
\begin{center}  \includegraphics[width=4.68in]{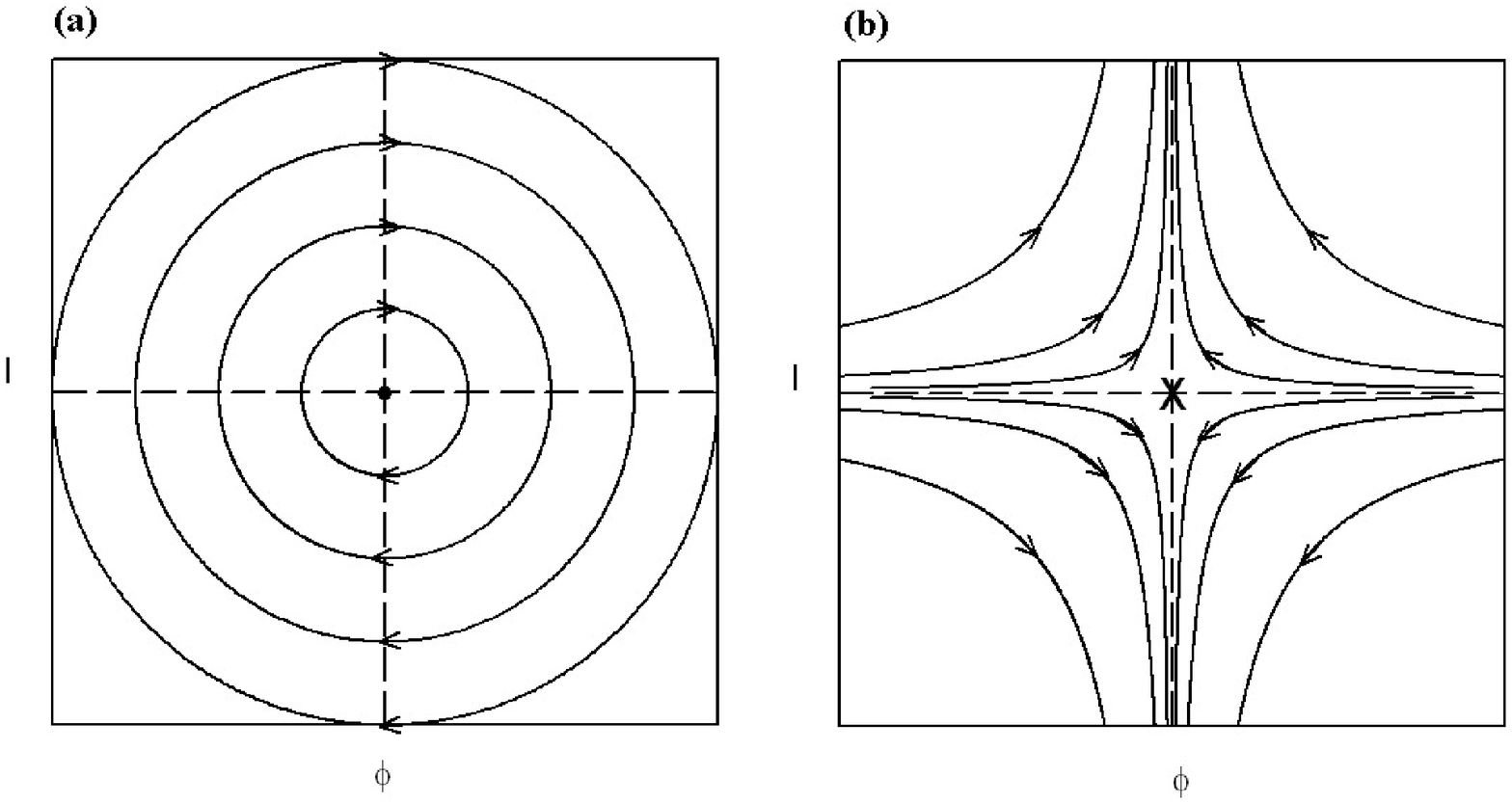}\end{center}  
\cpn{Dynamics near linearly stable and unstable critical points} {Dynamics near linearly stable and unstable critical points in a 1 DOF system.  Panel (a): trajectories near a linearly stable critical point. (b) trajectories near a linearly unstable critical point. \label{EH}}
\end{figure}

The {\it stability} of a critical point intuitively refers to the dynamics of trajectories in its neighborhood.  Near a stable (unstable) critical point, the dynamics can be compared to that near the minimum (maximum) of a classical potential.   In the illustration of Fig.~\ref{EH} (a), trajectories near a stable critical point are confined to the neighborhood, oscillating with small amplitude.  In panel (b), an unstable critical point behaves locally like a saddle point, with nearby trajectories deviating exponentially.  Mathematically, the stability of a critical point (or other invariant structures) has been defined in various ways to suit different purposes.  The most important ones are {\it Lyapunov}, {\it linear} and {\it spectral} stabilities.  Lyapunov stability implies linear stability, which in turn implies spectral stability \cite{Stability2}.  The linear stability has been widely used, as it is easy to calculate.  It will be the definition used in this thesis to characterize critical points; the exact derivation will be discussed in $\S$ 3.3.1.  

\newpage   
\newpage 

\noindent \underline {Periodic Orbits} \,\,\,  A {\it Periodic Orbit} (PO) is a trajectory that retraces itself with a finite period $T$:
\begin{eqnarray}
\{I_{i} (n T), \phi_{i} (n T) \} = \{I_i (0) , \phi_i (0) \} \mbox{\,\,\,\,\,\,with\,\,} n=1,2,3, \ldots  \label{PODefinition}
\end{eqnarray}
A PO is 1-dimensional invariant structure in phase space.   Unlike the critical points, the only general way to locate a PO is through an iterative numerical search \cite{FarantosReview}.  

\noindent \underline {Invariant Tori} \,\,\,   Another example of invariant phase space structures is the {\it invariant torus}.   In an $N$ DOF integrable system, if $I_i$ are the $N$ constants of motion, then their conjugate angles $\theta_i$ evolve at constant frequencies $\dot{\theta}_i=\frac{\partial H}{\partial I_i}$ for any initial condition.   An example with $N=2$ is illustrated below in Fig.~\ref{invarianttorus}.  If $\dot{\theta}_1 : \dot{\theta}_2 $ happens to be an integer ratio $p:q$ (this condition is known as being {\it commensurable}), a trajectory will close on itself after $q$ periods in direction $\theta_1$ and $p$ periods in direction $\theta_2$.  The surface of the torus is therefore covered by a family of PO \fn{See the 3D model on the accompanying CD-ROM.}.   

If the frequencies are not commensurable, a trajectory gradually fills the entire toroidal surface without closing on itself for any finite time.  These {\it quasiperiodic} trajectories form a set of nesting N-dimensional tori, filling the entire phase space.   

\newpage \begin{figure}[hbtp]
\begin{center} \includegraphics[width=3.6in]{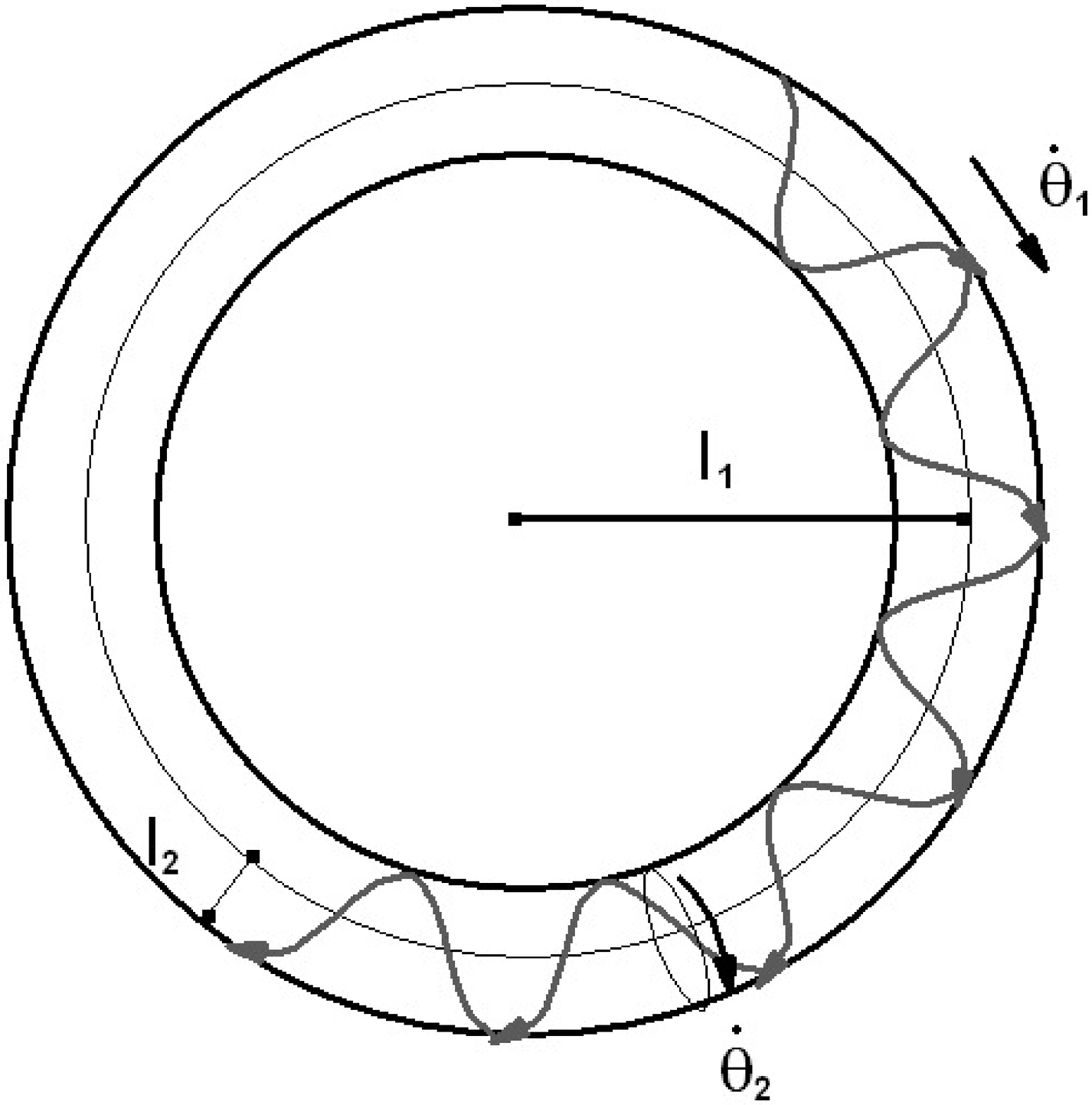}\end{center}
\cpn{Trajectory on an invariant 2-torus.} {Trajectory on an invariant 2-torus. \label{invarianttorus}}
\end{figure} 
\newpage 

When small perturbations are added to an integrable Hamiltonian, some of its invariant tori are destroyed and others become deformed.  This is the conclusion according to the Kolmogorov-Arnold-Moser theorem \cite{Tabor}.  The existence of invariant tori in a non-integrable system indicates regions where the local dynamics resembles an integrable one.

In higher dimensions, there are also the {\it normal hyperbolic invariant manifolds} \cite{EzraH2O}, which act as impenetrable barriers in the phase space \cite{PhaseSpaceBarriers}.   In an isolated system, the ($2N-1$)-dimensional constant energy shell is also an invariant structure.

\subsection{2.2.5 Bifurcations}
\addtocontents{toc}{\protect\vspace*{5pt}}

A bifurcation generally refers to any qualitative change in the phase portrait, as some external control parameters are being varied \cite{Kuznetsov}.  The ``qualitative change" is typically labeled by the change in the number and/or stability of the invariant phase space structures.  The ``external control parameters" may be either variable physical quantities (such as the energy), or the coefficients of the Hamiltonian.

\begin{figure}[hbtp]
\begin{center}  \includegraphics[width=4.98in]{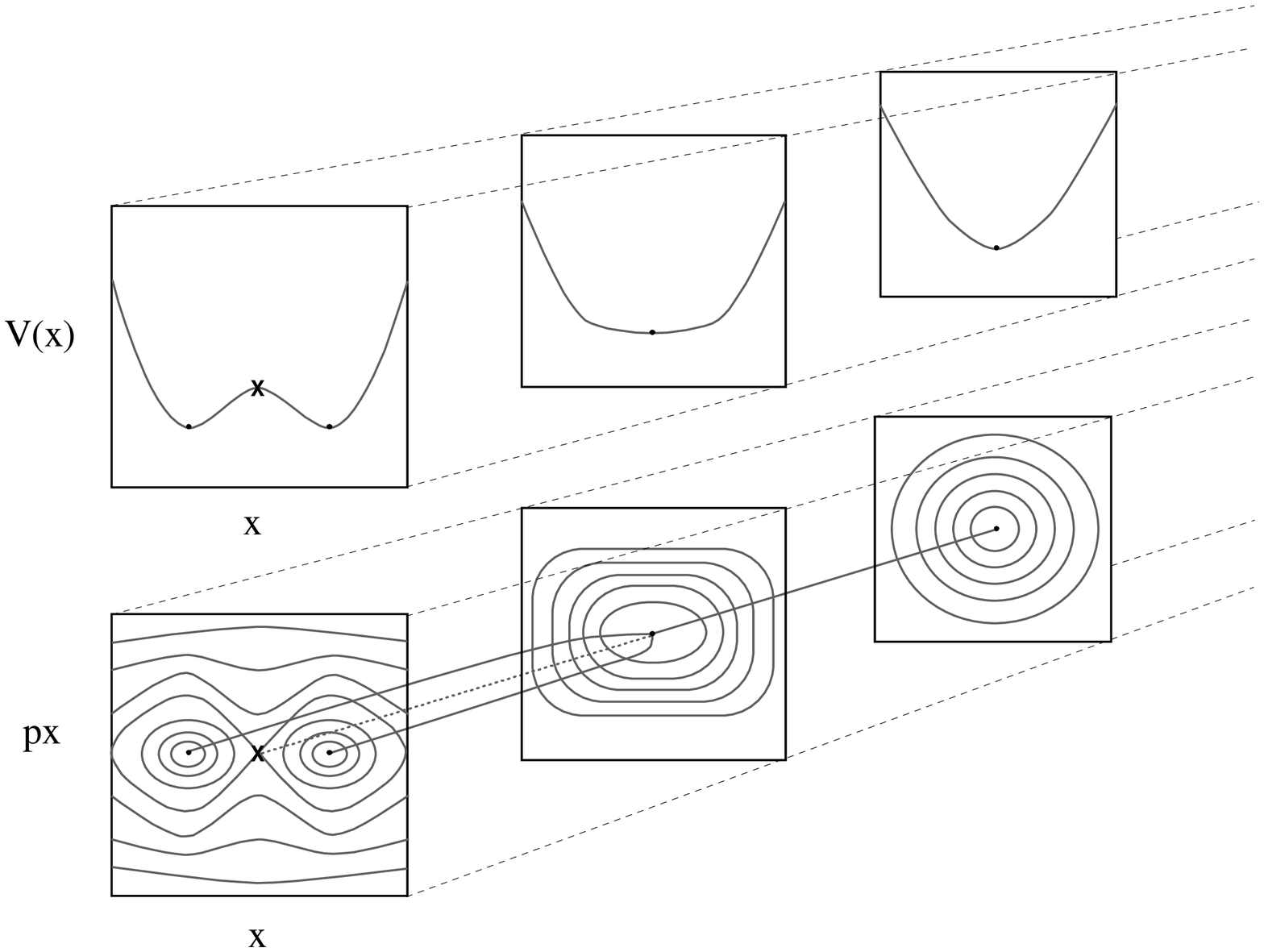}\end{center}  
\cpn{Pitchfork bifurcation in 1 DOF system} {Pitchfork bifurcation in 1 DOF system and the associated phase space change.  The right, middle and left panels are before, at and after the bifurcation point. \label{dbwellbifur}}
\end{figure}

Fig.~\ref{dbwellbifur} illustrates the {\it pitchfork bifurcation} of critical points in a 1 DOF system \cite{Kuznetsov}.    As the potential $V(x)$ in the Hamiltonian is continuously deformed, suddenly the single well (stable critical point) lifts to a barrier (unstable critical point), and two additional wells are born.   The overall phase portrait changes accordingly, adding two zones corresponding to trajectories ``trapped" in the two new minima.

\newpage   
\newpage 

In this thesis, qualitative changes in the classical phase space are tracked by bifurcations of critical points in the reduced phase space.  The parameters in the effective Hamiltonian are regarded as given, and the polyad number(s) is the variable control parameter.

\subsection{2.2.6 Poincar\'{e} Surface of Section}
\addtocontents{toc}{\protect\vspace*{7pt}}

The phase space of a 2 DOF Hamiltonian is 4-dimensional and cannot be directly graphed like in Fig.~\ref{dbwellbifur}.  However, it may be visualized as a series of 2-dimensional slices.  This technique is called Poincar\'{e} {\it Surface Of Section} (SOS) \cite{Tabor}.  

\begin{figure}[hbtp]
\begin{center}  \includegraphics[width=2.77in]{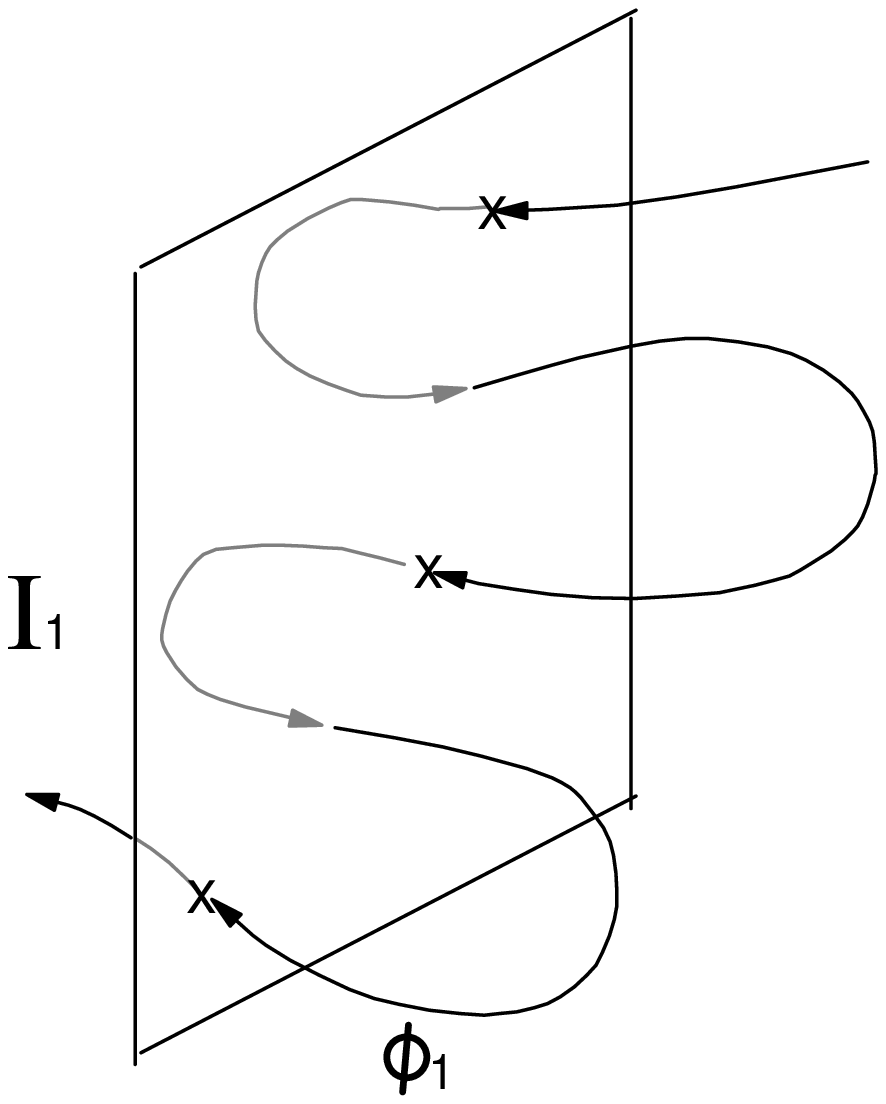}\end{center}  
\cpn{The construction of an SOS} {The construction of an SOS.  \label{SOS}}
\end{figure}

The typical construction of an SOS proceeds as follows.  Let the 4 canonical coordinates be ($I_1, \phi_1, I_2, \phi_2$).  First, an energy value of interest is determined \fn{It can also be some other constant of motion that is held fixed instead of energy-- see the footnote in $\S$ 3.1.2.}, as well as a 2-dimensional dividing surface (e.g. by setting $\phi_2$ at a constant value).  An ensemble of trajectories at this energy and starting on the dividing surface is then calculated.  Their intersections with the dividing surface are recorded by two of the other independent coordinates (e.g. $I_1, \phi_1$), as illustrated in Fig.~\ref{SOS}.  Due to time-reversal symmetry, it is sufficient to record crossings in one direction only, such as by letting ${d \phi _2}/{d t} >0$. 
\newpage   
\newpage 

\begin{figure}[hbtp]  
\begin{center}  \includegraphics[width=5.59in]{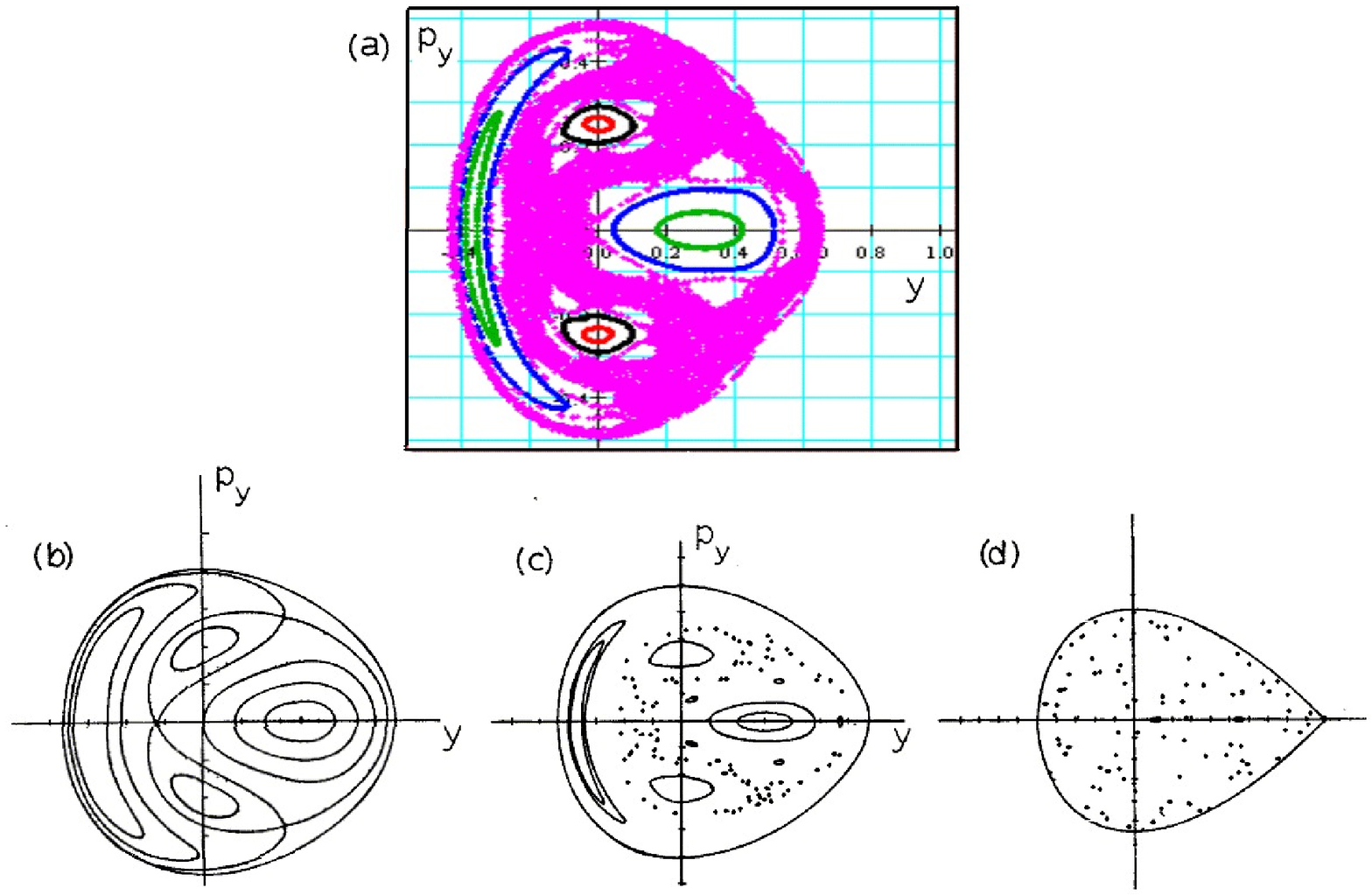}\end{center}  
\cpn{Regular, mixed and chaotic dynamics from an SOS} {Regular, mixed and chaotic dynamics from an SOS of the Henon-Heiles Hamiltonian.  Panel (a) was created in the online program at \cite{HHJava}.  It displays the marks from 5 individual trajectories coded by color.   Panels (b)-(d), adapted from \cite{Tabor}, are SOS taken at increasing energy.   The teardrop-shaped boundary in all panels is determined by the conservation of energy. \label{fig2.2}}
\end{figure}

The classical dynamics at this energy is reflected in patterns on the SOS.   Displayed in Fig.~\ref{fig2.2} are 4 SOS for the Henon-Heiles Hamiltonian \cite{HenonHeiles}, which consists of two coupled 1-dimensional oscillators
\begin{eqnarray}
H_{HH} = \frac{p_x^2+p_y^2}{2} +\frac{x^2+y^2}{2} + xy^2 -\frac{x^3}{3} 
\end{eqnarray}

\noindent In panel (a) there are two distinct types of trajectories.  For  each of the red, black, blue and green trajectories, the marks remain on a pair of closed curves.  It is because each of these trajectories lies on an invariant torus, and the two curves on the SOS are the result of slicing the torus with a plane (e.g. one that cuts along diameter $I_1$ in Fig.~\ref{invarianttorus}).  This kind of quasiperiodic motion is also known as {\it regular}.  The magenta trajectory, on the other hand, randomly fills an area complementary to the regular areas, indicating a lack of periodicity.  The corresponding random-looking trajectory is also known as {\it chaotic}.

\newpage    
\newpage 

Panels (b)-(d) of Fig.~\ref{fig2.2} are of the same Hamiltonian with increasing energy.  At the lowest energy (panel b), most of the area on the SOS has a regular pattern.  In panel (c) there are both regular and chaotic regions.  At the highest energy (d), most of the SOS is filled with chaotic trajectories.   This regular-to-chaotic trend is typical for non-integrable systems.   In this thesis, we focus on systems with mixed dynamics like that depicted in panel (c).  The important question here is how to distinguish the regular dynamics from a sea of chaos.

\section[Quantum-Classical Correspondence]{\underline{Quantum-Classical Correspondence}}
\addtocontents{toc}{\protect\vspace{0.15in}}

According to the Bohr correspondence principle, the behavior of a quantum system converges to that of the corresponding classical system when Planck's constant $\hbar \rightarrow 0$, or when the quantum number approaches infinity.   A more recent theorem by Helton and Tabor indicates that in the $\hbar \rightarrow 0$ limit, quantum eigenstates must localize into phase space regions supporting an ``invariant measure", i.e. the invariant phase space structures \cite{HeltonTabor}.    

\begin{figure}[hbtp]  
\begin{center}  \includegraphics[width=3.0in]{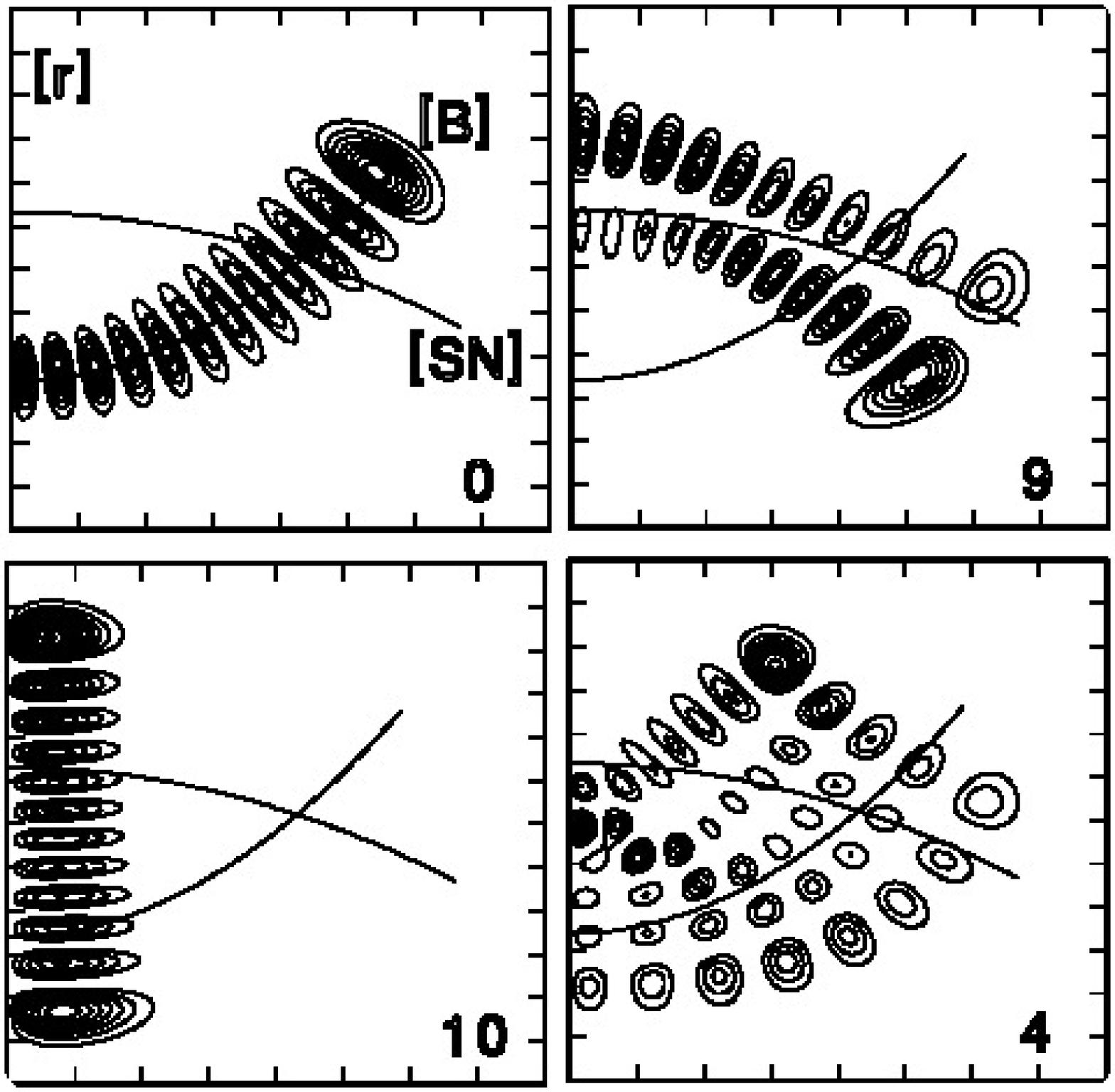}\end{center}  
\cpn{Localization of semiclassical wavefunctions} {Localization of semiclassical wavefunctions, adapted from \cite{HCP}.  The curves/lines labeled $\lbrack B \rbrack$, $\lbrack SN \rbrack$ and $\lbrack r \rbrack$ (which coincides with the left edge of upper left panel) are the periodic orbits that form the backbones of the wavefunctions.  The axes are the harmonic normal mode coordinates. \label{HCPWF}}  \end{figure}

The correspondence between classical invariant phase space structure and quantum wavefunction has also been observed at finite $\hbar$ in a phenomenon called {\it localization} \cite{QMlocalization}.  A regular region of classical phase space corresponds to eigenfunctions with nodes positioned according to the invariant phase space structures.  This alignment is illustrated in Fig.~\ref{HCPWF}.  The nodal backbones of wavefunctions closely follow the PO labeled $\lbrack r \rbrack$, $\lbrack B \rbrack$,  and $\lbrack SN \rbrack$.  Wavefunction 4 corresponds to a combination of modes $\lbrack r \rbrack$ and  $\lbrack B \rbrack$, and the nodes form a rectangular grid distorted along these directions.  

\newpage   
\newpage 

In a classically chaotic system, most eigenfunctions lack such regular patterns \cite{ReinhardtRegularChaotic}.  However, invariant phase space structures such as PO have been observed to have important influence, even when the overall dynamics is chaotic.   Examples with both stable and unstable PO are known.  The latter is called ``scarring" by Heller and co-workers \cite{HellerScarring}.   The wavefunctions could be assigned to new anharmonic modes by the PO.  

These intriguing observations of quantum-classical correspondence have led to renewed interest in using classical mechanics to understand molecular dynamics \cite{PollakReview}.  The goal is not only finding appropriate phase space structure to explain dynamics in an {\it a posteriori} manner, but also actively predicting the dynamics from analytically detected invariant phase space structures.
\addtocontents{toc}{\protect\vspace*{12pt}}
\chapter[\protect\uppercase{Methodology}]{Methodology}\label{ch.ch3}
\addtocontents{toc}{\protect\vspace*{0.2in}}

In this chapter, first we review two existing schemes of critical points analysis for systems with one polyad number in $\S$ 3.1, 3.2.  Then in $\S$ 3.3, additional questions concerning arbitrary DOF and multiple polyad numbers are addressed, in order to formulate a generalized version of the critical points analysis.   The method establishes that near a stable critical point in the reduced phase space, classical trajectories are quasiperiodic.  These critical points therefore indicate the existence of regular modes of vibration.
 
\section[Critical Points Analysis of Single $m:n$ Resonance]{\underline{Critical Points Analysis of Single $m:n$ Resonance}}
\addtocontents{toc}{\protect\vspace*{7pt}}

This analysis was developed by Kellman \et for the effective Hamiltonian consisting of two zero-order modes coupled by a single resonance.  Below, a brief overview is given on aspects that will be used in Chapter 4.  For a more detailed description, the reader is referred to \cite{KellmanReview,SvitakMNRes}.  

\subsection{3.1.1 The $m:n$ Resonance Hamiltonian}
\addtocontents{toc}{\protect\vspace*{5pt}}

In many triatomic molecules, the coupling between two vibrational modes $1$ and $2$ (not necessarily normal modes) can be approximated by an $m:n$ type resonance.  Both $m$ and $n$ are positive integers.  Eqn. (\ref{FermiExample}) is one example with $m:n=1:2$.  For the general $m:n$ resonance the effective Hamiltonian takes the following form:
\begin{eqnarray} 
\hat{H}_{mn}=\hat{H}_0(n_1, n_2) + V_{mn} \mbox{[} {(\hat{a}_1^\dagger)}^{m}{(\hat{a}_2)}^{n}+{(\hat{a}_2^\dagger)}^{n}{(\hat{a}_1)}^{m} \mbox{]}  \label{mnresquantum}
\end{eqnarray}
\noindent The second term on the right hand side corresponds to a matrix element between ZOS $\vert n_1, n_2 \rangle$ and $\vert n_1+m, n_2-n \rangle$.  This coupling destroys both $n_1$, $n_2$ as exact quantum numbers, but preserves one polyad number 
\begin{eqnarray}
P_{mn}=\frac{n_1}{m} +\frac{n_2}{n}
\end{eqnarray}

Using eqn. (\ref{Heisenberg}), a classical Hamiltonian in action-angle variables ($I_1, \phi_1, I_2, \phi_2$) is obtained from $\hat{H}_{mn}$.  Then let $\sigma$ be the largest common factor between $m$ and $n$, the following canonical transformation is carried out:
\begin{align}
I &   = \frac{\sigma}{2}\left(\frac{I_1}{m} + \frac{I_2}{n}\right), & \theta & = \frac{m\phi_1+n\phi_2}{\sigma}  \\
I_z & = \frac{\sigma}{2}\left(\frac{I_1}{m} - \frac{I_2}{n}\right), & \Psi   & = \frac{m\phi_1-n\phi_2}{\sigma}  \label{IzPsi}
\end{align}

$I$ is the constant of motion differing from $P_{mn}$ by a constant, while  $\theta$ is a cyclic angle.  $I_z$ can be regarded as a measure for the extent of mixing between the zero-order oscillators 1 and 2, and $\Psi$ as  their relative phase angle.  The classical Hamiltonian becomes
\begin{eqnarray} 
H_{mn}=H_0(I, I_z)+2V_{mn} (I+I_z)^{\frac{m}{2}} (I-I_z)^{\frac{n}{2}} \cos[\sigma \Psi] \label{mnres}
\end{eqnarray}

The 2 DOF Hamiltonian of (\ref{mnres}) is integrable, since both $H_{mn}$ and $I$ are constants of motion.  The classical phase space is reduced to 1 DOF with the equations of motion:
\begin{align}
\dot{I}_z & = -\frac{\partial H}{\partial \Psi}, & \dot{\Psi} & = \frac{\partial H}{\partial I_z}
\end{align}

\subsection{3.1.2 The Polyad Phase Sphere and Critical Points}
\addtocontents{toc}{\protect\vspace*{5pt}}

\begin{figure}[hbtp] 
\begin{center}\includegraphics[width=5.98in]{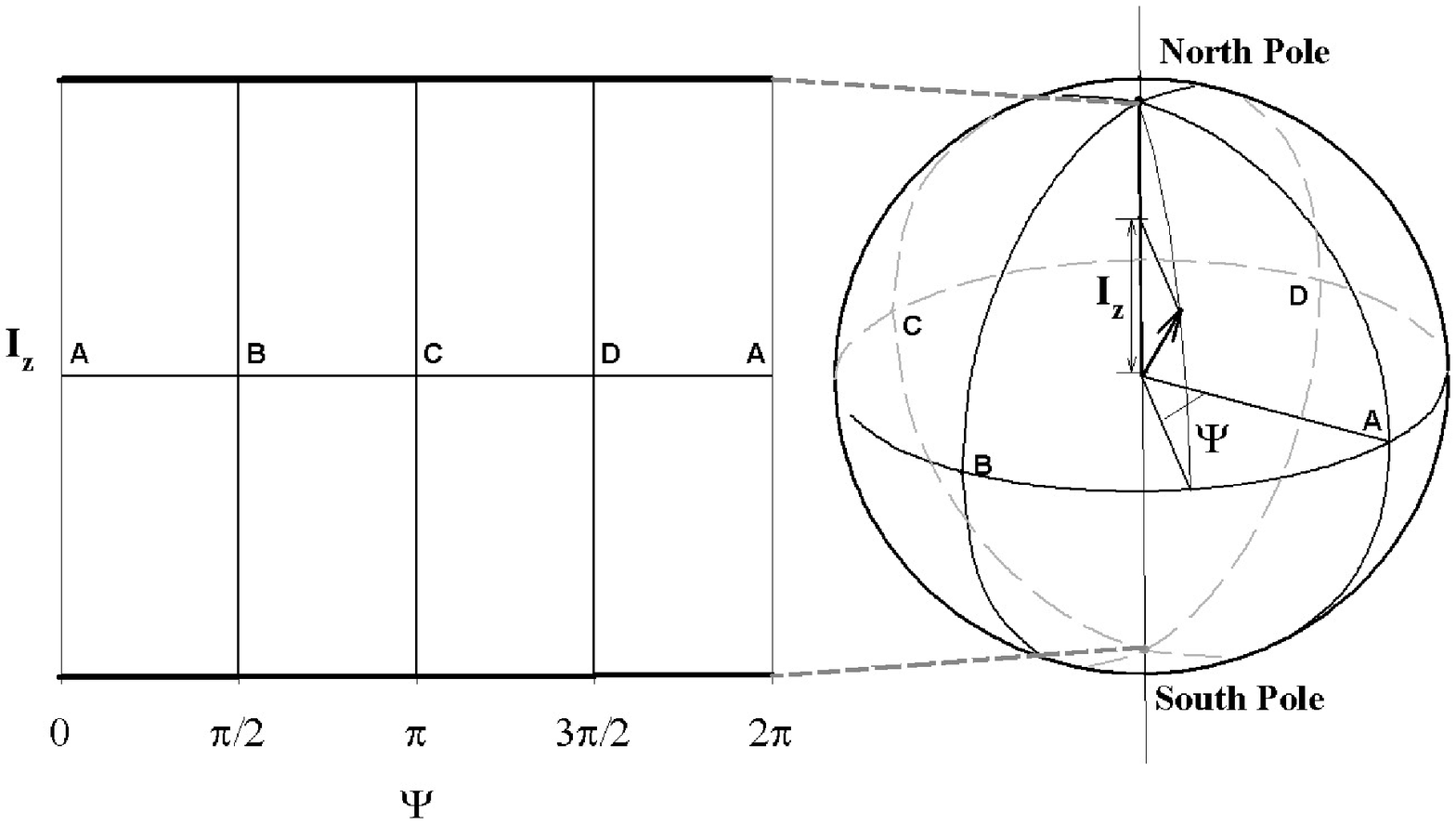}\end{center}
\cpn{Coordinates on the PPS}{Coordinates on the PPS.  The rectangular plane in panel (a) is a Mecartor projection of the spherical surface in panel (b).  Identical points are labeled by A-D in both panels to aid visualization. \label{phasesphere}} 
\end{figure}

The reduced phase space ($I_z, \Psi$) has the same topology as the surface of a 2-dimensional sphere \cite{XiaoPre}.   The sphere is called the {\it Polyad Phase Sphere} (PPS) \fn {Alternatively, ($I_z, \Psi$) could be regarded as a special SOS in the full phase space ($I, \theta, I_z, \Psi$).  Instead of energy, here $I$ is held constant.  The dividing surface is defined by a constant $\theta$.}.  As shown in Fig.~\ref{phasesphere}, the angle arccos$\lbrack {I_z}/{I} \rbrack$ is the longitude of the PPS, while $\Psi$ is its latitude.  The north pole ($I_2=0$) and the south pole ($I_1=0$) are the mode 1 and 2 overtones, respectively.  At these two points, $\Psi$ becomes unphysical, since the phase angle of an oscillator is ill defined when the action vanishes.  

\newpage   \newpage

In an integrable Hamiltonian, each eigenstate of $\hat{H}_{mn}$ is associated with an invariant torus in ($I$, $\theta$, $I_z$, $\Psi$) via {\it Einstein-Brillion-Keller} (EBK) quantization \cite{EBK}.  On the PPS (which eliminates $I$ and $\theta$), each torus appears as a closed semiclassical trajectory.  In practice, this trajectory can be well approximated (typically within 1 cm$^{-1}$) by simply solving for points on the PPS with the same energy as the quantum state \cite{HCP}.  

On the PPS, all semiclassical trajectories are organized by the critical points in the reduced phase space:
\begin{align}
\dot{I}_z & =-\frac{\partial H}{\partial \Psi}=0 \\
\dot{\Psi} &= \frac{\partial H}{\partial I_z} = 0
\end{align}
When there is no resonance ($V_{mn}=0$), all trajectories are parallel to the equator of the PPS, because $H_{0}$ has no dependence on $\Psi$.  The only critical points are then the north and south poles.  This corresponds to the trivial case where the eigenstates are assigned by the zero-order quantum numbers ($n_1,n_2$).  When the resonance is turned on, new critical points may emerge and the old ones may change their stabilities.  The semiclassical trajectories, as well as the quantum states they represent, change accordingly.  A bifurcation of the critical points  signals the birth, death and/or transformation of the vibrational modes.  

\begin{figure}[hbtp] 
\begin{center}\includegraphics[width=5.07in]{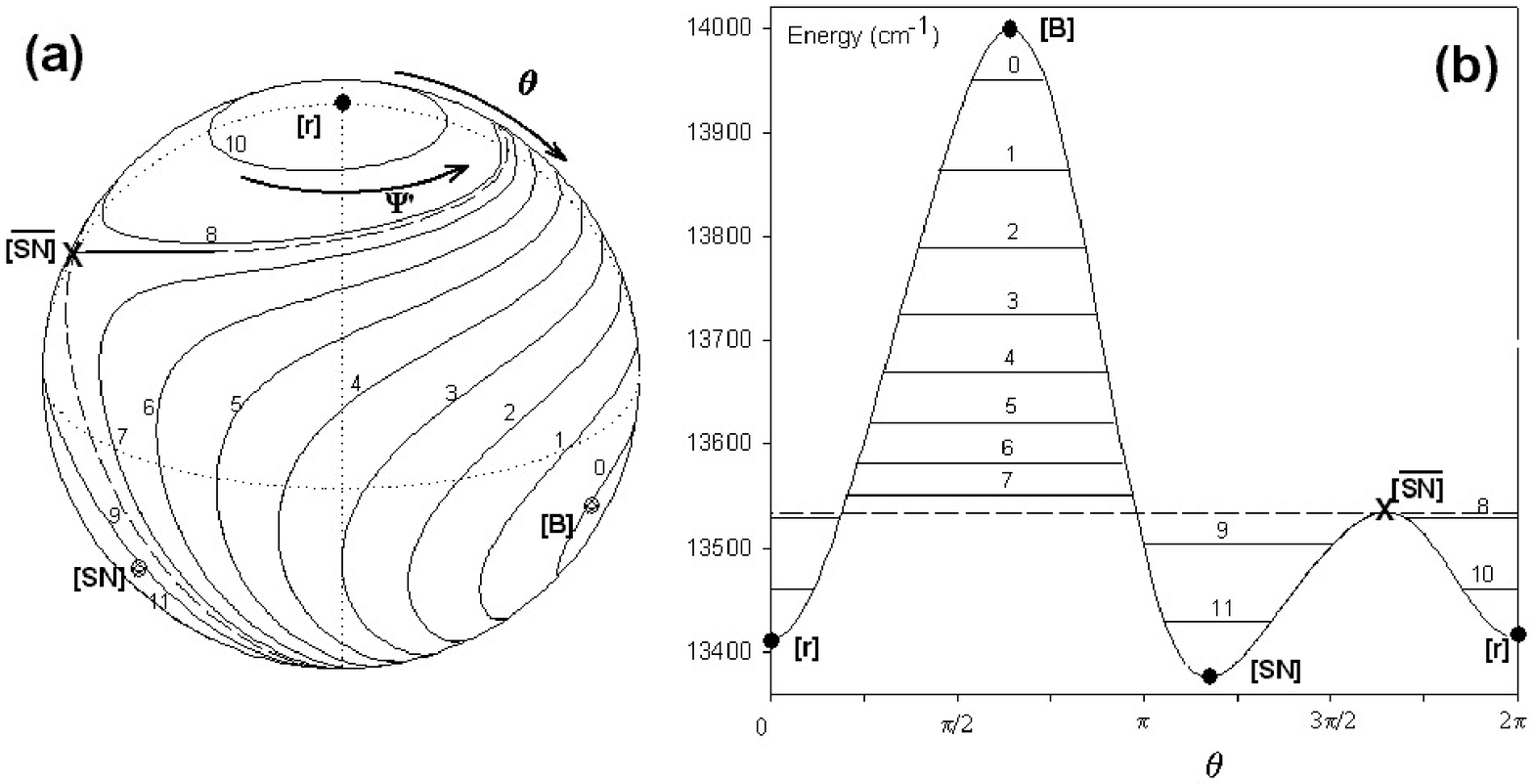}\end{center}
\cpn{PPS and semiclassical trajectories} {PPS and semiclassical trajectories from \cite{HCP}.  (a) a view of the PPS.  The points labeled $[r]$, $[B]$ and $[SN]$ are stable critical points, while $\overline{[SN]}$ is an unstable critical point.  The dashed line is the separatrix.  Panel (b) presents a cut along the great circle defined by $\Psi=0, \pi$, where $\theta \in [0,2\pi]$ is a parameter around the great circle.  Panel (b) also shows the relative energy of eigenstates in this polyad.   \label{sphereHCP}}
\end{figure}

Fig.~\ref{sphereHCP} shows a sample PPS for the HCP molecule ($m:n=1:2$).  Here modes 1 and 2 refer to the normal C-P stretch and normal H-C-P bend, respectively.   In this particular polyad $P=n_1+n_2/2 = 11$ there are 12 eigenstates, and their trajectories (labeled 0-11) are evenly spread over the surface of the PPS.  The most prominent structure on the PPS is a separatrix (dashed line) with the unstable critical point {\bf $\overline{[SN]}$} (``{\bf X}") at the center of its ``figure eight" shape.  The separatrix is so-named because it separates the phase space into three regions:  (1) occupied by trajectories 0-7 which surround the stable critical point {\bf $[B]$}; (2) occupied by trajectories 9,11 which  surround the stable critical point {\bf $[SN]$}; and (3) occupied by levels 8,10 which surround the stable critical point {\bf $[r]$} at the north pole.  Each level can be assigned to two quantum numbers: one is the polyad number $P$, the other determined by the critical point its trajectory surrounds.  Since the surface of the PPS is divided, the assignment is not uniform for all 12 states.   

\newpage    

\newpage \begin{figure}[hbtp] 
\begin{center}\includegraphics[width=4.77in]{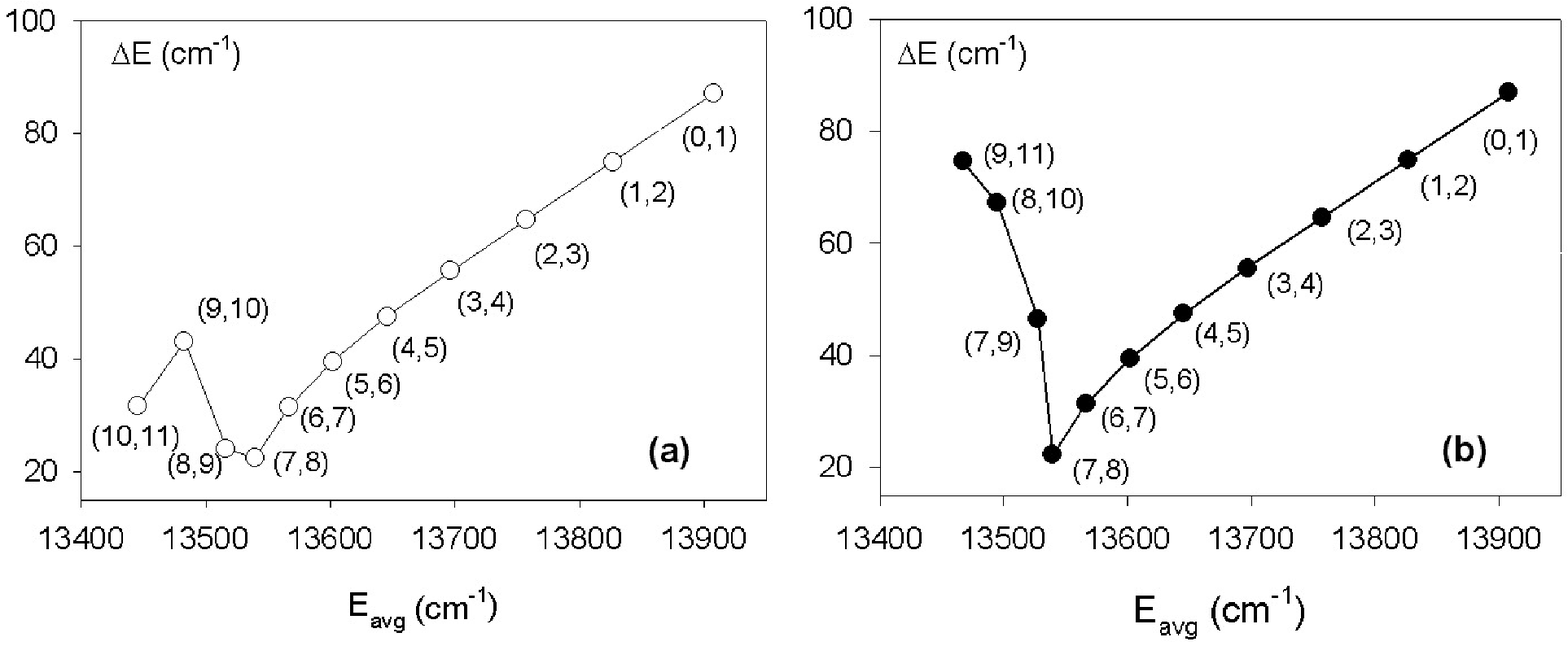}\end{center}
\cpn{Gap in the energy pattern} {Gap in the energy pattern for the eigenstates in Fig~\ref{sphereHCP}, from \cite{HCP}.  Plotted are $\Delta E = E_1-E_2$ vs. $E_{avg} = (E_1+E_2)/2$ between pairs of levels whose label are in the parentheses on the graph. \label{fig3.4}} 
\end{figure}  \newpage

In the full phase space, although $I$, $I_z$ and $\Psi$ are fixed at the critical points, the cyclic angle $\theta$ is not.  Instead, its value (modulo $2\pi$) precesses between $\lbrack 0, 2\pi \rbrack$ at a constant frequency.  Hence, a critical point in the reduced phase space corresponds to a PO in the full phase space.   

\subsection{3.1.3 Spectral Patterns}
\addtocontents{toc}{\protect\vspace*{5pt}}

Patterns in the quantum spectra are reflected in the semiclassical trajectories on the PPS.  First, the ratio between $n_1$ and $n_2$ for each state corresponds to the time-averaged $I_z$ of its trajectory.  In Fig.~\ref{sphereHCP}, for example, levels 9,11 with $I_1 \ll I_2$ are expected to have a strong bending character.  These levels are identified in experimental spectra by their large rotational constant \cite{HCPMikami}.  

Second, the separatrix on the PPS acts like a barrier in phase space.  The classical frequency traversing the top of the barrier is expected to drop to zero.  The quantum equivalence of this frequency is the energy difference $\Delta E$ between adjacent levels.  The pattern of $\Delta E$ therefore exhibits a dip when a separatrix is crossed. 

When there are more than 2 regions on the PPS (e.g. Fig.~\ref{sphereHCP}), $\Delta E$ should be taken only between eigenstates within the same region on the PPS.  If the states are sorted by energy alone, levels 9, 11 are intermingled with 8,10.  As shown in Fig.~\ref{fig3.4}, this choice creates a ``zigzag" pattern in panel (a).  The smooth dip is recovered in panel (b), when the energy differences are taken within the same zone.  This resorting procedure was first discussed by Svitak \et in \cite{SvitakPattern1}.

\subsection{3.1.4 Catastrophe Map}
\addtocontents{toc}{\protect\vspace*{5pt}}

\begin{figure}[hbtp] 
\begin{center}\includegraphics[width=5.66in]{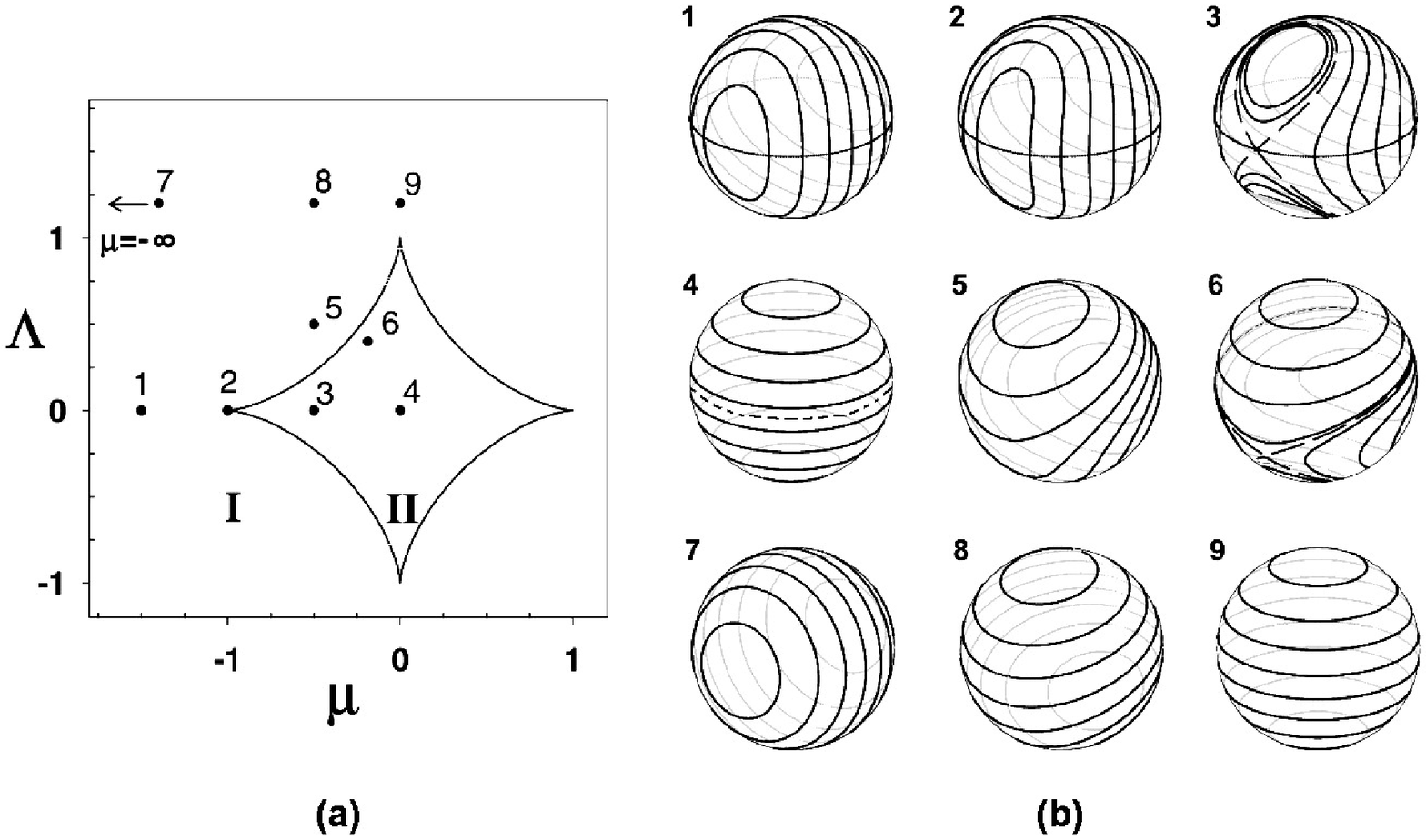}\end{center}
\cpn{Catastrophe map of $1:1$ resonance system} {Catastrophe map of $1:1$ resonance system and associated PPS, adapted from \cite{BEC}.  Panel (a) is the catastrophe map with two independent parameters being $\mu$ and $\Lambda$.  Panel (b) displays the PPS corresponding to the representative points labeled 1-9 on panel (a).   \label{catmap}}
\end{figure}

If $H_{eff}$ includes up to quadratic terms in $H_0$ and $V_{mn}$ is a constant, all possible PPS structures for a given $m:n$ can be further summarized by just two independent parameters with the help of catastrophe theory in mathematics \cite{catastrophe}.  The PPS up to a scaling factor can be reconstructed from these parameters.  This 2-parameter space (called {\it catastrophe map}) is divided into zones for any $m:n$ system \cite{SvitakMNRes}, and within each zone the PPS have the same {\it qualitative} structure.  As an example, Fig.~\ref{catmap} displays the catastrophe map and representative PPS for $m:n=1:1$ \fn{3D models of these spheres are also included on the accompanying CD-ROM.}.  In zone {\bf I}, (cases 1, 5, 7, 8 and 9) the spheres share an undivided structure, while in zone {\bf II} (cases 3, 4 and 6) the spheres are each divided by a separatrix.  In going from spheres 1-2-3-4, the bifurcation occurs at sphere 2 where its representative point crosses from {\bf I} (normal mode dynamics) to {\bf II} (local mode dynamics).  

\newpage   \newpage

One limitation of the catastrophe map is that its extension to include high-order terms is nontrivial.  As shown in $\S$ 5.2 of \cite{SvitakThesis}, the addition of a single cubic term in $H_0$ adds substantial complexity to the catastrophe map.  When the high-order terms are indeed not ignorable, a simpler alternative using the PPS and spectral patterns alone, since they contain the same amount of information.

\subsection{3.1.5 Summary}
\addtocontents{toc}{\normalfont\normalsize{\noindent Chapter \hfill Page}\protect\vspace*{30pt}}

\begin{figure}[hbtp] 
\begin{center}\includegraphics[width=3.98in]{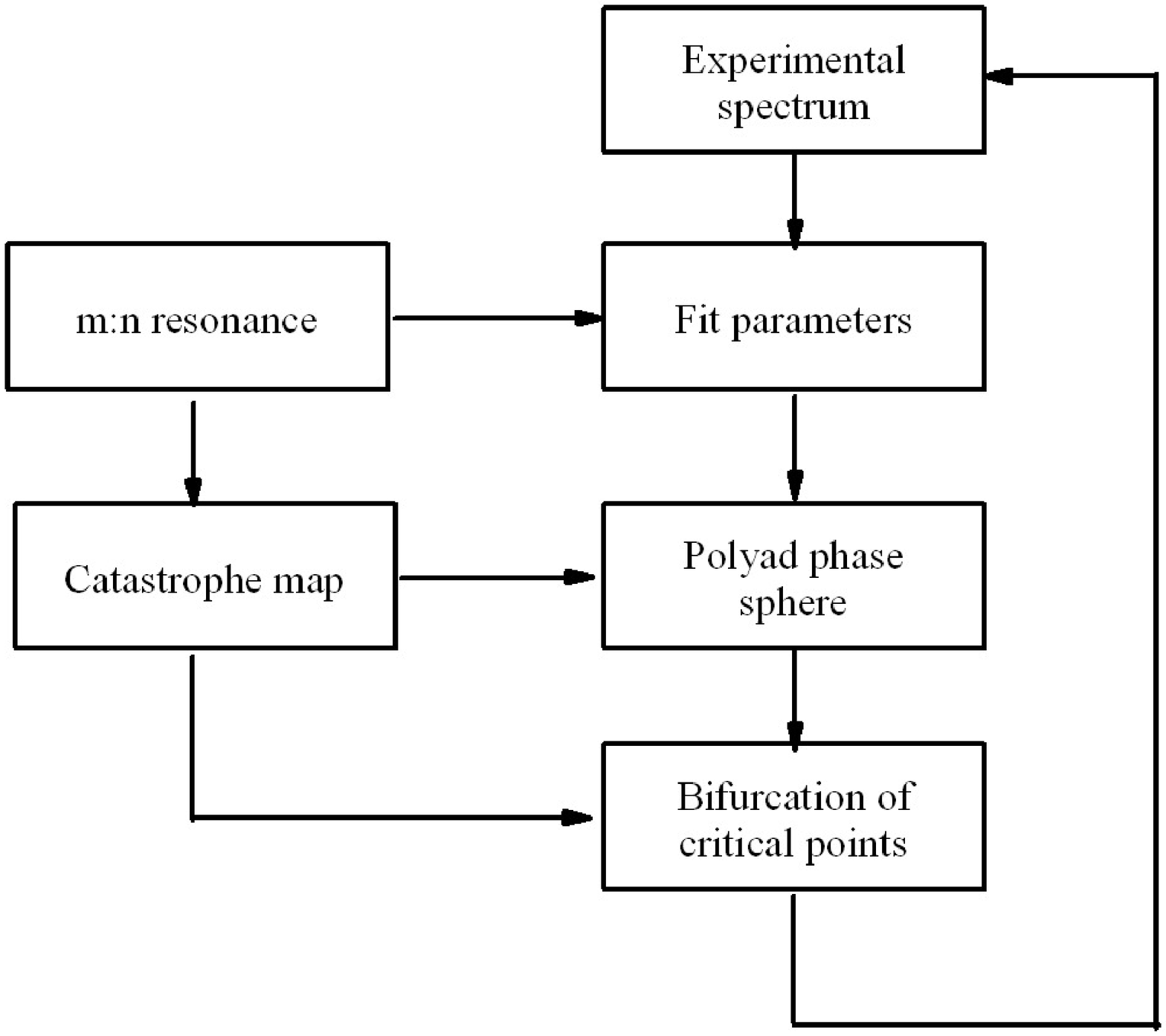}\end{center}
\cpn{Critical points analysis of the $m:n$ resonance Hamiltonian} {Critical points analysis of the $m:n$ resonance Hamiltonian, adapted from Fig.~2 of \cite{xiaocatmap} with modifications. \label{integrableflow}}
\end{figure}

The steps discussed in $\S$ 3.1.1 - 3.1.4 for the single resonance analysis are summarized in Fig.~\ref{integrableflow}.  The 2-dimensional reduced phase space is directly visualized with the PPS.  On the PPS, each quantum state corresponds to a semiclassical trajectory.  The trajectory can be assigned quantum numbers by the stable critical point it surrounds.  A separatrix (associated with an unstable critical point) causes a ``dip" in the energy gap pattern ($\Delta E$ versus $E_{avg}$) when the trajectory traverses the separatrix.  All possible divisions on the PPS with the same $m:n$ resonance can be further classified by two parameters on the catastrophe map.  

\newpage   \newpage

\section[Large-Scale Bifurcation Analysis]{\underline{Large-Scale Bifurcation Analysis}}
\addtocontents{toc}{\protect\vspace*{7pt}}


In a non-integrable Hamiltonian, the main distinction in the classical phase space structure is between the regular and the chaotic regions.  Even today, it remains a poorly understood field.  The most difficult cases have multiple resonances acting together, preventing reduction of the dynamics to less than 3 DOF.  Lu and Kellman proposed the {\it large-scale bifurcation analysis} as an extension to these non-integrable systems with one polyad number \cite{Zi-MinH2O1,Zi-MinH2O2}.   

The main assumption here is that {\it ``The large-scale bifurcation structure is defined by the lowest-order periodic orbits and their bifurcations"} \cite{Zi-MinH2O1}.  Especially, when the reduced phase space has 2 DOF, regions with different types of dynamics can be visually recognized on an SOS.   Each regular region surrounds a ``periodic orbit" on the SOS.  Here the word ``periodic" should not be confused with the continuous $T$ in eqn. (\ref{PODefinition}) for a PO. It refers to the trajectory that appears on the SOS at a few discrete points (as opposed to filling a continuous curve/area).  The period is the integer number of steps between the returns.   Those with period 1 are also known as {\it fixed points} on the SOS \fn{In existing literature, ``fixed points" and ``critical points" are often used interchangeably.  In this thesis, ``fixed points" is used in the context of a discrete mapping (such as an SOS).  ``Critical points" refer to stationery points in a {\it continuous}  dynamical system.}.    

Consider a 3 DOF system with one polyad number, such as the Baggott H$_2$O Hamiltonian in \cite{Baggott}.  The polyad number enables one to rewrite the Hamiltonian in a 4-dimensional reduced phase space ($I_1, \psi_1, I_2, \psi_2$), plus a conserved action $I_3$ and a cyclic angle $\psi_3$.  Dynamics in the reduced phase space can be visualized using a series of SOS.  Without loss of generality, let the energy and coordinate $\psi_2$ be held constant in the construction of this SOS, and the crossings of trajectories be recorded in ($I_1, \psi_1$) space.  A fixed point on the resulting SOS has all four action-angle variables ($I_1, \psi_1, I_2, \psi_2$) constant -- therefore it must be a critical point of the reduced phase space:
\begin{eqnarray}
\dot{I}_1=\dot{\psi}_1=\dot{I}_2=\dot{\psi}_2=0 \label{largescalebifur}
\end{eqnarray}
\noindent  Unless $\psi_3$ has zero frequency, these critical points are closed PO in the full phase space.   

These fixed points on the SOS therefore can be found by solving the simultaneous analytic equations (\ref{largescalebifur}).  It avoids numerical integration of many individual trajectories, as well as the subsequent problem of classifying their behavior.  

\begin{figure}[hbtp] 
\begin{center}\includegraphics[width=3.92in]{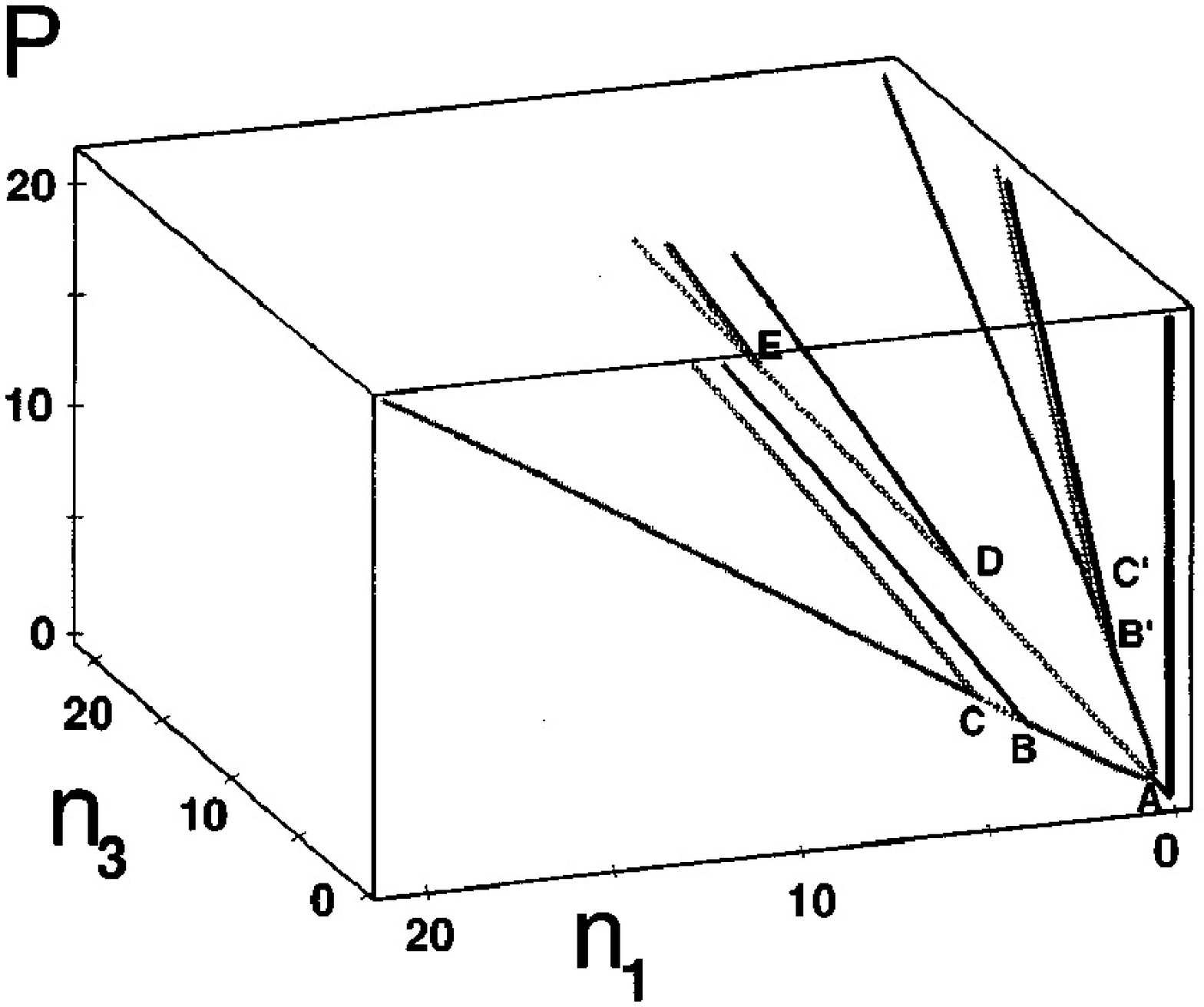}\end{center}
\cpn{Large-scale bifurcation structure in H$_2$O} {Large-scale bifurcation structure in H$_2$O, reproduced from Fig.~3 in \cite{Zi-MinH2O1}.   \label{LuKellman}} \end{figure}

Ref.~\cite{Zi-MinH2O1} solved the bifurcation structure of critical points for several triatomic systems. Fig.~\ref{LuKellman} shows the results in H$_2$O.  In the limit $P \rightarrow 0$, there are 3 branches of critical points corresponding to the 3 normal modes in H$_2$O.  In the lower right corner of Fig.~\ref{LuKellman}, the normal bend family is oriented vertically from the origin, while the two normal O-H stretch families (on top of each other in this figure) are along the short diagonal segment between the origin and point A.  As $P$ is increased, resonances cause the normal modes to bifurcate (at points A, B, B', etc.) into new families of critical points.  These critical points were then used to successfully assign all eigenstates in polyad $P=8$ to quantum numbers consistent with their vibrational dynamics \cite{Zi-MinH2O2}.  

\newpage    \newpage

Two subsequent studies \cite{EzraH2O,WigginsH2O} examined the finer details of the phase space structures in the same system.  Their agreement with \cite{Zi-MinH2O1} showed the validity of using critical points to identify the large-scale phase space structures.    

\section[Generalized Critical Points Analysis]{\underline{Generalized Critical Points Analysis}}
\addtocontents{toc}{\protect\vspace*{7pt}}

In $\S$ 3.1 and 3.2, the importance of critical points is illustrated for both integrable and nonintegrable Hamiltonians.  Two aspects are worthy of emphasizing:

\noindent \,\,\, (1) \,\,\, The existence of at least one polyad number is crucial for this analysis.  In reducing the DOF of the classical Hamiltonian,  the cyclic angle(s) not explicit in the reduced phase space provides time evolution for the critical points in the full phase space.  In contrast, a critical point defined in the {\it full} phase space usually conveys little information about the dynamics.   For example, although two coupled anharmonic oscillators may exhibit a rich range of dynamical behavior, this is not apparent from examination of the equilibrium point (usually with no motion in either oscillator).  With a {\it single} polyad number, critical points in the reduced phase space are PO in the full phase space.  These POs form the ``skeletons" of phase space \cite{EzraH2O}.  With {\it multiple} polyad numbers, the critical points are expected to have the same importance, although they now correspond to invariant tori in the full phase space. 

\noindent \,\,\, (2) \,\,\, The critical points are found by solving analytically defined equations.  Because the method does not rely on numerical integration of Hamilton's equations, it circumvents the problem induced by unstable/chaotic trajectories.  Unlike most existing nonlinear methods, the equations can be extended to arbitrary number of DOF without significant change.

Nevertheless, so far we only considered systems with 1 polyad number and up to 3 DOF.  The following three points need to be addressed in order to extend the analysis to multiple polyad numbers and arbitrary number of DOF.

\begin{enumerate}
\item In a 2 DOF system, the consistency between critical points and large-scale phase space structure may be verified by direct inspection, such as through SOS.  These visual aids are increasingly costly in higher dimensions.  Although it was suggested that the large-scale bifurcation analysis could be extended to $> 3$ DOF systems with one polyad number \cite{KellmanReview}, it remains unclear how the ``periodic orbits" defined on a SOS (see $\S$ 3.2) can be extended to arbitrary DOF. A dimensionality-independent description of the dynamics related to a critical point is strongly preferred.

\item With multiple polyad numbers, the critical points generally have multiple non-commensurable frequencies associated with the cyclic angles.  Motion at these critical points is quasiperiodic in the full phases space, instead of being closed PO.  To what extent would this difference affect the classical and quantum dynamics of the molecule?

\item In references \cite{EzraH2O,Zi-MinH2O2}, the eigenstate assignment was performed through visually identifying the localization of the Husimi distribution function of the eigenstates.   As both the computation of these semiclassical wavefunctions and the visual assignment become impractical in higher dimensions, a more general consideration of how to assign wavefunction localization is necessary.
\end{enumerate}

The next three subsections $\S$ 3.3.1-3.3.3 discuss these questions in their order.  The result is a more generalized version of the critical points analysis, which will be used in Chapter 4 on the pure bending subsystem of C$_2$H$_2$.

\subsection{3.3.1 Reduced Phase Space Trajectory Near Critical Points}
\addtocontents{toc}{\protect\vspace*{5pt}}

First, we consider an effective Hamiltonian of the most general form.   Let the Hamiltonian have a total of $N$ modes, $M$ linearly independent resonance vectors, and ($N-M$) polyad numbers.  The classical Hamiltonian after a suitable canonical transformation has $2M$ action-angle variables spanning the reduced phase space: 
$$\vec{X}=\{ x_i \} = \{ J_1, \ldots, J_M, \Psi_1, \ldots, \Psi_M \}$$
\noindent and ($N-M$) constants of motion and their conjugate cyclic angles 
$$\{ P_{M+1}, \ldots , P_N, \theta_{M+1}, \ldots , \theta_{N} \}$$

Hamilton's equations of motion in the reduced phase space can be written in the following matrix form
\begin{eqnarray}
\frac{d}{dt} \vec{X} = \left( \begin{array}{c} -\frac{\partial H}{\partial x_{M+1}} \\ \cdots \\  -\frac{\partial H}{\partial x_{2M}} \\ \, \\  \frac{\partial H}{\partial x_1} \\ \cdots \\ \frac{\partial H}{\partial x_M} \end{array} \right) = \left( \begin{array}{cc} 0 & -E_M \\ E_M & 0 \end{array} \right)  \left( \begin{array}{c} \frac{\partial H}{\partial x_1} \\ \cdots \\ \frac{\partial H}{\partial x_M} \\ \, \\ \frac{\partial H}{\partial x_{M+1}} \\ \cdots \\ \frac{\partial H}{\partial x_{2M}} \end{array} \right) \label{matrixEOM}  \end{eqnarray}
\noindent with $E_M$ being the $M \times M$ unit matrix.  A critical point $\vec{X}_0$ in the reduced phase space is defined by $2M$ simultaneous equations:
\begin{eqnarray}
\left( \frac{\partial H}{\partial x_i} \right)_{X_0} = 0 \label{GeneralCP}
\end{eqnarray}

The linear stability of $\vec{X}_0$ is defined by the behavior of the {\it linearized} equations of motion in the neighborhood.  Let the point be
\begin{eqnarray}
\vec{X}= \vec{X}_0 + \{ d x_1, \ldots, d x_i,\ldots, d x_{2M} \} = \vec{X}_0+ d \vec{X} 
\end{eqnarray}
\noindent The linearized equations of motion are obtained by expanding ${\partial H}/{\partial x_i}$ on the right hand side of eqn. (\ref{matrixEOM}) into a Taylor series, and keeping only terms linear to the displacement
\begin{eqnarray}
\left( \frac{\partial H}{\partial x_i} \right)_{X} = \left(\frac{\partial H}{\partial x_i} \right)_{X_0} + \sum_j \left( \frac{\partial^2 H}{\partial x_i \partial x_j} \right)_{X_0} d x_j  = \sum_j \left( \frac{\partial^2 H}{\partial x_i \partial x_j} \right)_{X_0} d x_j  \label{Taylor}
\end{eqnarray}
\noindent Then eqn. (\ref{matrixEOM}) is reduced to the linearized form:
\begin{eqnarray}
\frac{d}{dt} \vec{X} = \left( \begin{array}{cc} 0 & -E_M \\ E_M & 0 \end{array} \right) \left( \frac{\partial^2 H}{ \partial x_i \partial x_j } \right)_{X_0} d \vec{X} = A \cdot d \vec{X}  \label{linearEOM}
\end{eqnarray}
\noindent which is a set of homogeneous ordinary differential equations.  The standard procedure of solving them requires first finding the $2M$ eigenvalues $\lambda_i$ and eigenvectors $\vec{V}_i$ of matrix $A$ \cite{Diff}.  The $\lambda_i$ and their respective $\vec{V}_i$ satisfy
\begin{eqnarray}  A \cdot \vec{V}_i = \lambda_i \ \vec{V}_i  \end{eqnarray}
\noindent If none of the $\lambda_i$ is zero, the solutions have the following form:  
\begin{eqnarray}  \vec{X}(t)= \vec{X}_0 + \sum_{i=1}^{2M} a_i \mbox{\,\,} e^{\lambda_i t} \mbox{\,\,} \vec{V}_i  \label{linearEOMsolution} \end{eqnarray}
\noindent With $a_i$ being arbitrary complex coefficients.  The time evolution of $\vec{X}(t)$ therefore is separable into $2M$ directions, each indicated by the vector $\vec{V}_i$. 

The linear stability of $\vec{X}_0$ is defined in terms of eqn. (\ref{linearEOM}), through the eigenvalues $\lambda_i$.   In a Hamiltonian system, the conservation of phase space volume (Liouville's theorem) leads to the result that $\lambda_i$ always appear in the form of conjugate quadruplets ($\pm a \pm b i$), for which there are four cases described below.  

\noindent \,\,\, {\bf i.}  \,\,\, When a pair of $\lambda_i$ is purely imaginary ($a=0$), all solutions in eqn. (\ref{linearEOMsolution}) would oscillate in the subspace spanned by $\vec{V}_i$ with a characteristic frequency determined by $\vert \lambda_i \vert$.  The linear stability in this direction is known as {\it stable}, {\it elliptic} or (E).  

\noindent \,\,\, {\bf ii.}  \,\,\, When a pair of $\lambda_i$ is real ($b=0$), in the $\vec{V}_i$ subspace all solutions in eqn. (\ref{linearEOMsolution}) would be attracted to or repelled from $\vec{X}_0$ exponentially with time.   This direction is known as linearly {\it unstable}, {\it hyperbolic} or (H).  The names elliptic and hyperbolic originated from the shape of these linearized trajectories (Fig.~\ref{EH}).

\noindent \,\,\, {\bf iii.}  \,\,\, When $a \neq 0, b \neq 0$, the solution contains {\it both} oscillating and exponential attraction/repulsion components in the subspace spanned by the four $\vec{V}_i$ corresponding to the quadruple $\lambda_i$.  In two of the four directions the nearby trajectory ``spirals" into the critical point, while in the other two directions it ``spirals" out of the critical point.  This stability type is called {\it mixed} or (M) \cite{Zi-MinH2O1}.  

\noindent \,\,\, {\bf iv.}  \,\,\, When a pair of $\lambda_i = 0$, the stability type is degenerate (D).  In this case, the linearized equations eqn. (\ref{linearEOM}) become insufficient, and higher-order terms in the Taylor expansion are needed to evaluate the stability near a critical point.

If all the eigenvalues fall into category {\bf i.}, then the linearized trajectories defined by eqn. (\ref{linearEOMsolution}) oscillate with $M$ distinctive frequencies.   {\it Hence, near an all-stable critical point, the linearized equations of motion are quasiperiodic.}   These linearized trajectories are expected to resemble the trajectories of the nonlinear Hamiltonian $H_{eff}$ for at least a finite time.   


\subsection{3.3.2 The Presence of Multiple Cyclic Angles}
\addtocontents{toc}{\protect\vspace*{5pt}}

\begin{figure}[hbtp] 
\begin{center}\includegraphics[width=5.13in]{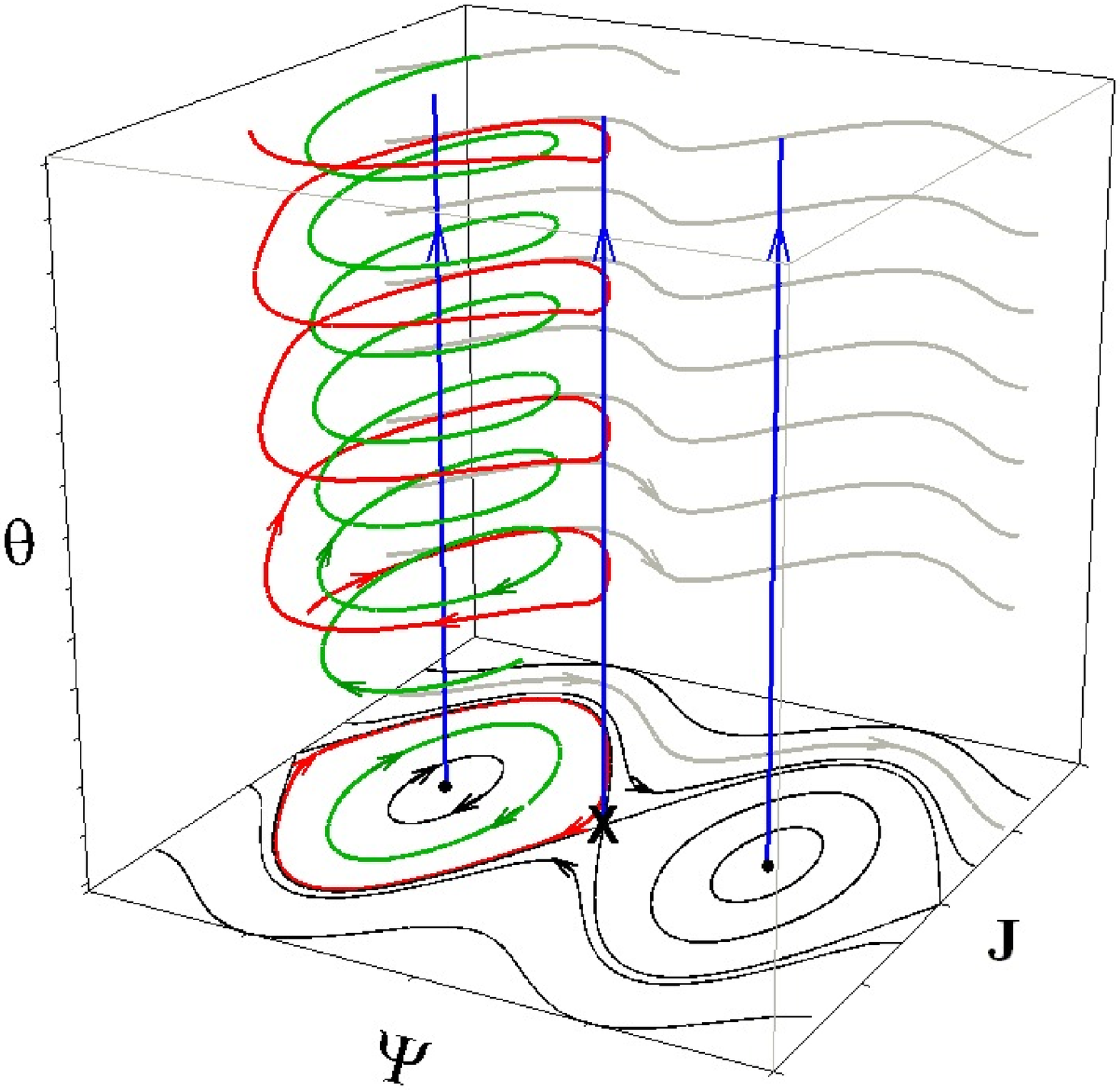}\end{center}
\cpn{Dynamics in the reduced and full phase spaces} {Dynamics in the reduced and full phase spaces, a schematic illustration.  The reduced Hamiltonian is taken to resemble Fig.~\ref{dbwellbifur} in action-angle variables ($J, \Psi$).  The angle $\theta$ is the cyclic angle, which evolves (modulo $2\pi$) between $\lbrack 0, 2\pi \rbrack$.  \label{fullphasespace}}
\end{figure}

At a critical point, all the canonical variables are fixed except the ($N-M$) cyclic angles $\theta_i$.  When there is more than one polyad number, the trajectory does not close onto itself within a finite time.  Otherwise, unless any of their frequencies becomes zero or commensurable with another, the full phase space trajectory is quasiperiodic and restricted to an ($N-M$) dimensional invariant torus.  Fig.~\ref{fullphasespace} illustrates the case with $N=2, M=1$, which is integrable.  The reduced phase space ($J, \Psi$) is a projection of the full phase space ($J, \Psi, \theta$).  Critical points in ($J, \Psi$) trace out PO in the full phase space (blue line).  A trajectory near the stable critical point (green oval at bottom) is a PO in the reduced phase space,  and a quasiperiodic motion in the full phase space.

\newpage    \newpage

Intuitively, the role these critical points play in the phase space should not change whether there are one or more cyclic angles.  As an example, consider the case of HCP where only 2 of the 3 normal modes are coupled by a Fermi resonance \cite{HCP}.  Excitation in the spectator mode $3$ (C-H stretching) can be treated as a parameter in the effective Hamiltonian. So strictly speaking, there are {\it two} polyad numbers and cyclic angles:
\begin{align}
P_1 & =n_1 + \frac{n_2}{2}, & \theta_1 &= 2\phi_1+\phi_2 \\
P_2 & =n_3, & \theta_2 &= \phi_3
\end{align}
\noindent Yet, in assigning e.g. the $n_3=1$ states, one could simply use the critical points found in this manifold, in spite of the fact that the frequency of $\theta_2$ is not zero in these polyads.

We argue that the cyclic angles $\theta_i$ in general represent a trivial aspect of the dynamics.  As far as quantum assignment is concerned, this is evident if one considers the semiclassical quantization procedure.  There $\theta_i$ appear only in a pre-factor with the form $\Pi e^{i P_i \theta_i}$ in the resulting wavefunctions \cite{TaylorCHBrClF}.  In classical mechanics, also note that $\theta_i$ do not have physical meaning on their own, since the polyad number $P_i$ are not uniquely defined (Appendix A).


Critical points in a reduced phase space, especially those with non-zero frequencies in the cyclic coordinate, have been known as {\it relative equilibria} in mathematical literature \cite{Marsden,REDyn}.  Near a relative equilibrium, classical dynamics in the full phase space can be separated into two parts: the {\it group orbit}, which is motion along the cyclic angles, and motion in the reduced phase space \cite{Gaeta}.  The latter is a  multidimensional ``slice" transverse to the group orbit \cite{REHamiltonian}.  The slice contains all the ``essential dynamics" \cite{Marsden}, in the sense that the full dynamics can be reconstructed from a point on the slice and appropriate initial conditions.  This provides a further argument against making a distinction between systems with one and multiple polyad numbers.


In the field of chemistry, relative equilibria theory has been used to classify rotationally excited molecular spectra \cite{REZhilinskii,RERoberts}.  The total angular momentum $J$ plays the same role as the polyad numbers in this thesis.  A stable relative equilibrium corresponds to the molecule rotating with a fixed shape.  Vibrational modes are defined by the {\it normal form} of the Hamiltonian in the neighborhood.  As $J$ is increased, the bifurcations of relative equilibria correspond to predictions of the (as yet unobserved) rovibrational spectral patterns. 

\subsection{3.3.3 Semiclassical Localization Near Critical Points}
\addtocontents{toc}{\protect\vspace*{5pt}}

\begin{figure}[hbtp] 
\begin{center}\includegraphics[width=4.55in]{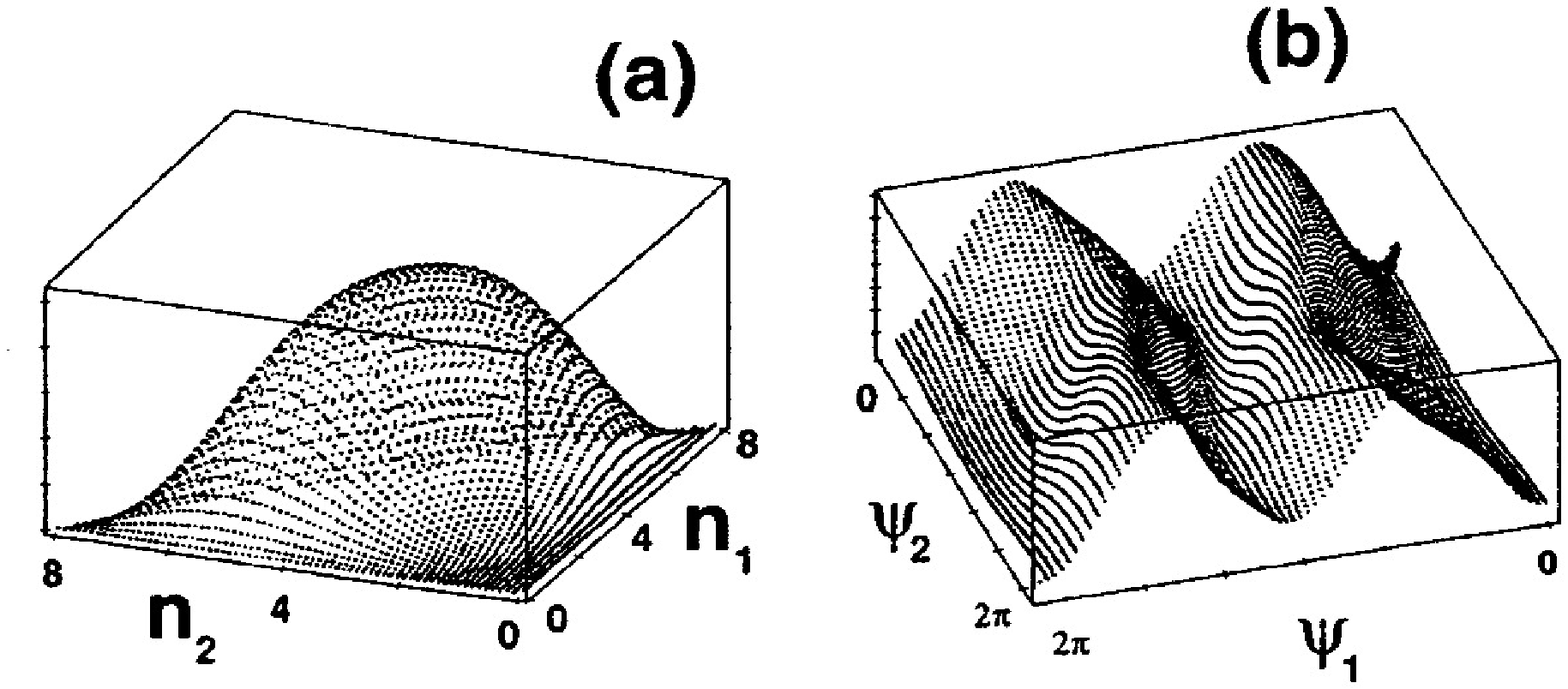}\end{center}
\begin{center}\includegraphics[width=4.55in]{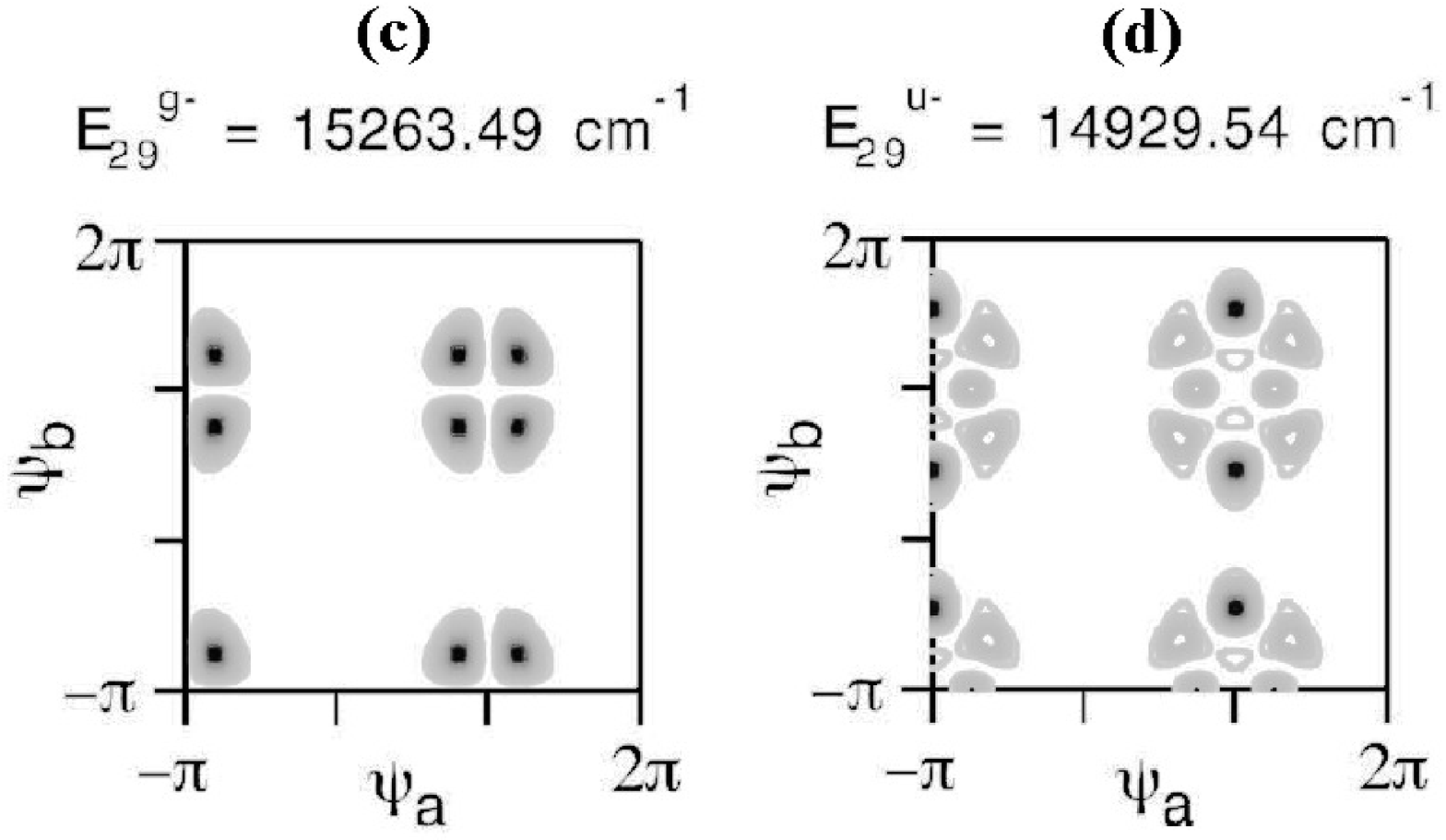}\end{center}
\cpn{Semiclassical localization near critical points} {Semiclassical localization near critical points.   Panels (a), (b) are reproduced from Fig.~3 of \cite{Zi-MinH2O2}, displaying the two projections of the Husimi function of the same eigenstate in action ($n_1, n_2$) and angle ($\psi_1, \psi_2$) space, respectively.  Panels (c) and (d) are reproduced from Fig.~5 of \cite{Jacobson15000}, which display the angle-space representation of two different wavefunctions both localized around ($\psi_a = \pm \pi, \psi_b=\pm \pi$).  \label{localization}}
\end{figure}

An eigenstate may be assigned meaningful quantum numbers based on the critical point, if its representation in the same ($\Psi_i, J_i$) space is localized near the critical point with a well-ordered nodal pattern.  The semiclassical wavefunctions can be obtained through either phase space representations (e.g. the Wigner or Husimi function \cite{Wigner}), or the $\Psi_i$ space {\it semiclassical quantization} proposed by Voth and Marcus in \cite{Marcus1985}.    Examples from both methods are illustrated in Fig.~\ref{localization}.  In panels (a) and (b), the localization occurs around the normal mode critical points at ($n_1=n_2=3.5, \psi_1=\frac{\pi}{2}$ and $\frac{3\pi}{2}$).  In panels (c) and (d), the localization is around the critical points at ($\psi_a = \pm \pi, \psi_b=\pm \pi$).

\newpage   \newpage

\begin{figure}[hbtp] 
\begin{center}\includegraphics[width=3.94in]{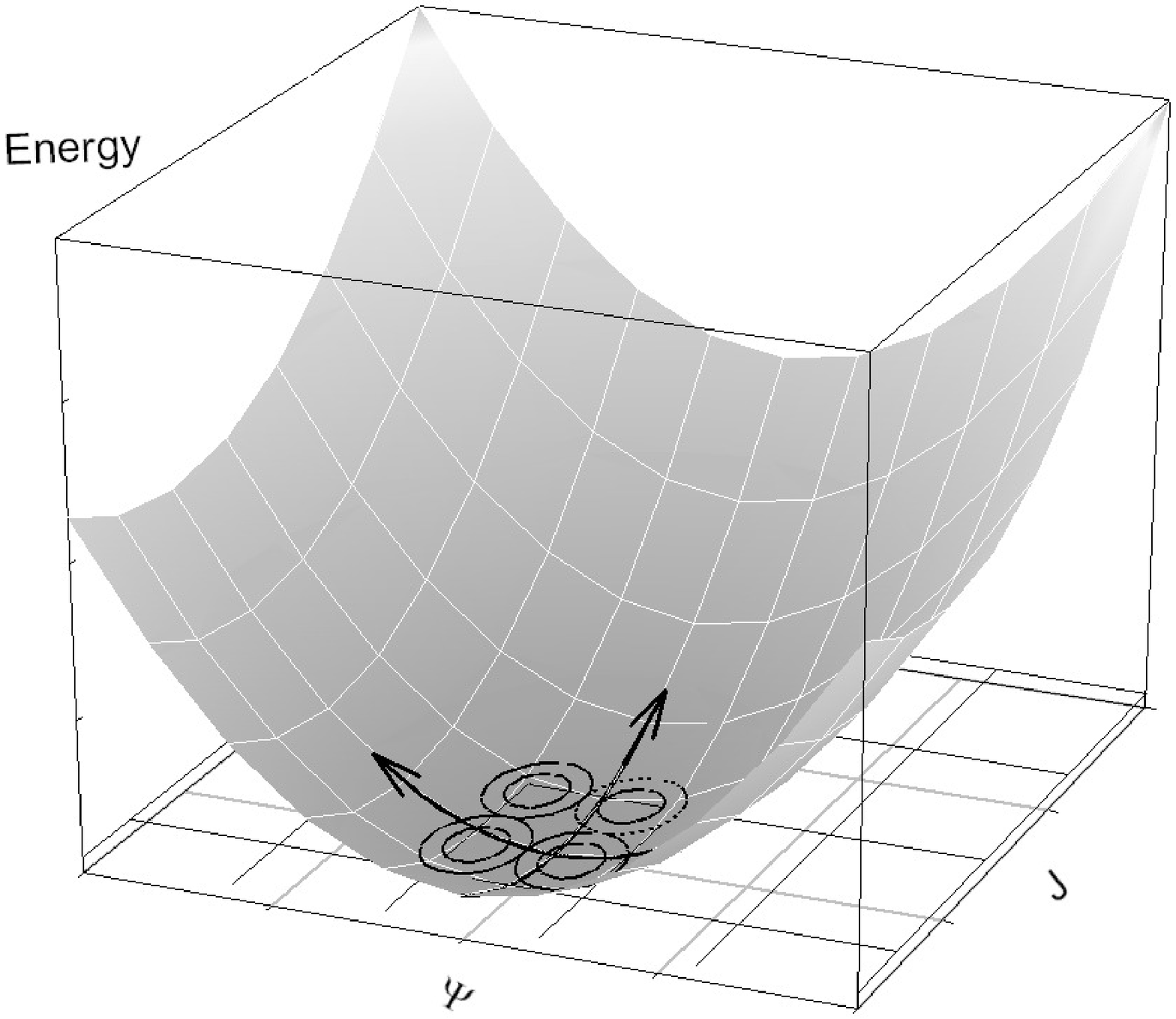}\end{center}
\cpn{Localization near a minimum in the reduced phase space} {Localization near a minimum in the reduced phase space, which is also an elliptic critical point of the Hamiltonian.  An eigenstate whose energy is close to the minimum must be localized in the nearby phase space.  \label{localillu}}
\end{figure}

Consider a local minimum or maximum (together referred to as {\it extremum}) in the reduced phase space ($\Psi_i, J_i$).  This extremum point is necessarily a critical point.   Then if there is a quantum eigenstate whose energy is nearby,  intuitively one expects the semiclassical representation of the eigenstate in either ($\Psi_i, J_i$) or ($\Psi_i$) space to localize near the critical point, simply because of the limited volume of accessible phase space into which it can expand.   This is illustrated schematically in Fig.~\ref{localillu} in the case of a minimum.  

\newpage    \newpage

There is no apparent reason why the same argument should not be valid for all choices of semiclassical representation as well as for arbitrary DOF, except for the following two scenarios.   The localization may be disrupted by quantum tunneling when the local extremum is not sufficiently prominent, or there are other local extrema nearby with similar energy. 

In $\S$ 3.3.2, it was shown that the linearized motion near an all-stable critical point is quasiperiodic.  If (1) it is a good approximation for the real classical trajectories in this region and (2) the region is large enough to support one quantizing invariant torus, then the semiclassical wavefunctions may be localized around the torus, with $M$ quantum numbers assigned by EBK quantization.    Therefore, one could expect the all-stable critical points to correspond to quantum modes around which semiclassical wavefunctions localize.   Other critical points with partial linear stability (while unstable in some directions) may also become the center of localization under favorable circumstances.

\subsection{3.3.4  Summary}
\addtocontents{toc}{\protect\vspace*{5pt}}

From $\S$ 3.3.1 - 3.3.3, we can draw the following conclusions about critical points in the reduced phase space:

\begin{enumerate}
\item Along the stable directions of a critical point, the linearized classical trajectories nearby are quasiperiodic.
\item The presence of multiple cyclic angles is not expected to affect the essential part of the classical dynamics or semiclassical localization.
\item If a critical point is also a local extremum in the reduced phase space of $H_{eff}$, then it is expected to be a center of localization for semiclassical eigenfunctions.
\end{enumerate}

Therefore, the critical points can be used to assign vibrational modes to the quantum spectra.  The change in their number and/or stability should correspond to the change in birth, death and transformations of the vibrational modes.  

\addtocontents{toc}{\protect\vspace*{12pt}}
\chapter[\protect\uppercase{Bifurcation Analysis of C$_2$H$_2$ Bends}]{Bifurcation Analysis of C$_2$H$_2$ Bends}\label{ch.ch4}%
\addtocontents{toc}{\protect\vspace{0.25in}}%

\section[Introduction]{\underline{Introduction }}
\addtocontents{toc}{\protect\vspace*{7pt}}

Acetylene (C$_2$H$_2$) is among the most-studied polyatomic molecules in spectroscopy.  Its normal mode constants on the ground electronic state (S$_0$) have been refined over a long period of time \cite{Pliva}.  At increased energy, however, its vibrational dynamics is rather complex.  Recently the highly excited vibrational states acetylene became more accessible due to new techniques such as Stimulated Emission Pumping.  The observations include additional spectral features under enhanced resolution \cite{SEPReview}, signatures of both regularity and chaos in level statistics \cite{Field1985}, and the emergence of local modes in H-C-C bending \cite{SibertC2H22,Jacobson15000,SibertC2H21,Holme-Levine,FieldLocalMode,C2H2Oss} and C-H stretching dynamics \cite{SmithWinn1,C2H2HCAO,SmithWinn2}. The stretch-bend system has also been investigated in a few studies \cite{C2H2Farantos2,RosePre,Gaspard,WuSB,DerivState,Rose2345,RoseC2H22000}.  Theoretical analysis of these results, however, remains challenging: while the coupled vibrational DOF are too numerous for many analytical methods, they are not enough to warrant a statistical approach.    

\begin{figure}[hbtp] 
\begin{center}\includegraphics[width=5.86in]{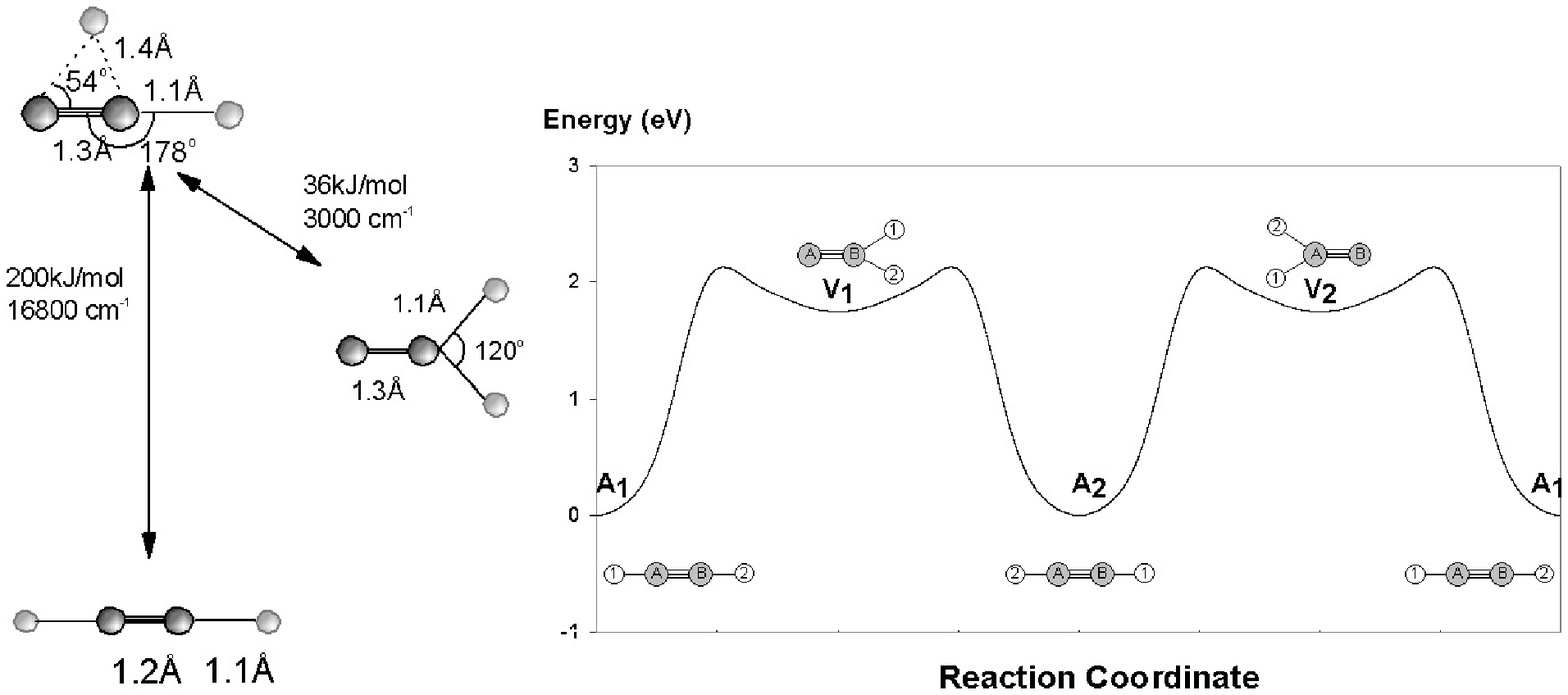}\end{center}
\cpn{Acetylene-vinylidene isomerization}{Acetylene-vinylidene isomerization on the S$_0$ electronic surface.  The energy and configurations are averaged over compiled {\it ab initio} results in Table~1 of \cite{C2H2AbInitio}.  Note that $1$ eV = $8080$ cm$^{-1}$. \label{isomerization}} 
\end{figure}

The dynamics of acetylene vibration has an important role in the combustion processes.  Its interconversion with vinylidene, a marginally stable isomer, affects the outcome of reaction rate modeling \cite{C2H2internalrotor}.  Fig.~\ref{isomerization} illustrates the geometry and energy changes during the isomerization process.  The transition state is planar and involves mostly one C-H bond in acetylene bending over.  The highest bending levels recorded by Field \et are believed to be within 5,000 cm$^{-1}$ of the transition state \cite{C2H2FieldHeff}.  Decoding the dynamics hidden in these spectra is expected to shed light on this isomerization process.  

\newpage   \newpage

For modeling highly excited states ($\approx$ 10,000 cm$^{-1}$), the effective Hamiltonians from fitting spectra are more reliable than any current {\it ab initio} PES.  The best-known PES \cite{HalonenChildCarter} is only qualitative in reproducing the experimental spectra in this energy range \cite{C2H2Guo}.  A refined PES recently published by Bowman \et is still not as accurate as a direct fit to the spectra \cite{C2H2Bowman,BowmanScaled}.

\begin{figure}[hbtp] 
\begin{center}\includegraphics[width=5.37in]{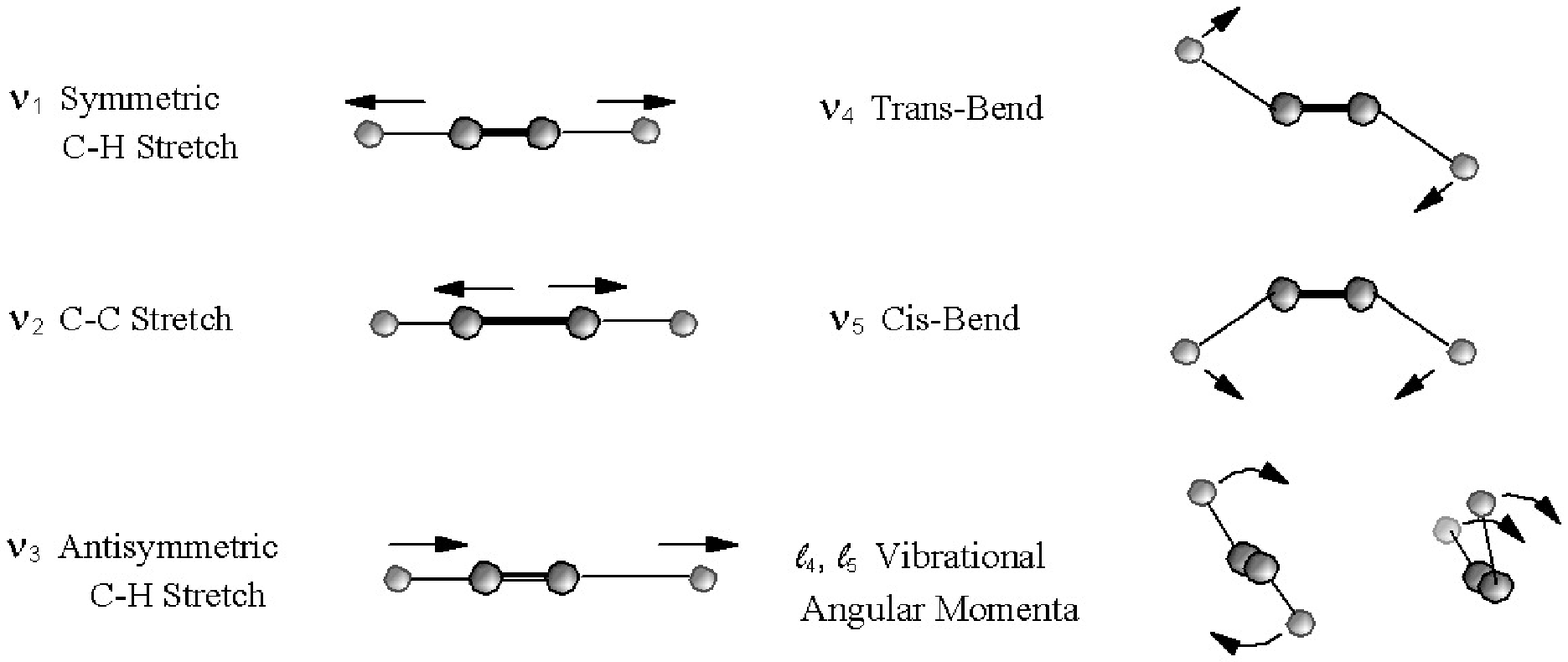}\end{center}
\cpn{Normal vibrational modes of C$_2$H$_2$}{Normal vibrational modes of C$_2$H$_2$. \label{normalmodes}} 
\end{figure}

Near its linear equilibrium configuration, acetylene has $3N-5=7$ normal modes as shown in Fig.~\ref{normalmodes}.  Both $\nu_4, \nu_5$ bends are doubly degenerate.  Two additional quantum numbers $\ell_4, \ell_5$ are used to label the respective {\it vibrational angular momenta}.  

\newpage  \newpage

All existing stretch-bend effective Hamiltonians conserve three polyad numbers: 
\begin{subequations} \label{3polyadnumbers}
\begin{align} 
N_t & = 5n_1+3n_2+5n_3+n_4+n_5 \label{3polyadnumbers1} \\ 
N_s & = n_1+n_2+n_3 \label{3polyadnumbers2} \\ 
\ell & = \ell_4 + \ell_5 \label{3polyadnumbers3}
\end{align} \end{subequations}

\noindent $N_{t}$ is the ``total" quantum number, representing the approximate integer ratio among the normal mode frequencies.  Using the frequencies of \cite{Herman1995} in units of cm$^{-1}$, the ratio is  
\begin{eqnarray}
\omega_1 : \omega_2 : \omega_3 : \omega_4 : \omega_5 = 3371.66 : 1974.76 : 3288.75 : 608.50 : 729.18  \approx 5:3:5:1:1 \nonumber
\end{eqnarray}
\noindent $N_s$ is the total number of stretching quanta.  In the absence of rotational excitation ($J=0$), $\ell$ denotes the total angular momentum of the molecule, which is always conserved.  The inversion symmetry (gerade/ungerade) and parity (+/-) are also conserved \cite{JacobsonThesis}.  Each polyad can therefore be labeled as $\lbrack N_t, N_s, \ell \rbrack ^{g/u}$.

Incidentally, all three polyad numbers remain good for the isotopomer $^{13}$C$_2$H$_2$ \cite{13C2H2}.  The doubly deuterated C$_2$D$_2$ conserves $N_s$ and $\ell$ \cite{C2D2Herman}.  Experimental spectra also suggest the existence of polyad structure in the monodeuterated C$_2$HD \cite{HCCD}.

\section[C$_2$H$_2$ Pure Bending System]{\underline{C$_2$H$_2$ Pure Bending System}}
\addtocontents{toc}{\protect\vspace*{7pt}}

\subsection{4.2.1 Quantum Effective Hamiltonian}
\addtocontents{toc}{\protect\vspace*{5pt}}

Among the $\lbrack N_t, N_s, \ell \rbrack ^{g/u}$ polyads, the pure bending states with ($N_s=0$) form a separate subsystem.  In the remainder of this chapter, these polyads are labeled by $\lbrack N_b, \ell \rbrack ^{g/u}$ with 
\begin{equation}  
N_b = N_t = n_4+n_5   
\end{equation}

The latest pure bending effective Hamiltonian was produced by Field \et  The data used in their fit include both energy and intensity information from FTIR and Dispersed Fluorescence spectra \cite{C2H2FieldHeff}.  The high-lying levels were obtained from DF spectra with a frequency resolution of 2 cm$^{-1}$, and unresolved $\ell=0/2$ states.  These levels are as much as 15,000 cm$^{-1}$ above the ground vibrational state.  The parameters in this Hamiltonian are listed in Table~\ref{bendingparams}.

\begin{table}[hbt]
\cpn{Parameters in pure bending effective Hamiltonian} {Parameters in pure bending effective Hamiltonian, from \cite{C2H2FieldHeff}.  The parameters are in units of cm$^{-1}$. \label{bendingparams}}  \begin{center} \vspace{0.2in}
\begin{tabular}{|c|c|c|c|} \hline\hline  \, & \, & \, & \, \\
$\omega_4$ & 608.657 & $y_{555}$ & 0.00955 \\  \, & \, & \, & \, \\
$\omega_5$ & 729.137 & $g_{44}$ & 0.677 \\  \, & \, & \, & \, \\
$x_{44}$ & 3.483 & $g_{45}$ & 6.670 \\  \, & \, & \, & \, \\
$x_{45}$ & -2.256 & $g_{55}$ & 3.535 \\  \, & \, & \, & \, \\
$x_{55}$ & -2.389 & $S_{45}$ & -8.574 \\  \, & \, & \, & \, \\
$y_{444}$ & -0.03060 & $r_{45}^0$ & -6.193 \\  \, & \, & \, & \, \\
$y_{445}$ & 0.0242 & $r_{445}$ & 0.0304 \\  \, & \, & \, & \, \\
$y_{455}$ & 0.0072 & $r_{545}$ & 0.0110 \\   \, & \, & \, & \, \\ \hline \hline
\end{tabular}  \end{center}  \end{table}

Using the normal mode basis $\vert n_4^{\ell_4}, n_5^{\ell_5} \rangle$ ($n_i \geq |\ell_i|$) as the ZOS, the fitting Hamiltonian has a diagonal part $\hat{H}_0$ and three resonance couplings.
\begin{align}  
\hat{H}_{bend} = \hat{H}_0 + \hat{V}_{DDI} + \hat{V}_\ell + \hat{V}_{DDII}  \label {qham}  
\end{align}
\begin{align}  
\hat{H}_{0} & = \omega_4 n_4 + \omega_5 n_5+x_{44} n_4^2+x_{45}n_4 n_5+x_{55}n_5^2 + y_{444} n_4^3 +y_{445}n_4^2 n_5 +y_{455} n_4 n_5^2  \nonumber\\
& \quad + y_{555} n_5^3 + g_{44}\ell_4^2+g_{45}\ell_4 \ell_5 + g_{55}\ell_5^2  \label{quantumzos}
\end{align}

\noindent 1. A Darling-Dennison \cite{DarlingDennison} resonance (DD-I):
{\small \begin{eqnarray}
\langle n_4^{\ell_4},n_5^{\ell_5} \vert \hat{V}_{DDI} \vert (n_4-2)^{\ell_4},(n_5+2)^{\ell_5} \rangle = \frac{S_{45}}{4} {\left[ (n_4^2-\ell_4^2)(n_5+\ell_5+2)(n_5-\ell_5+2) \right]}^{1/2}
\end{eqnarray} }

\noindent 2. An $\ell$-resonance:
{\small \begin{align}
\langle n_4^{\ell_4}, n_5^{\ell_5} \vert \hat{V}_\ell \vert n_4^{\ell_4 \mp 2},n_5^{\ell_5 \pm 2} \rangle = \frac{R_{45}}{4} \left[ \left(n_4 \mp \ell_4\right)\left(n_4 \pm \ell_4+2\right) \left( n_5 \pm \ell_5 \right) \left( n_5 \mp \ell_5+2 \right) \right]^{1/2} \label{ell}
\end{align} }
\noindent with $R_{45}=r_{45}^0+r_{445}(n_4-1)+r_{545}(n_5-1)$.

\noindent 3. Another Darling-Dennison resonance (DD-II), with matrix elements smaller than those of the previous two: 
\begin{align}
&\langle n_4^{\ell_4},n_5^{\ell_5} \vert\hat{V}_{DDII}\vert (n_4-2)^{\ell_4 \mp 2}, (n_5+2)^{\ell_5 \pm 2} \rangle = \frac{R_{45}+2g_{45}}{16} \times \nonumber\\
& \quad \quad \quad  \quad {\left[ (n_4\pm\ell_4)(n_4\pm\ell_4-2)(n_5\pm\ell_5+2)(n_5\pm\ell_5+4) \right] }^{1/2} \label{DDII}
\end{align} 

\begin{figure}[hbtp] 
\begin{center}\includegraphics[width=5.47in]{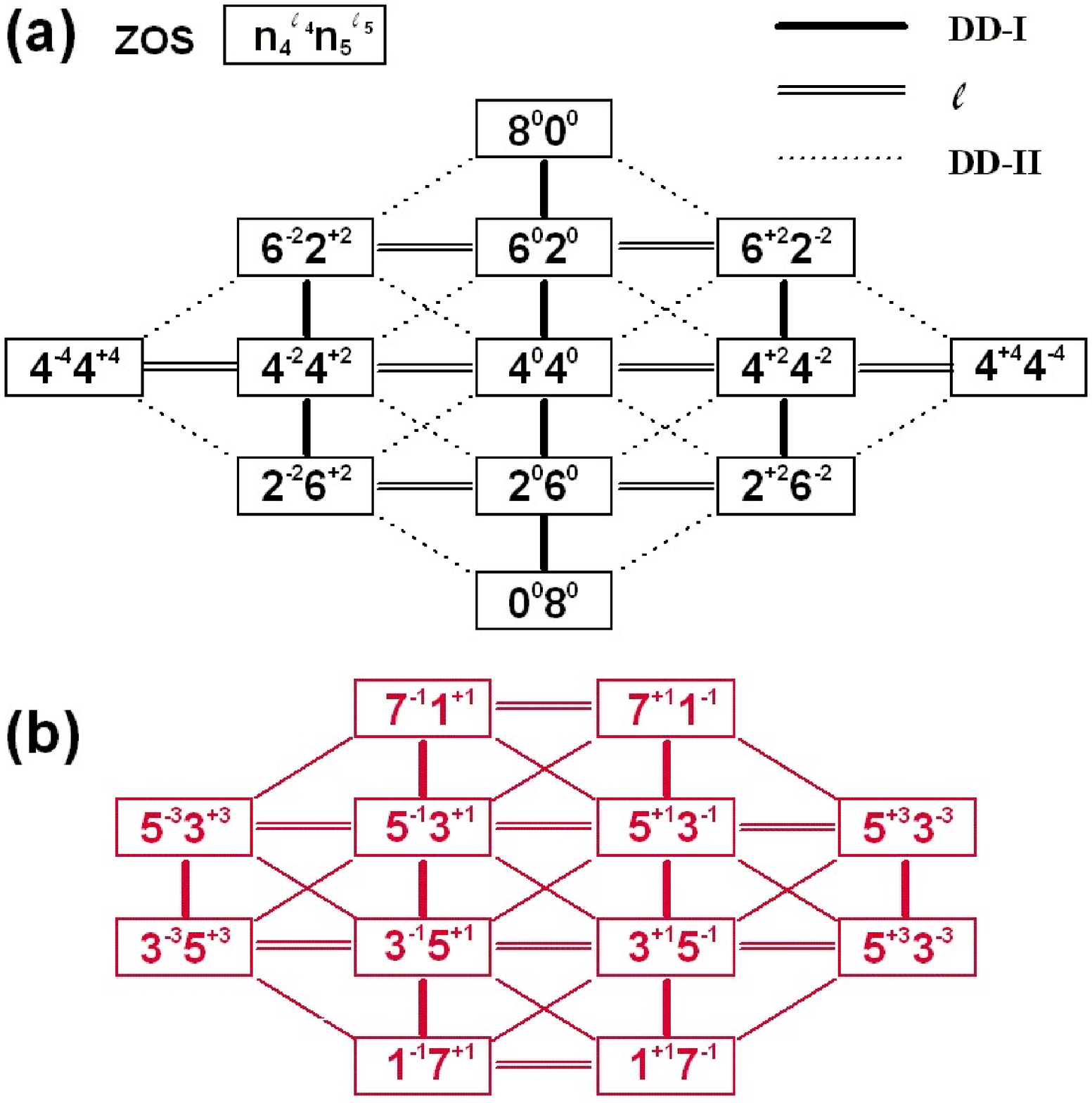}\end{center}
\cpn{Resonance couplings within a pure bending polyad}{Resonance couplings within a pure bending polyad $\lbrack 8, 0 \rbrack$.  Panel (a) displays the $g$ states and panel (b) the $u$ states. \label{diamond}} 
\end{figure}

Fig.~\ref{diamond} illustrates the manner these resonances act within a polyad.  The resonances are shown as lines connecting pairs of ZOS.  $\hat{V}_{DDI}$ couples within each column (same $\ell_4, \ell_5$), while $\hat{V}_\ell$ couples within each row (same $n_4, n_5$).  $\hat{V}_{DDII}$ in eqn. (\ref{DDII}) contains couplings along the diagonal directions.  With both $\hat{V}_{DDI}$ and $\hat{V}_\ell$, or $\hat{V}_{DDII}$ alone, all ZOS in a polyad $\lbrack N_b, \ell \rbrack ^{g/u}$ are connected into an inseparable network.   
 
\newpage    \newpage

Alternatively, $\hat{H}_{bend}$ can be written with raising/lowering operators, which act on the normal mode trans- and cis- 2-dimensional oscillators. The symmetry-adapted operators $\hat{a}_{4d}^\dagger,\hat{a}_{4g}^\dagger,\hat{a}_{4d},\hat{a}_{4g}$ and $\hat{a}_{5d}^\dagger,\hat{a}_{5g}^\dagger,\hat{a}_{5d},\hat{a}_{5g}$ \fn{The $g$ (gauche/left) and $d$ (droit/right) subscripts are named according to the manner they affect the vibrational angular momentum $\ell_i$.} are defined according to \cite{Cohen} as
\begin{align}
\hat{a}_{id}^\dagger |n_i^{\ell_i}\rangle &=\sqrt{\frac{n_i+\ell_i+2}{2}}|(n_i+1)^{\ell_i+1} \rangle,  & \hat{a}_{ig}^\dagger |n_i^{\ell_i}\rangle &=\sqrt{\frac{n_i-\ell_i+2}{2}}|(n_i+1)^{\ell_i-1} \rangle   \\
\hat{a}_{id} |n_i^{\ell_i}\rangle &=\sqrt{\frac{n_i+\ell_i}{2}}|(n_i-1)^{\ell_i-1} \rangle,  &  \hat{a}_{ig} |n_i^{\ell_i}\rangle &=\sqrt{\frac{n_i-\ell_i}{2}}|(n_i-1)^{\ell_i+1} \rangle   
\end{align}

The zero-order part of $\hat{H}_{bend}$ can be expressed with
\begin{align}
n_i &= \hat{a}_{id}^\dagger \hat{a_{id}} + \hat{a}_{ig}^\dagger \hat{a}_{ig}, &  \ell_i & = \hat{a}_{id}^\dagger \hat{a}_{id} - \hat{a}_{ig}^\dagger \hat{a}_{ig}  \label{operatorH0}
\end{align}
\noindent and the resonance terms as
\begin{align}
\hat{V}_{DDI} & = S_{45} [ \hat{a}_{4d}^\dagger \hat{a}_{4g}^\dagger \hat{a}_{5d}\hat{a}_{5g} + \hat{a}_{4d}\hat{a}_{4g}\hat{a}_{5d}^\dagger\hat{a}_{5g}^\dagger ] \\
\hat{V}_\ell & = R_{45} [ \hat{a}_{4d} \hat{a}_{4g}^\dagger \hat{a}_{5d}^\dagger \hat{a}_{5g} + \hat{a}_{4d}^\dagger \hat{a}_{4g}\hat{a}_{5d}^\dagger \hat{a}_{5g} ] \\
\hat{V}_{DDII} & = \frac{R_{45}+2g_{45}}{4} [ \hat{a}_{4d}^\dagger \hat{a}_{4d}^\dagger \hat{a}_{5d}\hat{a}_{5d} +  \hat{a}_{4g}^\dagger \hat{a}_{4g}^\dagger \hat{a}_{5g}\hat{a}_{5g} \nonumber\\  
& \quad \quad + \hat{a}_{4d} \hat{a}_{4g} \hat{a}_{5d}^\dagger \hat{a}_{5d}^\dagger + \hat{a}_{4g} \hat{a}_{4g} \hat{a}_{5g}^\dagger \hat{a}_{5g}^\dagger ] \label{operatorV}
\end{align}

\subsection{4.2.2 Classical Hamiltonian}
\addtocontents{toc}{\protect\vspace*{7pt}}

Next, a classical Hamiltonian is obtained from $\hat{H}_{bend}$ using Heisenberg's Correspondence Principle of eqn. (\ref{Heisenberg}).   The result is expressed in canonical action-angle type variables $I_i, \phi_i$ ($i=4d, 4g, 5d, 5g$).  The actions are related to the zero-order quantum numbers by
\begin{align}
I_{4d} &=(n_4+1+\ell_4)/2,   & I_{4g} &=(n_4+1-\ell_4)/2   \\
I_{5d} &=(n_5+1+\ell_5)/2,   & I_{5g} &=(n_5+1-\ell_5)/2
\end{align}
\noindent Corresponding to $N_b$ and $\ell$, there are two conserved classical actions: 
\begin{align}
I_{4d}+I_{4g}+I_{5d}+I_{5g} & = n_4+n_5+2=N_b+2 \\
I_{4d}-I_{4g}+I_{5d}-I_{5g} & = \ell_4+\ell_5 = \ell \end{align}

In order to reduce the dimensionality of the phase space, the following canonical transformation is useful \fn{These new coordinates are defined in the same way as Jacobson \et in \cite{Jacobson15000}.  The only difference is that our actions are expanded by a factor of 2, while the angles are reduced by a factor of $1/2$.  Such a difference is trivial, except ours are more compatible with the single-resonance analysis of $\S$ 4.4.1.}: 
\begin{align}
K_a & = \frac{I_{4d}+I_{4g}+I_{5d}+I_{5g}}{2}=\frac{N_b+2}{2},  & \theta_a &=\frac{\phi_{4d}+\phi_{4g}+\phi_{5d}+\phi_{5g}}{2} \nonumber \\
K_b & = \frac{I_{4d}-I_{4g}+I_{5d}-I_{5g}}{2}=\frac{\ell}{2}, & \theta_b &=\frac{\phi_{4d}-\phi_{4g}+\phi_{5d}-\phi_{5g}}{2} \nonumber \\
J_a &=\frac{I_{4d}+I_{4g}-I_{5d}-I_{5g}}{2}= \frac{n_4-n_5}{2}, & \psi_a &= \frac{\phi_{4d}+ \phi_{4g}-\phi_{5d}-\phi_{5g}}{2}  \nonumber \\
J_b &=\frac{I_{4d}-I_{4g}-I_{5d}+I_{5g}}{2}= \frac{\ell_4-\ell_5}{2}, & \psi_b &=\frac{\phi_{4d}-\phi_{4g}-\phi_{5d}+\phi_{5g}}{2}    \label{actang}
\end{align}

The resulting classical Hamiltonian is
{\small \begin{align}
&H_{bend}^\ell (K_a,K_b,J_a,J_b,\psi_a,\psi_b)= \omega_4(K_a+J_a-1)+\omega_5(K_a-J_a-1) \nonumber\\
& \quad +x_{44}(K_a+J_a-1)^2 +x_{45}(K_a+J_a-1)(K_a-J_a-1)+x_{55}(K_a-J_a-1)^2 \nonumber\\
& \quad +y_{444}(K_a+J_a-1)^3+y_{445}(K_a+J_a-1)^2(K_a-J_a-1) \nonumber\\
& \quad+y_{455}(K_a+J_a-1)(K_a-J_a-1)^2 +y_{555}(K_a-J_a-1)^3 \nonumber\\
& \quad +g_{44}{(K_b+J_b)}^2+g_{45}({K_b}^2-{J_b}^2)+g_{55}{(K_b-J_b)}^2 \nonumber\\
& \quad +\frac{S_{45}}{2}{\left[ {(K_a^2-K_b^2)}^2+(J_a^2-J_b^2)^2-2(K_a^2+K_b^2)(J_a^2+J_b^2)-8K_a K_b J_a J_b \right] } ^{\frac{1}{2}} \cos[2\psi_a] \nonumber\\
& \quad +\frac{R_{45}^*}{2}{ \left[ {(K_a^2-K_b^2)}^2+{(J_a^2-J_b^2)}^2-2(K_a^2+K_b^2)(J_a^2+J_b^2)-8K_a K_b J_a J_b \right]}^{\frac{1}{2}} \cos[2\psi_b] \nonumber\\ 
& \quad +\frac{1}{8}[R_{45}^*+2g_{45}] \{[(K_a+K_b)^2-(J_a-J_b)^2]\cos[2(\psi_a-\psi_b)]+\nonumber\\ 
& \quad \quad \quad \quad [(K_a-K_b)^2-(J_a+J_b)^2]\cos[2(\psi_a+\psi_b)] \} \label{Classical2} 
\end{align}}
\noindent with 
\begin{align}
R_{45}^* =r_{45}^0+r_{445}(K_a+J_a-2)+r_{545}(K_a-J_a-2)
\end{align}
\noindent In order for all the actions $I_{4d}$, $I_{4g}$, $I_{5d}$ and $I_{5g}$ in eqn. (\ref{actang}) to be non-negative, additional restriction are applied to the values of $J_a, J_b$:
\begin{align}
K_a - K_b  & \geq \vert J_a +J_b \vert, &  K_a + K_b  & \geq \vert J_a -J_b \vert \label{JaJbRestric}
\end{align}

Since $K_a$ and $K_b$ are constants of motion, their conjugate angles $\theta_a, \theta_b$ are the cyclic variables and therefore absent from the Hamiltonian.  However, they do evolve in time with 
\begin{align}
\dot{\theta_a}= & \frac{\partial H}{\partial {K_a}} \label{thetaa}\\
\dot{\theta_b} = & \frac{\partial H}{\partial {K_b}}  \label{thetab}
\end{align}

The reduced phase space ($J_a, \psi_a, J_b, \psi_b$) is four-dimensional.  The equations of motion in it are:
\begin{align}
&\dot{\psi}_a = \frac{\partial H}{\partial {J_a}}, & \dot{J}_a =&-\frac{\partial H}{\partial {\psi_a}} \\ 
&\dot{\psi}_b = \frac{\partial H}{\partial {J_b}}, & \dot{J}_b =&-\frac{\partial H}{\partial {\psi_b}}  \label{eom}
\end{align}

The parameters in Table~\ref{bendingparams} apply to $\ell=0, 2$, which from eqns. (\ref{actang}) correspond to $K_b=0, \frac{1}{2}$, respectively.  This chapter will focus primarily on these cases.  For $\ell=K_b=0$, eqn. (\ref{Classical2}) becomes
\begin{eqnarray}
H_{bend}(K_a,J_a,J_b,\psi_a,\psi_b)= H_0+V_{DDI}+V_{\ell}+V_{DDII}+V_{DDII}^{'} \label{Classical3} 
\end{eqnarray}
\noindent with
\begin{align}
H_0  & = \omega_4(K_a+J_a-1)+\omega_5(K_a-J_a-1)+x_{44}(K_a+J_a-1)^2 \nonumber\\ 
&  \quad +x_{45}(K_a+J_a-1)(K_a-J_a-1)+x_{55}(K_a-J_a-1)^2+y_{444}(K_a+J_a-1)^3\nonumber\\ 
&  \quad +y_{445}(K_a+J_a-1)^2(K_a-J_a-1)+y_{455}(K_a+J_a-1)(K_a-J_a-1)^2 \nonumber\\
&  \quad +y_{555}(K_a-J_a-1)^3+(g_{44}-g_{45}+g_{55}) J_b^2  \label{ClassicalH0}
\end{align}
\noindent and
\begin{align}
V_{DDI} & = \frac{S_{45}}{2} { \left[ K_a^4+(J_a^2-J_b^2)^2-2K_a^2(J_a^2+J_b^2) \right]}^{1/2}\cos[2\psi_a]  \label{DDIterm} \\
V_{\ell} & = \frac{R_{45}^*}{2}{ \left[ K_a^4+(J_a^2-J_b^2)^2-2K_a^2(J_a^2+J_b^2) \right]}^{1/2} \cos[2\psi_b] \label{ellterm} \\ 
V_{DDII} & = \frac{1}{8}(R_{45}^*+2g_{45})[K_a^2-(J_a-J_b)^2]\cos[2(\psi_a-\psi_b)]  \\ 
V_{DDII}^{'} & = \frac{1}{8}(R_{45}^*+2g_{45}) [K_a^2-(J_a+J_b)^2] \cos[2(\psi_a+\psi_b)]
\end{align}

\section[Critical Points Analysis]{\underline{Critical Points Analysis}}
\addtocontents{toc}{\protect\vspace*{7pt}}

Following the method outlined in Chapter 3, we now explicitly solve for the critical points for $\ell=0$ in eqn. (\ref{Classical3}) and their bifurcations with variation of $N_b$.   Four new families of critical points are found at increasing $N_b$ when the initially stable normal modes become unstable in distinct bifurcations.  Two of the new families, namely the Local and Counter Rotator critical points, are linearly bi-stable (EE).  They correspond to new stable modes of bending vibration.

Readers not interested in details of the calculation may skip the next subsection and go directly to $\S$ 4.3.2 for the results. 

\subsection{4.3.1 Computational Details}
\addtocontents{toc}{\protect\vspace*{5pt}}

The critical points of eqn. (\ref{Classical3}) are defined by four simultaneous equations:
\begin{align}
\dot{J_a} &=-\frac{\partial H_{bend}}{\partial \psi_a} =0   \label{deriYa}  \\
\dot{J_b} &=-\frac{\partial H_{bend}}{\partial \psi_b} =0   \label{deriYb}  \\
\dot{\psi_a} &=\frac{\partial H_{bend}}{\partial J_a}  =0   \label{deriJa} \\
\dot{\psi_b} &=\frac{\partial H_{bend}}{\partial J_b}  =0   \label{deriJb} 
\end{align}
\noindent To simplify the notation we let  
\begin{align}
\Lambda  = [ K_a^4-2K_a^2(J_a^2+J_b^2)+(J_a^2-J_b^2)^2 ]^{1/2}  
\end{align}
\noindent The partial derivatives in eqns. (\ref{deriYa}-\ref{deriJb}) become 
{\small \begin{align}
\frac{\partial H_{bend}}{\partial \psi_a} = & -\Lambda S_{45} \sin[2 \psi_a] - \frac{1}{4}(R_{45}^*+2g_{45}) \{ [ K_a^2-(J_a-J_b)^2]\sin[2(\psi_a-\psi_b)]+ \nonumber \\
& \quad  [K_a^2-(J_a+J_b)^2] \sin[2(\psi_a+\psi_b)] \}  \label{greatcircle1}  \\
\frac{\partial H_{bend}}{\partial \psi_b} = & - \Lambda R_{45}^*  \sin[2 \psi_b] + \frac{1}{4}(R_{45}^*+2g_{45}) \{ [K_a^2-(J_a-J_b)^2] \sin[2(\psi_a-\psi_b)]- \nonumber \\
& \quad  [K_a^2-(J_a+J_b)^2] \sin[2(\psi_a+\psi_b)] \}  \label{greatcircle2} \\
\frac{\partial H_{bend}}{\partial J_a} = & \frac{\partial H_0}{\partial J_a}-\frac{J_a}{\Lambda} (K_a^2-J_a^2+J_b^2) (S_{45}\cos[2\psi_a]+R_{45}^*\cos[2\psi_b])+ \Lambda(r_{445}-r_{545})\cos[2\psi_b] \nonumber\\
& \quad -\frac{R_{45}^*+2g_{45}}{4} \{(J_a-J_b)\cos[2(\psi_a-\psi_b)] + (J_a+J_b) \cos[2(\psi_a+\psi_b)] \} \nonumber\\
& \quad + \frac{r_{445}-r_{545}}{2} \{ 2 J_a J_b \sin[2\psi_a] \sin[2\psi_b]+(K_a^2-J_a^2-J_b^2)\cos[2\psi_a]\cos[2\psi_b] \}  \label{greataction1} \\
\frac{\partial H_{bend}}{\partial J_b} = & 2(g_{44}-g_{45}+g_{55}) J_b -\frac{J_b}{\Lambda} (K_a^2+J_a^2-J_b^2)(S_{45} \cos[2\psi_a]+R_{45}^*\cos[2\psi_b]) \nonumber\\
& \quad +\frac{R_{45}^*+2g_{45}}{4} \{ (J_a-J_b)\cos[2(\psi_a-\psi_b)] - (J_a+J_b)\cos[2(\psi_a+\psi_b)] \} \label{greataction2}
\end{align} }

A visual inspection of the first two equations (\ref{greatcircle1}, \ref{greatcircle2}) reveals a {\it sufficient condition} \fn{An additional root search was carried out, and no extra critical points were found beyond the ones discussed here.} for both of them to vanish: 
\begin{align}
\sin[2\psi_a]  = \sin[2\psi_b]=0 \label{gc1} 
\end{align}
\noindent or 
\begin{align}
(\psi_{a} , \psi_{b} )=(\frac{m \pi}{2}, \frac{n \pi}{2}) \mbox{\,\,\,\,\,\,\,\,\,\, with } m,n=0,1,2,3 \label{greatcircles2}
\end{align}

Eqn. (\ref{greatcircles2}) leads to the 16 combinations of ($\psi_a$, $\psi_b$) in Table~\ref{16cond}.  The remaining two equations (\ref{greataction1}, \ref{greataction2}) are then solved for $J_a$ and $J_b$, with ($\psi_a$, $\psi_b$) held at these discrete values.  Also, note that eqns. (\ref{greataction1}, \ref{greataction2}) are invariant with regard to one or both the following two transformations 
\begin{align}
\psi_a & \rightarrow \psi_a + \pi &  \psi_b & \rightarrow \psi_b + \pi
\end{align}
\noindent It is sufficient to consider only one out of the four sets of values on each row in Table~\ref{16cond}.  Here we use:
\begin{align}  
\left( \psi_a, \psi_b \right) = \left(0,0 \right), \left(0,\frac{\pi}{2} \right), \left(\frac{\pi}{2},0 \right), \left(\frac{\pi}{2},\frac{\pi}{2} \right) \label{CR} 
\end{align}  

\begin{table}[htb] 
\cpn{($\psi_a$, $\psi_b$) values of bending critical points} {($\psi_a$, $\psi_b$) values of bending critical points. \label{16cond}}
\begin{center} \vspace{0.2in}  \begin{tabular}{|c|c|c|c|} \hline\hline
($\psi_a$, $\psi_b$) & ($\psi_a$, $\psi_b$) & ($\psi_a$, $\psi_b$) & ($\psi_a$, $\psi_b$) \\ \hline   & \, & \, & \, \\
(0, 0) & (0, $\pi$) & ($\pi$, 0) & ($\pi$, $\pi$)   \\ \hline \, & \, & \, & \, \\
(0, $\frac{\pi}{2}$) & (0, $\frac{3\pi}{2}$) & ($\pi$, $\frac{\pi}{2}$) & ($\pi$, $\frac{3\pi}{2}$) \\ \hline  \, & \, & \, & \, \\
($\frac{\pi}{2}$, 0) & ($\frac{3\pi}{2}$, $\pi$) & ($\frac{\pi}{2}$, 0) & ($\frac{3\pi}{2}$, $\pi$) \\ \hline  \, & \, & \, & \, \\
($\frac{\pi}{2}$, $\frac{\pi}{2}$) & ($\frac{\pi}{2}$, $\frac{3\pi}{2}$) & ($\frac{3\pi}{2}$, $\frac{\pi}{2}$) & ($\frac{3\pi}{2}$, $\frac{3\pi}{2}$) \\  \hline \hline  \end{tabular}  \end{center}  \end{table}

\begin{figure}[hbtp]   
\begin{center}\includegraphics[width=5.68in]{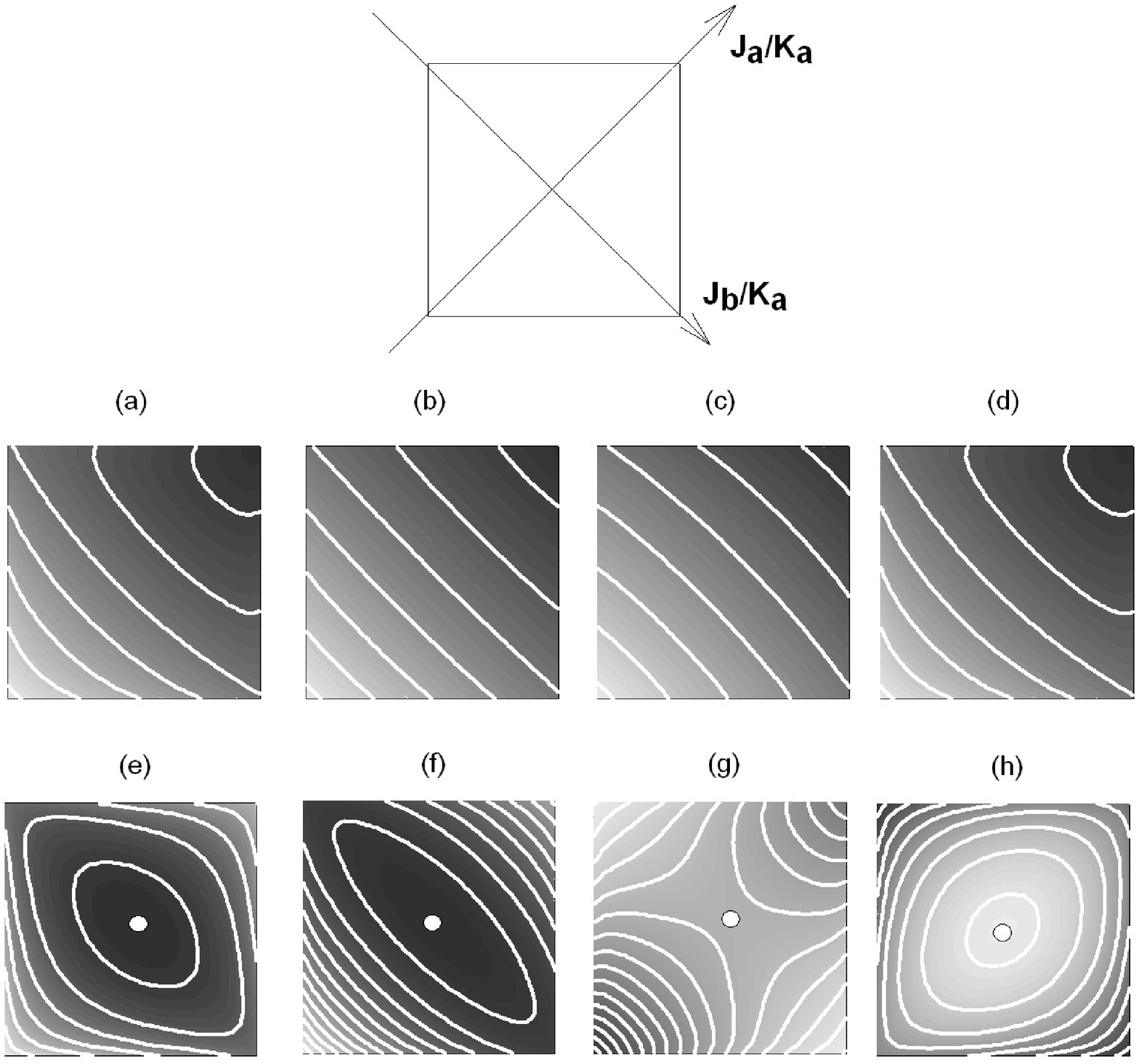}\end{center}
\cpn{Pseudopotentials of $\lbrack 6, 0 \rbrack$ and $\lbrack 22, 0\rbrack$ }{Pseudopotentials of $\lbrack 6, 0 \rbrack$ and $\lbrack 22, 0 \rbrack$, as contour plots of $H_{bend}$.   The gray scale shading is the lightest where $H_{bend}$ is at the largest.  Panels (a-d) depict the pseudopotentials in $N_b=6$ with the four values of ($ \psi_a, \psi_b$) in eqn. (\ref{CR}), respectively.  Panels (e), (f), (g) and (h) depict the pseudopotentials in $N_b=22$ with the same order in ($ \psi_a, \psi_b$).   \label{pseudo}} 
\end{figure}

The simultaneous eqns. (\ref{greataction1}, \ref{greataction2}) may be visually solved using the following ``pseudopotential" approach.   Here it is generalized from the 1-dimensional case in \cite{HCP} to 2-dimensional.   When $H_{bend}$ is viewed as a function of $J_a, J_b$ with the discrete ($\psi_a, \psi_b$) values in eqn. (\ref{CR}), the solutions satisfying both (\ref{greataction1}, \ref{greataction1}) are the ``flat spots" (e.g. minima, maxima and saddle points), and can be identified visually on plots of $H_{bend}$ (the {\it pseudopotential}).   Fig.~\ref{pseudo} depicts the pseudopotentials for polyads $\lbrack 6, 0 \rbrack$ and $\lbrack 22, 0\rbrack$.  The $J_a$ and $J_b$ values (already constrained by eqn. \ref{JaJbRestric}) are divided over $K_a$ to scale their range to $\lbrack -1, 1 \rbrack$.   In panels (a)-(d), there are no flat spots in the interior of the ($J_a/K_a, J_b/K_b$) space when $N_b=6$.    In panels (e)-(h), under each of the four ($\psi_a, \psi_b$) conditions there is a flat spot (white dot with black rim) corresponding to a critical point.   This indicates the existence of at least four critical points when $N_b=22$.

\newpage   \newpage

However, when $K_a$ is treated as a {\it continuously varied} parameter, it is more straightforward to solve (\ref{greataction1}, \ref{greataction2}) analytically, instead of locating them graphically through the pseudopotentials.  Given the ($\psi_a, \psi_b$) values in eqn. (\ref{CR}), eqn. (\ref{greataction2}) is further simplified to:
\begin{align}
\frac{\partial H_{bend}}{\partial J_b} &  =J_b \{ 2(g_{44}-g_{45}+g_{55})-\frac{K_a^2+J_a^2-J_b^2}{\Lambda} (S_{45} \cos[2\psi_a] + R_{45}^* \cos[2\psi_b]) \nonumber\\
& \quad -\frac{R_{45}^*+2g_{45}}{2} \cos[2\psi_a]\cos[2\psi_b] \} =0\label{derivJb2}
\end{align}
\noindent All terms on the right-hand side are proportional to $J_b$. The other multiplier (the sum of terms between the curly brackets) can be numerically shown to be always positive.  Hence eqn. (\ref{deriJb}) has the trivial solution:
\begin{align}
J_b=0
\end{align}
\noindent The last equation (\ref{deriJa}) is then solved analytically on a computer using the software {\it Mathematica} \cite{Mathematica}.   

At each critical point, the linear stability is determined by the eigenvalues of the stability matrix described in $\S$ 3.3.1.  Here the matrix is $4\times4$ in size, and the possible stability types are bi-stable (EE), bi-unstable (HH), stable-unstable (EH), mixed (MM) and degenerate (D).  

Special consideration of the critical points is required where $|J_a|+|J_b|=K_a$.  At these points, the values of $\psi_a$ and $\psi_b$ are indeterminate as the denominator $\Lambda$ in eqns. (\ref{greataction1}, \ref{greataction2}) vanishes.  The ($J_a, \psi_a,  J_b, \psi_b$) coordinate system becomes singular at these locations.  An alternative coordinate system is required to evaluate (1) whether a point is critical point or not, and if yes, (2) its linear stability.   The technical details are discussed in Appendix B.   Only the following four are critical points with $|J_a|+|J_b|=K_a$: 
\begin{align}
J_a & = \pm K_a, J_b =0 \mbox{\,\,\,\,\,\, with (EE) stability}\label{specialAB} \\
J_a & = 0,  J_b = \pm K_a \mbox{\,\,\,\,\,\, with (MM) stability}\label{specialCD} 
\end{align} 

\subsection{4.3.2 Results of the $\lbrack N_b, 0 \rbrack$ Polyads}
\addtocontents{toc}{\protect\vspace*{5pt}}

\noindent \underline{Normal Modes at Low $N_b$} \,\,\, Up to $N_b=6$, there are two families of critical points at $J_a= K_a$ and $J_a=-K_a$ with (EE) stability.  They correspond to the normal {\bf Trans} and {\bf Cis} modes, respectively.  This is consistent with the usual assumption that  vibrations near equilibrium are dominated by the normal modes.  

\noindent \underline{Bifurcation at Higher $N_b$} \,\,\, With increasing polyad number $N_b$ (as well as increasing energy), four bifurcations occur to the {\bf Trans}  and {\bf Cis} critical points.  Four new families of critical points called the {\it Local} ({\bf L}), {\it Precessional} ({\bf Pre}), {\it Orthogonal} ({\bf Orth}) and {\it Counter Rotator} ({\bf CR}) emerge out of the normal mode families at the points of the bifurcations.   The analytical solutions of these families are listed in Table~\ref{ana.bifur} and eqns. (\ref{4family1}-\ref{4family4}).
  
\begin{table}[hbt]
\cpn{New critical points in $\lbrack N_b, 0 \rbrack$ polyads} {New critical points in $\lbrack N_b, 0 \rbrack$ polyads. \label{ana.bifur}} 
\begin{center} \vspace{0.2in}   \begin{tabular}{|c|c|c|c|} \hline \hline \, & \, & \, & \, \\
\, & Name & $(\psi_a, \psi_b)$ & Stability \\ \hline \, & \, & \, & \, \\  \, & \, & \, & \, \\
{\bf L} & Local  & $(0, 0)$ & (EE) \\  \, & \, & \, & \, \\ \hline \, & \, & \, & \, \\ \, & \, & \, & \, \\
{\bf Orth} & Orthogonal & $\left(0, \frac{\pi}{2} \right)$ & (EH) \\ \, & \, & \, & \, \\ \hline \, & \, & \, & \, \\ \, & \, & \, & \, \\
{\bf Pre} & Precessional & $\left(\frac{\pi}{2}, 0 \right)$ & (HH) \\ \, & \, & \, & \, \\  \hline \, & \, & \, & \, \\  \, & \, & \, & \, \\
{\bf CR} & Counter Rotator & $\left(\frac{\pi}{2}, \frac{\pi}{2} \right)$ & (EE) \\  \, & \, & \, & \, \\ \hline \hline \end{tabular}  \end{center}  \end{table}

\begin{align}
\mbox{{\bf L}:\,\,\,\,\,\,} J_a & = 21.166-0.584K_a-0.2681 \sqrt{4091.902+42.376K_a-K_a^2} \label{4family1} \\
\mbox{{\bf Orth}:\,\,\,\,\,\,} J_a &= 24.876-0.496K_a-0.8227\sqrt{531.865+32.598K_a-K_a^2} \label{4family2} \\
\mbox{{\bf Pre}:\,\,\,\,\,\,} J_a &= 10.920-0.564K_a-0.4603\sqrt{-276.741+93.490K_a-K_a^2} \label{4family3} \\
\mbox{{\bf CR}:\,\,\,\,\,\,} J_a &= -18.312-0.536K_a + 0.6344 \sqrt{309.356 + 143.457 K_a-  K_a^2} \label{4family4}
\end{align}
At each of the four bifurcation points, either {\bf Trans} or {\bf Cis} changes its stability, and gives birth to one new family of critical points  in the following manner:
\begin{subequations} \label{4bifurcations}\begin{align}
\mbox{{\bf Trans} (EE)} & \xrightarrow{N_b=7.63} \mbox{{\bf Trans} (EH)} + \mbox{{\bf L} (EE)} \label{bif1} \\
\mbox{{\bf Cis} (EE)} & \xrightarrow{N_b=9.56} \mbox{{\bf Cis} (EH)} + \mbox{{\bf CR} (EE)} \label{bif2} \\
\mbox{{\bf Trans} (EH)} & \xrightarrow{N_b=9.77} \mbox{{\bf Trans} (HH)} + \mbox{{\bf Orth} (EH)} \label{bif3} \\
\mbox{{\bf Trans} (HH)} & \xrightarrow{N_b=14.56} \mbox{{\bf Trans} (EH)} + \mbox{{\bf Pre} (HH)} \label{bif4} \end{align} \end{subequations}

\begin{figure}[hbtp]  
\begin{center}\includegraphics[width=5.49in]{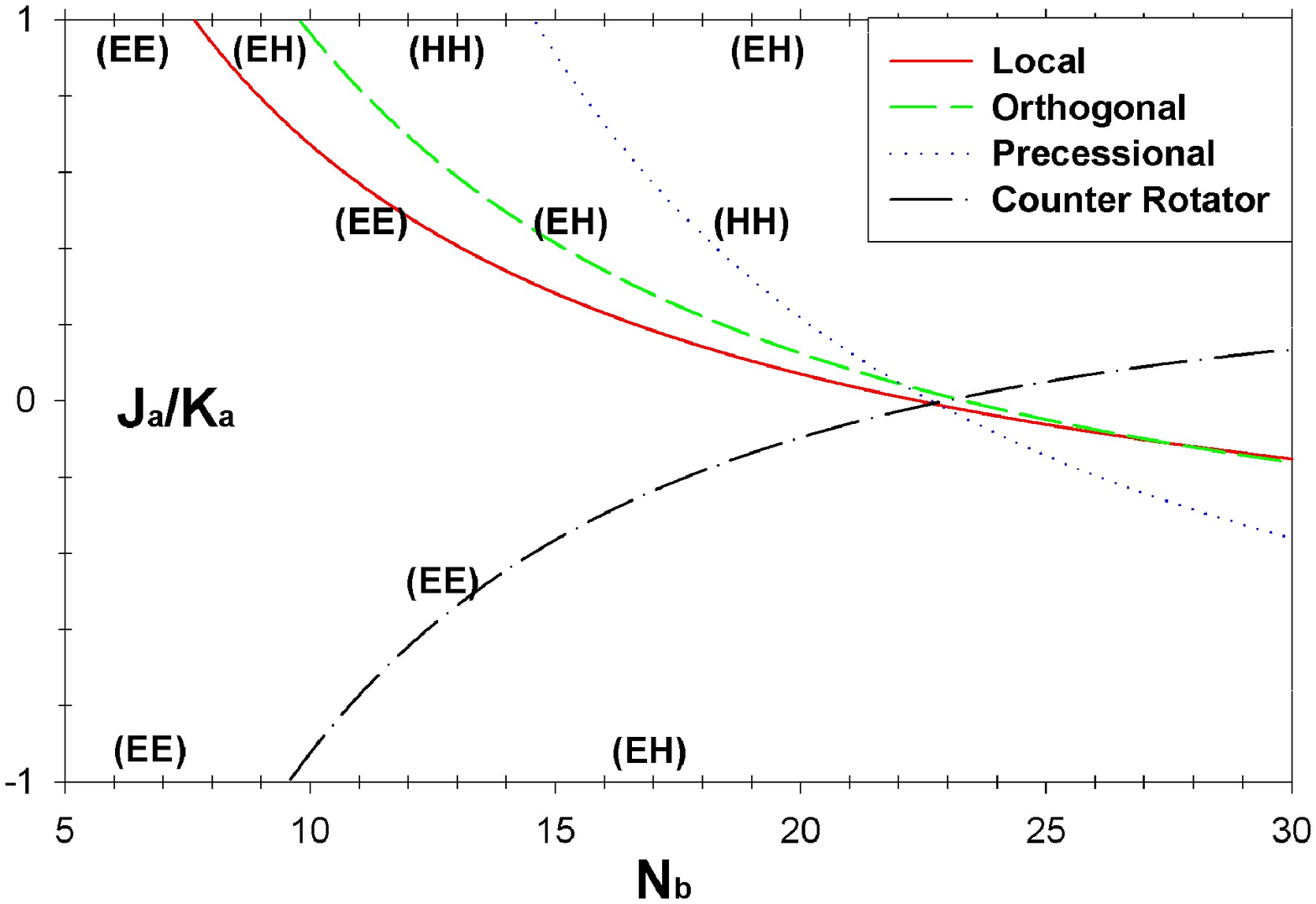}\end{center}
\cpn{Critical points and their bifurcations in $\lbrack N_b, 0 \rbrack$ polyads}{Critical points and their bifurcations in $\lbrack N_b, 0 \rbrack$ polyads.  The four new families of critical points are shown on the figure, while the {\bf Trans} and {\bf Cis} families coincide with the upper and lower edges of the figure, respectively.  The stability of critical points is indicated on the figure. \label{fullbifs}} 
\end{figure}

The {\bf Trans} family undergoes three consecutive bifurcations while the {\bf Cis} family undergoes one.  All six families involved in these bifurcations have $J_b=0$, and thus $\ell_4=\ell_5=0$.  The ratio ($J_a/K_a$) is presented in Fig.~\ref{fullbifs} as a function of $N_b$ \fn{In Fig.~\ref{fullbifs}, the bifurcation is calculated up to $N_b=30$, although the effective Hamiltonian is based on spectra up to only $22$.}.   At each bifurcation point, the new family appears at the respective $J_a=\pm K_a$, then migrates towards $J_a/K_a=0$ with increasing $N_b$.  This trend reflects increasingly equal mixing between the {\bf Trans} and {\bf Cis} in the new critical points, which will be explained in detail below.

\newpage   \newpage

\noindent \underline{Trajectories In the Full Phase Space} \,\,\,\,\,\,\,\,\, The critical points in the reduced phase space have two cyclic angles ($\theta_a$, $\theta_b$) that are absent from $H_{bend}$.  $\theta_a$ corresponds to an overall translation in time, while $\theta_b$ corresponds to a rotation of the system around the C-C axis \cite{Jacobson15000}.  According to $\S$ 3.3.2, in the full phase space such a critical point is a quasiperiodic trajectory on a 2-dimensional invariant torus. 

However, the $\lbrack N_b, 0 \rbrack$ polyads have $K_b=0$ and thus the frequency of $\theta_b$ is no longer physical.  A proper analogy is that when one ``shrinks" to zero the tube thickness of the invariant torus, quasiperiodic trajectories on its surface are reduced to {\it periodic orbits} with a single frequency $\dot{\theta}_a$.  At the critical points the frequency $\dot{\theta}_b$ is zero when we substitute in $K_b=J_b=0$:
\begin{align}
\dot{\theta_b} =& \left(\frac{\partial H_{bend}}{\partial K_b} \right)_{K_b=0} = 2J_b(g_{44}+g_{45}+g_{55})- \frac{S_{45} J_b \cos[2\psi_a]+R_{45}^* J_b \cos[2\psi_b]}{\Lambda} \times \nonumber\\
& \quad \quad \quad (K_a^2+J_a^2-J_b^2)- \frac{R_{45}^*+2g_{45}}{2} J_b \cos[2\psi_a] \cos[2\psi_b] =0
\end{align}
\noindent This leaves only $\theta_a$ with non-zero frequency.  Therefore, critical points found in the $\lbrack N_b, 0 \rbrack $ polyads correspond to  PO in the full phase space.

\begin{figure}[hbtp]   
\begin{center}\includegraphics[width=3.82in]{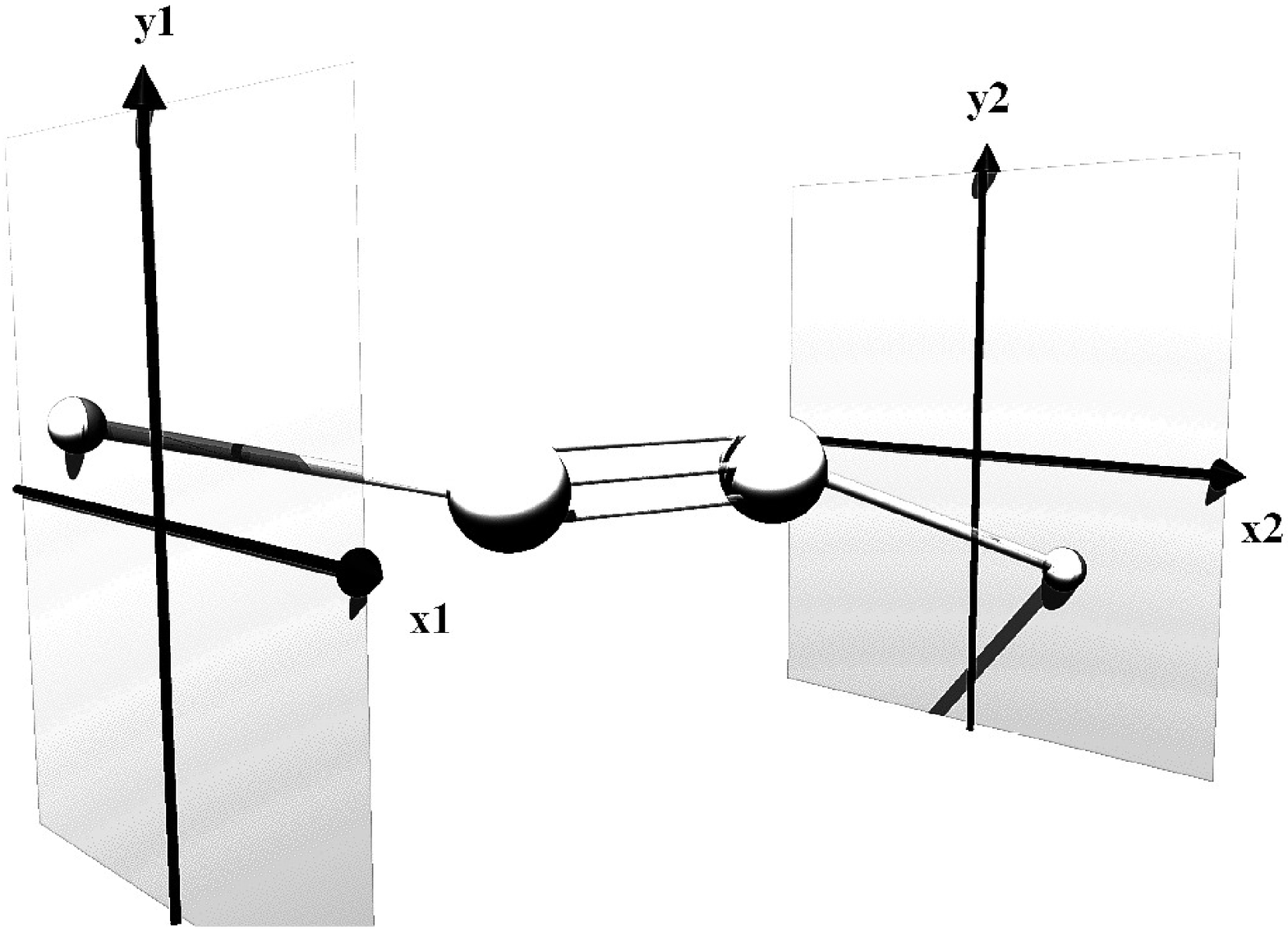}\end{center}
\cpn{Cartesian bending coordinates}{Cartesian bending coordinates, as defined in eqns. (\ref{xy1},\ref{xy2}). \label{x1y1}} 
\end{figure}

\noindent \underline{Visualization of Critical Point PO} \,\,\,\,\,\,\,\,\,  In order to understand the critical points in a more intuitive manner, their corresponding PO are calculated in the Cartesian coordinate.  This is done by assuming each normal mode to be a 2-dimensional harmonic oscillator with frequencies $\omega_4$ or $\omega_5$.  Following the method in \cite{Jacobson15000}, the normal Cartesian coordinates are:
\begin{subequations}\label{lifting}\begin{align}
x_4 = & [(K_a +K_b +J_a-J_b)/\omega_4]^{1/2} \cos[(\theta_a+\theta_b+\psi_a-\psi_b)/2]+ \nonumber\\
& \quad [(K_a-K_b+J_a+J_b)/\omega_4]^{1/2} \cos[(\theta_a-\theta_b+\psi_a+\psi_b)/2] \label{liftingx4} \\
y_4 = & [(K_a+K_b+J_a-J_b)/\omega_4]^{1/2} \sin[(\theta_a+\theta_b+\psi_a-\psi_b)/2]-\nonumber\\
& \quad [(K_a-K_b+J_a+J_b)/\omega_4]^{1/2} \sin[(\theta_a-\theta_b+\psi_a+\psi_b)/2] \label{liftingy4} \\
x_5 = & [(K_a+K_b-J_a+J_b)/\omega_5]^{1/2} \cos[(\theta_a+\theta_b-\psi_a+\psi_b)/2]+\nonumber\\
& \quad [(K_a-K_b-J_a-J_b)/\omega_5]^{1/2} \cos[(\theta_a-\theta_b-\psi_a-\psi_b)/2] \label{liftingx5} \\
y_5 = & [(K_a+K_b-J_a+J_b)/\omega_5]^{1/2} \sin[(\theta_a+\theta_b-\psi_a +\psi_b)/2]-\nonumber\\
& \quad [(K_a-K_b-J_a-J_b)/\omega_ 5]^{1/2} \sin[(\theta_a-\theta_b-\psi_a-\psi_b)/2] \label{liftingy5}  \end{align} \end{subequations}
\noindent ($x_4,y_4,x_5,y_5$) are related to the local Cartesian coordinates ($x_1,y_1,x_2,y_2$) shown in Fig.~\ref{x1y1} by
\begin{align}
x_1 & = x_4+x_5,  &  y_1 & = y_4+y_5 \label{xy1} \\
x_2 & = -x_4+x_5, & y_2 & = -y_4+y_5 \label{xy2}
\end{align}

\newpage   \newpage

\begin{figure}[hbtp]  
\begin{center}\includegraphics[width=5.46in]{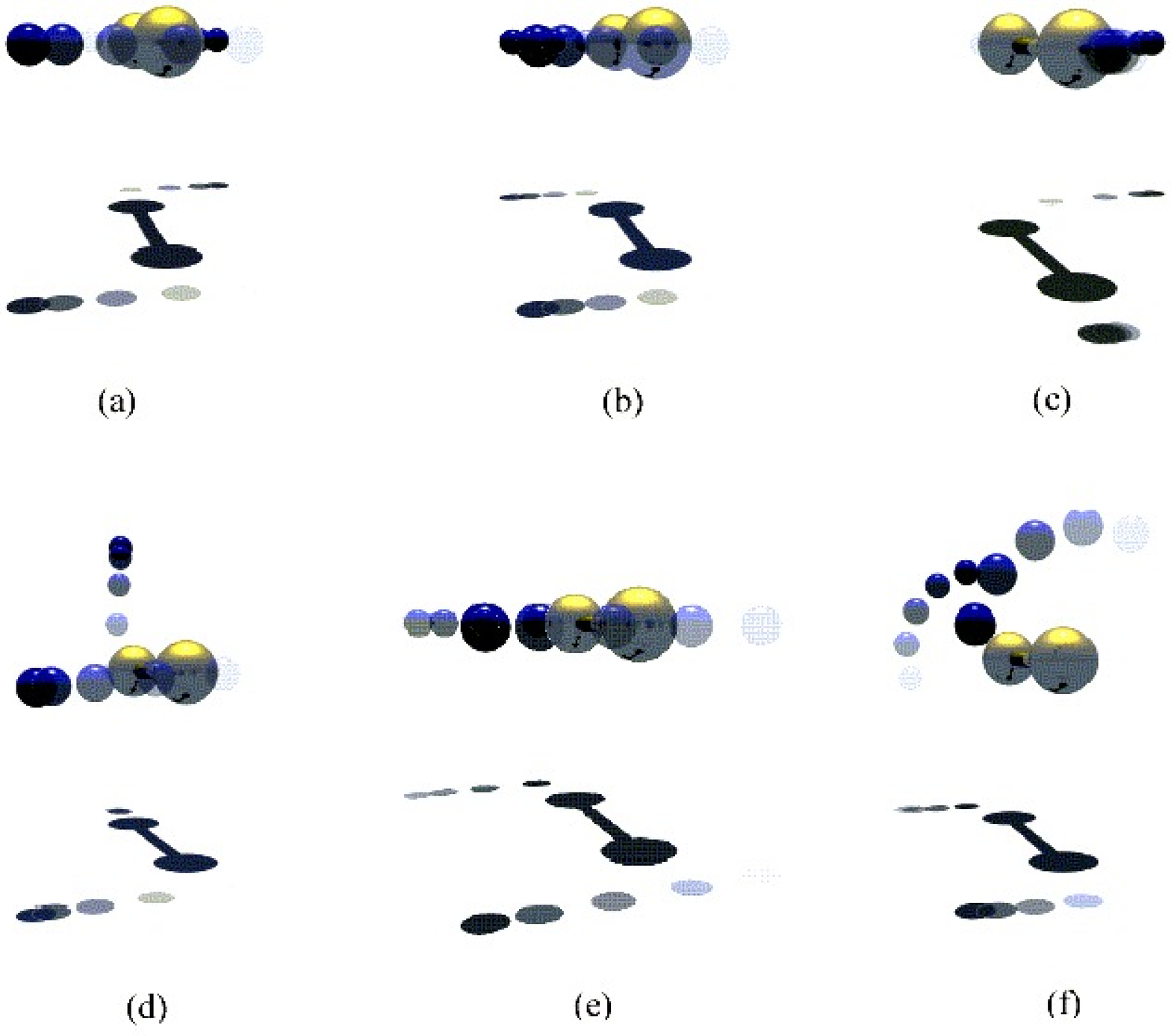}\end{center}
\cpn{Cartesian PO corresponding to critical points}{Cartesian PO corresponding to critical points in $\lbrack 22, 0 \rbrack$ polyad, produced from still frames in the computer animation.  The carbon and hydrogen atoms are represented by the large and small spheres, respectively.  The panels refer to (a) {\bf Trans}; (b) {\bf Cis}; (c) {\bf L}; (d) {\bf Orth}; (e) {\bf Pre}; (f) {\bf CR} critical points. \label{animationstills}} \end{figure}

In order to visualize the resulting 3-dimensional bending motions, they are converted into animations with the modeling software {\it Bryce 4} \cite{Bryce}.   The animations for $N_b=$22 are included on the accompanying CD-ROM in {\it QuickTime 4} format \cite{Animation}.   Fig.~\ref{animationstills} displays some of the still frames of these animations.  The trajectories of {\bf L} and {\bf Pre} are both planar and resemble their namesakes in \cite{RosePre}.  {\bf L} has most of the amplitude of bending in one C-H bending oscillator (bender), and very little in the other.  {\bf Pre} has the two C-H benders at approximately equal amplitude but out of phase by $\frac{\pi}{2}$.  For {\bf Orth}, the two benders vibrate on planes orthogonal to each other, and reach their turning points in phase.  Finally, in {\bf CR} the two hydrogens rotate in ellipses (or circles when near $J_a=0$) in opposite directions at the two ends of the C-C unit.  

\newpage    \newpage

Immediately after each bifurcation, in Fig.~\ref{fullbifs} the nascent family of critical points remains close to $J_a = \pm K_a$. The corresponding Cartesian trajectory resembles the ``parent" normal mode motions.   As the family migrates towards $J_a/K_a=$0, the Cartesian trajectory looks increasingly like the respective ideal motion in Fig.~\ref{animationstills}.  For example, the {\bf L} trajectory initially resembles a slightly asymmetrical trans- bend.  Then the imbalance of amplitude between the two C-H benders increases with $N_b$.   Finally, at $N_b=$22 where $J_a/K_a \approx 0$, almost all the amplitude is in one of the two oscillators.      

The {\bf L} and {\bf CR} critical points are the global extrema points of the reduced Hamiltonian: {\bf L} has the lowest energy while {\bf CR} has the highest.  This is apparent from an examination of the $V_{DDI}$ and $V_{\ell}$ terms in (\ref{DDIterm}, \ref{ellterm}) reveals that both of them contribute most negatively to the Hamiltonian when $(\psi_a, \psi_b)=(0, 0)$, and most positively when $(\psi_a, \psi_b)=(\frac{\pi}{2}, \frac{\pi}{2}$).

In summary, critical points analysis of the $\lbrack N_b, 0 \rbrack$ polyads yields four new families of critical points, in addition to the {\bf Trans} and {\bf Cis} normal families.  The low-energy (EE) stability of {\bf Trans} and {\bf Cis} is destroyed in their first bifurcations.  Due to the special condition $K_b=\ell=0$, the frequency of one cyclic angle $\dot{\theta}_b$ vanishes for all critical points.  Therefore, the critical points correspond to PO in the full phase space.

\subsection{4.3.3 Results of the $\lbrack N_b, \ell \rbrack$ Polyads}
\addtocontents{toc}{\protect\vspace*{7pt}}

Although the $\ell \neq 0$ cases have not been explicitly considered by any other research groups, our analysis can be extended to arbitrary $\ell$ values \fn{Nevertheless, the validity of the effective Hamiltonian beyond $\ell=2$ is questionable due to a lack of supporting experimental data.}.  Here we present critical points found in $\lbrack N_b,2 \rbrack$, $\lbrack N_b,6 \rbrack$ and $\lbrack N_b,10 \rbrack$ polyad series as preliminary predictions, and as a demonstration of this significant extension of our methods.    These series correspond to $K_b=\ell/2=1, 3, 5$, respectively. 

Similar to the $\ell=0$ case except using $H_{bend}^\ell$ in eqn. (\ref{Classical2}) as the Hamiltonian, two of the critical points equations are solved by holding ($\psi_a, \psi_b$) fixed at the four values in (\ref{CR}).  However, it is no longer true the second of the two remaining equations 
\begin{align}
\frac{\partial H_{bend}^\ell}{\partial J_a} = 0  \label{deriJaL} \\
\frac{\partial H_{bend}^\ell}{\partial J_b} = 0 \label{deriJbL}
\end{align}
\noindent has the trivial solution $J_b=0$ as in the $\ell=0$ case.  Instead, (\ref{deriJaL}, \ref{deriJbL}) must be solved simultaneously for the ($J_a, J_b$) values. The solutions are numerically found by first transforming these equations into a polynomial form, then use the homotopy continuation package {\it PHCpack} \cite{PHC}.  This package has no limitation on the number and form of polynomials to solve, and can be used with no preliminary knowledge about the solutions.  

\begin{figure}[hbtp] 
\begin{center}\includegraphics[width=5.97in]{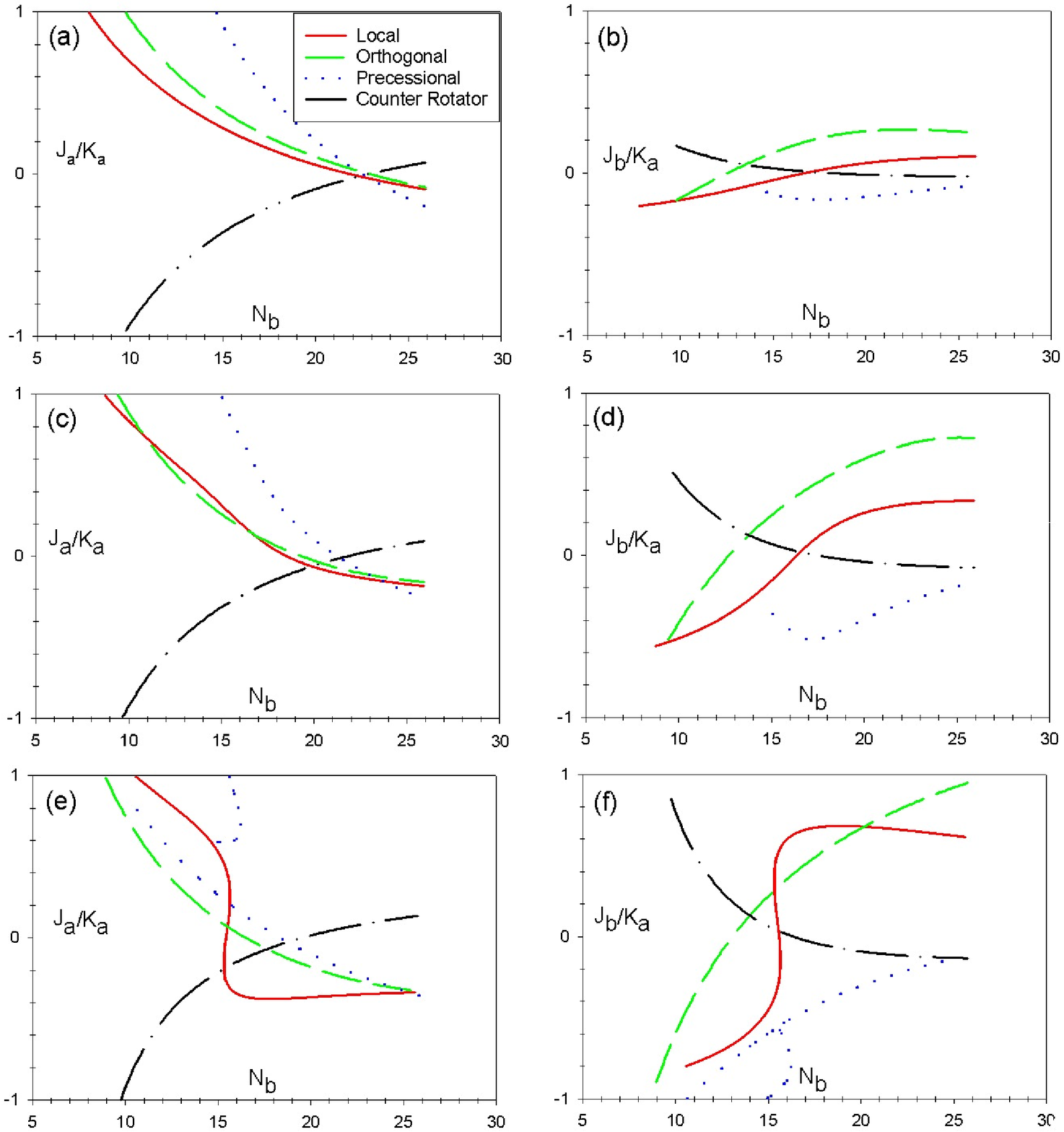}\end{center}
\cpn{Critical points and their bifurcations in $\lbrack N_b, \ell \rbrack$ polyads}{Critical points and their bifurcations in $\lbrack N_b, \ell \rbrack$  polyads.  The left panels display $J_a/K_a$ vs. $N_b$, while the right panels display $J_b/K_a$ vs. $N_b$ values.   Panels (a), (b): $\ell=2$;  (c), (d): $\ell=6$; (e), (f): $\ell=10$.    \label{Kb2610Bifur}} 
\end{figure}

The $J_a/K_a$ and $J_b/K_a$ values of the resulting critical points are plotted against $N_b$ in Fig.~\ref{Kb2610Bifur}.  The solutions for each of the four ($\psi_a, \psi_b$) conditions are named in the same way as Table~\ref{ana.bifur}.  In panels (a), (c) and (e),  all families branch out of the {\bf Trans} and {\bf Cis} normal modes in a manner similar to the $\ell=0$ results (Fig.~\ref{fullbifs}).  From the three panels on the right, these new critical points are clearly not restricted to $J_b=0$.  Instead, they diverge from $J_b=0$ with increasing $\ell$. 

The close resemblance between Fig.~\ref{fullbifs} and panel (a) of Fig.~\ref{Kb2610Bifur} is consistent with the observation that in DF spectra the $\ell=0/2$ states have nearly indistinguishable intensity patterns.  Were the dynamics vastly different for these two $\ell$ series, the fractionation patterns of the bright states would be different and therefore distinguishable.   At still higher $\ell$, our analysis reveals interesting bifurcation structures, which awaits further interpretation.  Even though the quantitative predictions here are limited by the validity of the effective Hamiltonian, we nevertheless expect the vibrational dynamics with sufficiently high $\ell$ to become qualitatively different from the $\ell=0$ case.  

\newpage     \newpage

In spite of their similarity, the $\ell=0$ and $\ell \ne 0$ critical points have different physical natures in the full phase space.  In the former case they are periodic orbits and in the latter quasiperiodic orbits.  The claim made by Jacobson \et that {\it ``a fixed point in the abstract action-angle space $\ldots$ must lift to a periodic orbit in the physical coordinates of the molecule $\ldots$ that lies on a full dimension torus" } \cite{Jacobson15000} is accurate only when applied to $\ell=0$.  Generally,  there exist as many independent frequencies as the number of additional constants of motion (polyad numbers).  It is the critical points in reduced phase space, not the PO in the full phase space, that are the centers of phase space organization.

\section[Discussions]{\underline{Discussions}}
\addtocontents{toc}{\protect\vspace*{7pt}}

\subsection{4.4.1 Comparison with Other Studies}
\addtocontents{toc}{\protect\vspace*{5pt}}

Previously, the same $H_bend$ Hamiltonian was analyzed by Jacobson \et in a numerical study.  They first visually examined sequences of SOS (of the 2 DOF reduced phase space) in order to identify stable fixed points at the center of regular regions.  These fixed points are then followed while the energy is varied within the same polyad.  A ``family tree" of these fixed points is then built for $\lbrack 22, 0 \rbrack$ \cite{Jacobson15000} and $\lbrack 16, 0 \rbrack$ \cite{Jacobson10000}.  In both cases, they found a {\it local} fixed point at the bottom and a {\it counter rotator} fixed point at the top energy end of the polyad.  The fixed points were used to assign eigenstates based on the nodal pattern of the semiclassical wavefunctions.  

\begin{figure}[hbtp]  
\begin{center}\includegraphics[width=4.42in]{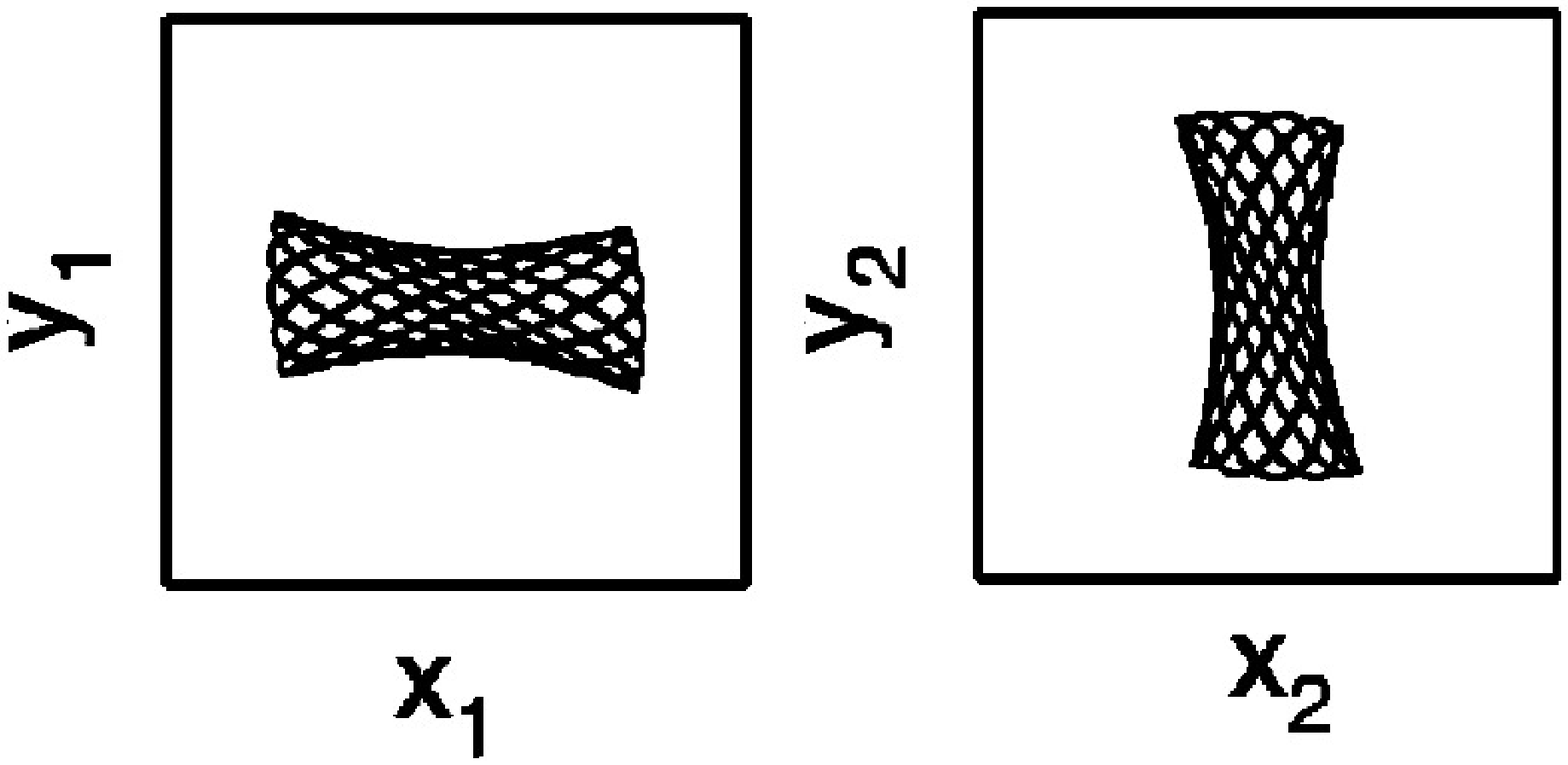}\end{center}
\cpn{The M$_2$ trajectory}{The M$_2$ trajectory in Fig.~3 of \cite{Jacobson15000}, with $N_b=22$ and $E=14,561$cm$^{-1}$.  \label{TaylorM2}} 
\end{figure}

The local and counter rotator fixed points correspond to our critical points with the same names, because they occupy exactly the same place in the reduced phase space \fn{Specifically, in \cite{Jacobson15000} the local mode states are localized around ($\psi_a, \psi_b$)=($0, 0$), and the counter rotator mode states are around ($\pi, \pi$).  Their $\psi_a$ and $\psi_b$ are defined as twice ours.}.   There is another fixed point Jacobson \et call the ``$M_2$ mode", which appears to be related to our {\bf Orth} family of critical points with (EH) stability.  Reproduced in Fig.~\ref{TaylorM2} is the stable $M_2$ PO trajectory they found in $N_b=22$.  The motion qualitatively resembles panel (d) of our animation in Fig.~\ref{animationstills}, with the two C-H benders perpendicular but in phase with each other.    Jacobson \et observed numerically that this PO family becomes unstable below $14,161$ cm$^{-1}$ within this polyad, while our {\bf Orth} critical point at $14,114$ cm$^{-1}$ has (EH) stability.   Currently, we believe that the $M_2$ family (both the stable and unstable segments) is formed by {\it secondary} PO surrounding the {\bf Orth} critical point, which corresponds to a {\it primary} PO \cite{EzraH2O}.   

\newpage   \newpage

Independently, Champion \et deduced the existence of {\bf L}, {\bf CR} and {\bf Orth} modes as idealized cases using Lie algebraic analysis of the same bending Hamiltonian \cite{C2H2Oss}, even though they did not perform a bifurcation analysis with variable polyad numbers.   Moreover, eigenstates corresponding to the {\bf L} and {\bf CR} modes are located in {\it ab initio} calculations \cite{C2H2Guo} as well as numerical PO search on a molecular PES \cite{C2H2Farantos2}. 

\begin{figure}[hbtp]  
\begin{center}\includegraphics[width=3.768in]{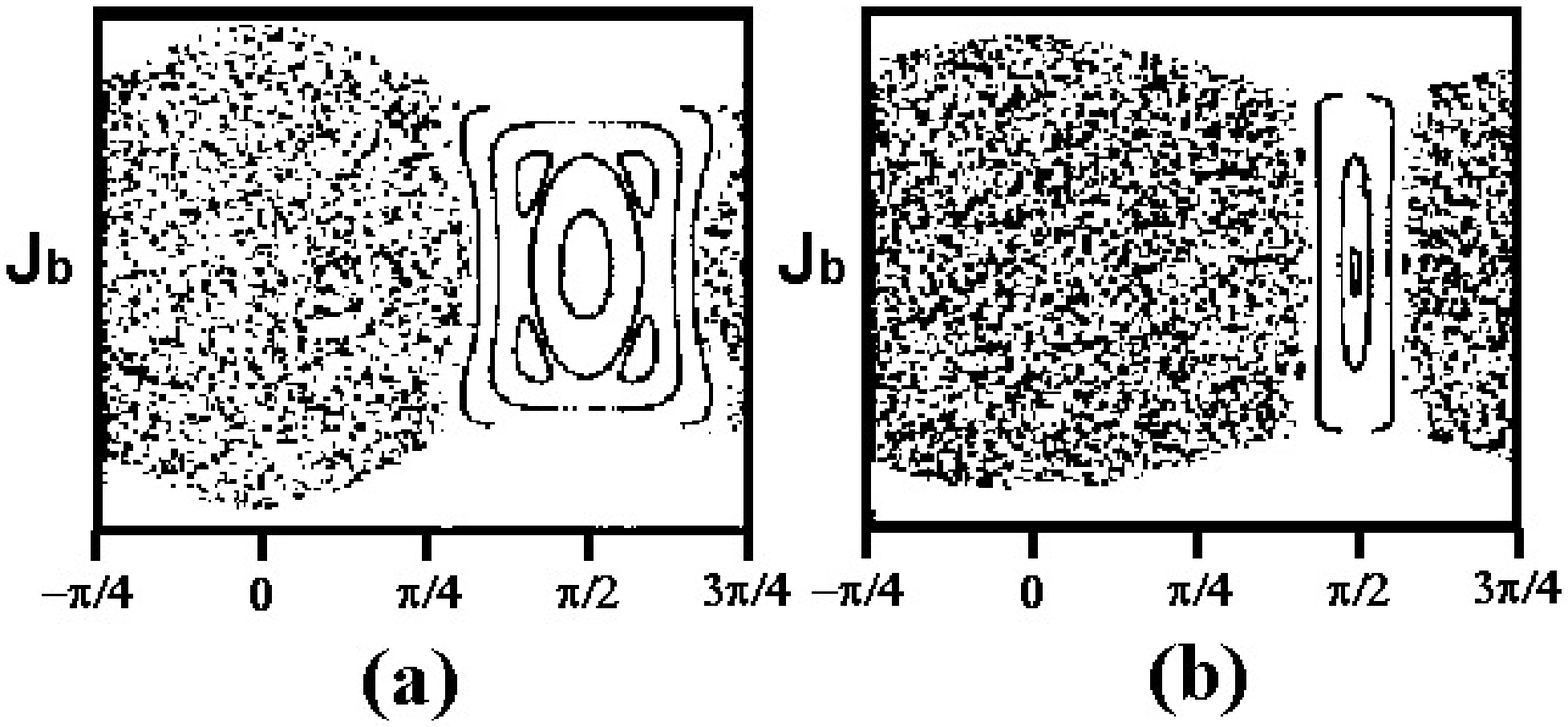}\end{center}
\cpn{SOS near the {\bf Pre} critical point}{SOS near the {\bf Pre} critical point calculated by Jacobson \et, and adapted from Fig.~1 of  \cite{Jacobson10000}.  The $\psi_a, \psi_b$ values are converted to the notation used in this thesis.  Both sections have $\psi_a$ fixed at $\frac{\pi}{2}$. \label{TaylorSOS}} 
\end{figure}

According to our analysis, the {\bf Pre} family is bi-unstable (HH) and therefore expected to be surrounded by chaos.  Thus, this family is not apparent in the study by Jacobson \et, which relies on inspection of SOS for {\it regular} regions.   Their SOS (reproduced below as Fig.~\ref{TaylorSOS}) at nearby energies shows only a strongly chaotic region.

\newpage   \newpage

\subsection{4.4.2 Summary of Method}
\addtocontents{toc}{\protect\vspace*{5pt}}

Compared to the numerical procedure of Jacobson \et , our analysis is more efficient in picking out major changes in the phase space structures of the 2 DOF reduced Hamiltonian here.  In the former, a large number of SOS with different directions and energy is examined in each polyad.  In contrast, finding critical points in the reduced phase space involves only the solution of simultaneous analytic equations for a continuously varied polyad parameter.  In the current case, almost all the significant observations from inspecting the SOS (specifically, those with manifestation in the quantum wavefunctions) can be obtained through four families of critical points.  

In the reduced Hamiltonian, time evolutions of the cyclic variable(s) are formally factored out and not considered as the essential part of dynamics.  The critical points are determined in a uniform manner, regardless of the number ($\ge 1$) of polyad numbers.  Hence, we are able to carry the same critical points analysis to both $\ell=0$ and $\ell \ne 0$ cases.  At high $\ell$ values, the latter cases display additional bifurcations that have not been observed previously.  

Moreover, the usage of SOS is almost exclusively reserved for 2 DOF systems.  With 3 DOF, the phase space is 6-dimensional.  A 2-dimensional surface does not have the correct dimensionality to divide it.  Usage of 2-dimensional surfaces therefore is at best limited to short time scales, before the trajectories stop intersecting the dividing surface \cite{Contopoulos}.  The generalized SOS is a 4-dimensional ``hyper slice" whose visualization is not trivial \cite{Patsis}.    Analytic detection (without visual inspection) of dynamics becomes essential for analyzing systems with 3 or more DOF.  The critical points analysis satisfied this requirement.   The number of equations scales {\it linearly} with the DOF.  In addition, all critical points are found in a comprehensive manner regardless of their stability property.   The resulting critical points then enable a guided exploration of the phase space using other techniques.


\section[Effect of Single DD-I and Single $\ell$ Resonances]{\underline{Effect of Single DD-I and Single $\ell$ Resonances}}
\addtocontents{toc}{\protect\vspace*{7pt}}

In this section, we propose an interpretation of the bifurcations in the full 4 DOF $H_{bend}$ in terms of its two strong resonances ( namely the DD-I and the $\ell$ resonances) taken individually.  With either one of these resonances, the $\lbrack N_b, \ell \rbrack$ polyad is further separated into subpolyads.  The DD-I resonance subpolyads are aligned along the columns in Fig.~\ref{diamond}, while the $\ell$ resonance subpolyads are along the rows.  

The corresponding single-resonance Hamiltonians 
\begin{align}
H_{DDI}  &=H_0+V_{DDI} \\
H_{\ell} &=H_0+V_{\ell}
\end{align} 
\noindent can be analyzed with the method described in $\S$ 3.1 with $m:n=2:2$. The $H_{DDI}$ Hamiltonian is parameterized by two reduced phase space variables ($J_a, \psi_a$) according to eqns. (\ref{actang}).  Within each DD-I subpolyad, these parameters define a DD-I PPS.  Similarly, the $H_{\ell}$ Hamiltonian is reduced to ($J_b, \psi_b$) space,  which defines an $\ell$ PPS for each $\ell$ subpolyad.  For simplicity, we only consider the $\lbrack N_b, 0 \rbrack $ polyads without the high-order coefficients $y_{ijk}, r_{445}$ and $r_{545}$.   Then each DD-I PPS has radius $K_a-|J_b|$ while each $\ell$ PPS has radius $K_a-|J_a|$.

\begin{landscape}  \begin{figure}[hbtp] 
\begin{center}\includegraphics[width=6.01in]{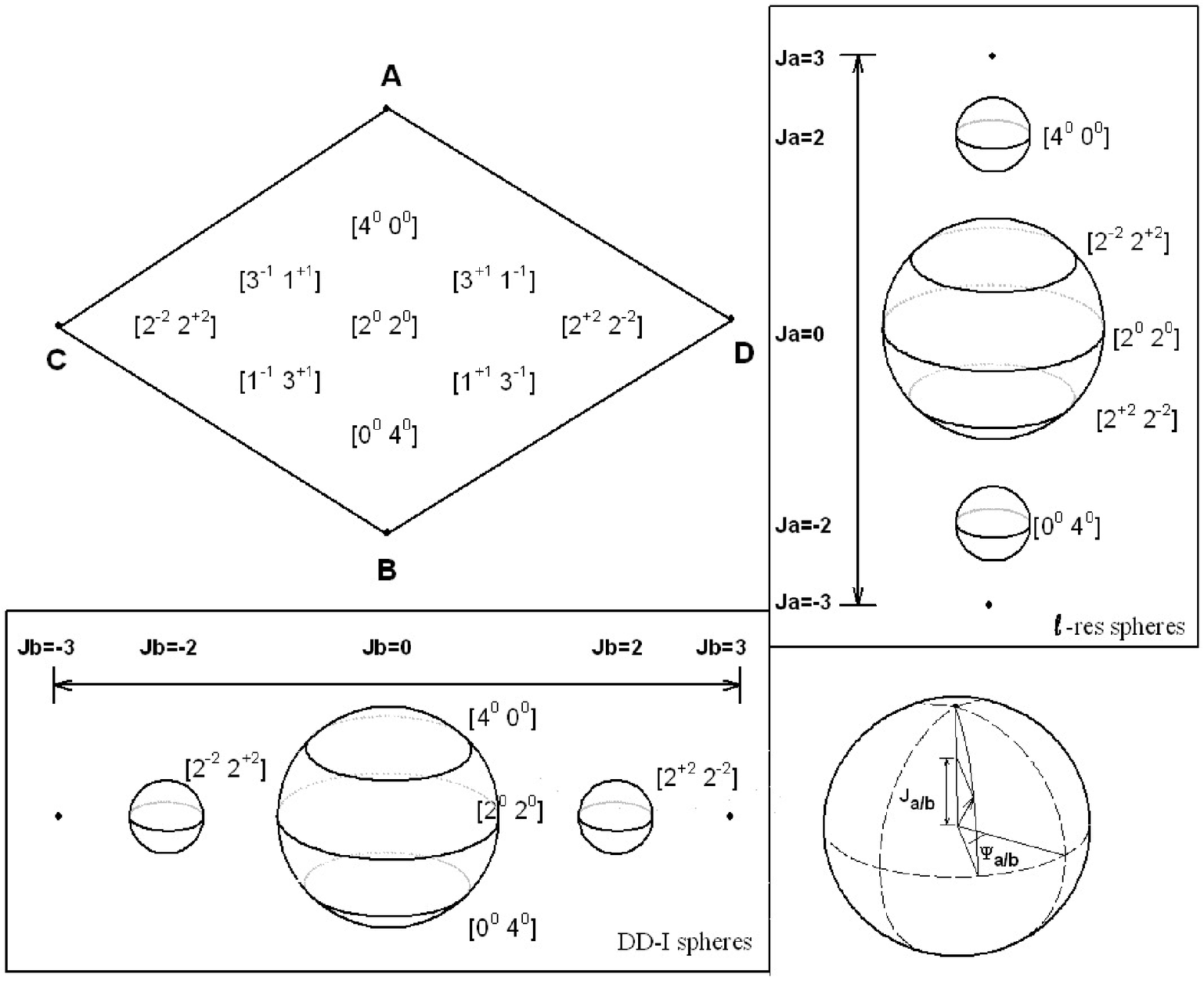}\end{center}
\cpn{Quantum states and PPS of the zero-order system}{Quantum states and PPS of the zero-order system $H_0$.  Corresponding to each column is a DD-I PPS parameterized by $(J_a, \psi_a)$.  Corresponding to each row is an $\ell$ PPS parameterized by $(J_a, \psi_a)$.  The sphere shrinks to a point at the corners A, B, C and D, where $J_a$ or $J_b$ vanishes.\label{spherediamond}} 
\end{figure}  \end{landscape}

Fig.~\ref{spherediamond} presents the zero-order  DD-I and $\ell$ PPS, and the semiclassical trajectories on them.  With only the $H_0$ term, the Hamiltonian does not depend on $\psi_{a}$ or $\psi_{b}$.  All semiclassical trajectories of $H_0$ can be regarded as the ``composition" between one DD-I trajectory and one $\ell$ trajectory, as labeled in the figure.  The critical points on the PPS are where
\begin{align}
\frac{ \partial H_{DDI} } { \partial J_a } & = \frac{\partial H_{DDI} } {\partial \psi_a }=0  & \mbox{or\,\,\,\,\,\,\,\,\,\,\,\,\,\,\,\,} \frac{ \partial H_{\ell} } { \partial J_b }  & = \frac{ \partial H_{\ell} } { \partial \psi_b }=0   
\end{align}
\noindent In the zero-order case, the only critical points are the north and south poles of each PPS.

\begin{landscape}   \begin{figure}[hbtp]  
\begin{center}\includegraphics[width=7.25in]{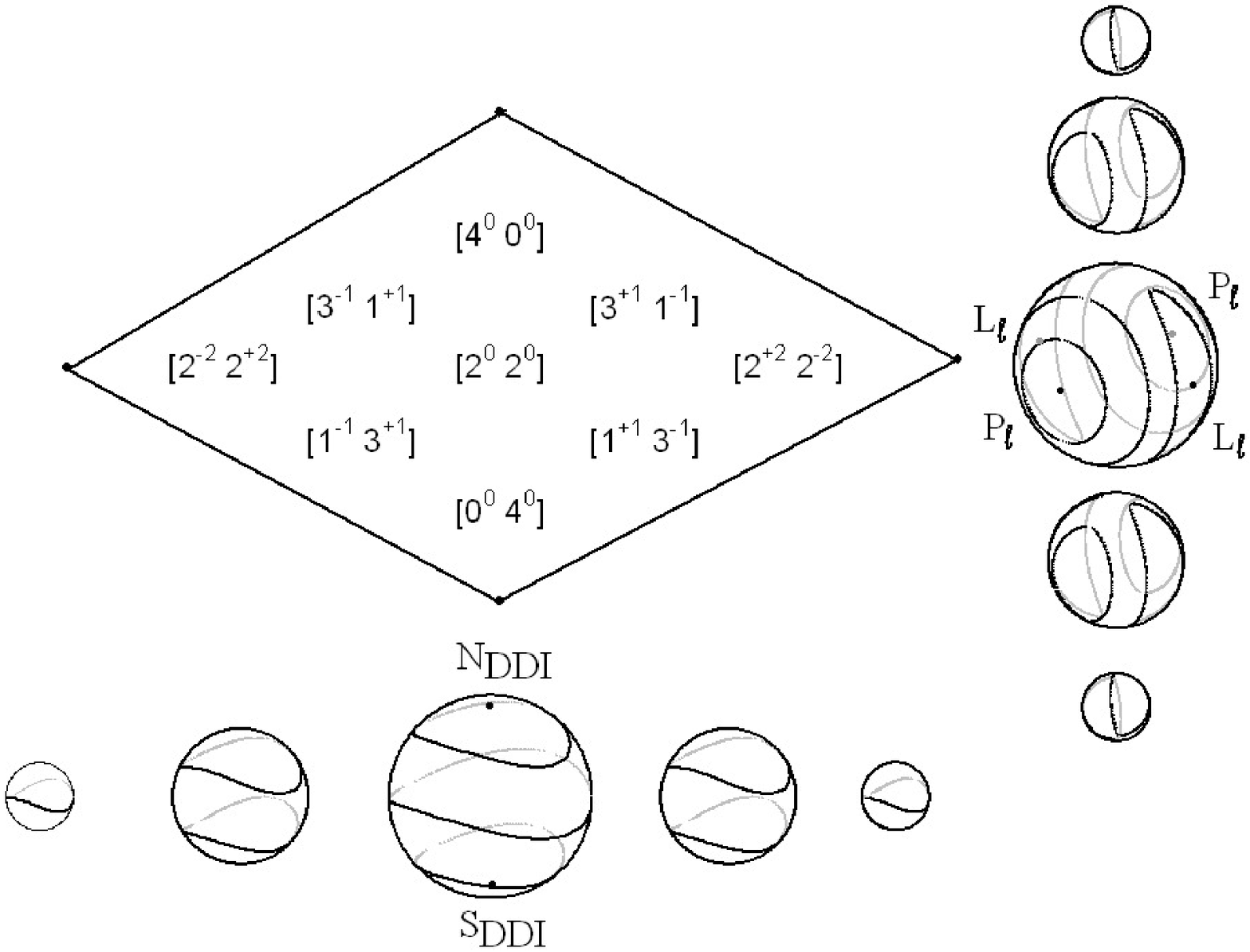}\end{center}
\cpn{DD-I and $\ell$ resonance PPS of $\lbrack 4, 0 \rbrack$ polyad}{DD-I and $\ell$ resonance PPS in $\lbrack 4, 0 \rbrack $  polyad. \label{diamond4}}  
\end{figure}  \end{landscape}

\begin{landscape}  \begin{figure}[hbtp]  
\begin{center}\includegraphics[width=7.25in]{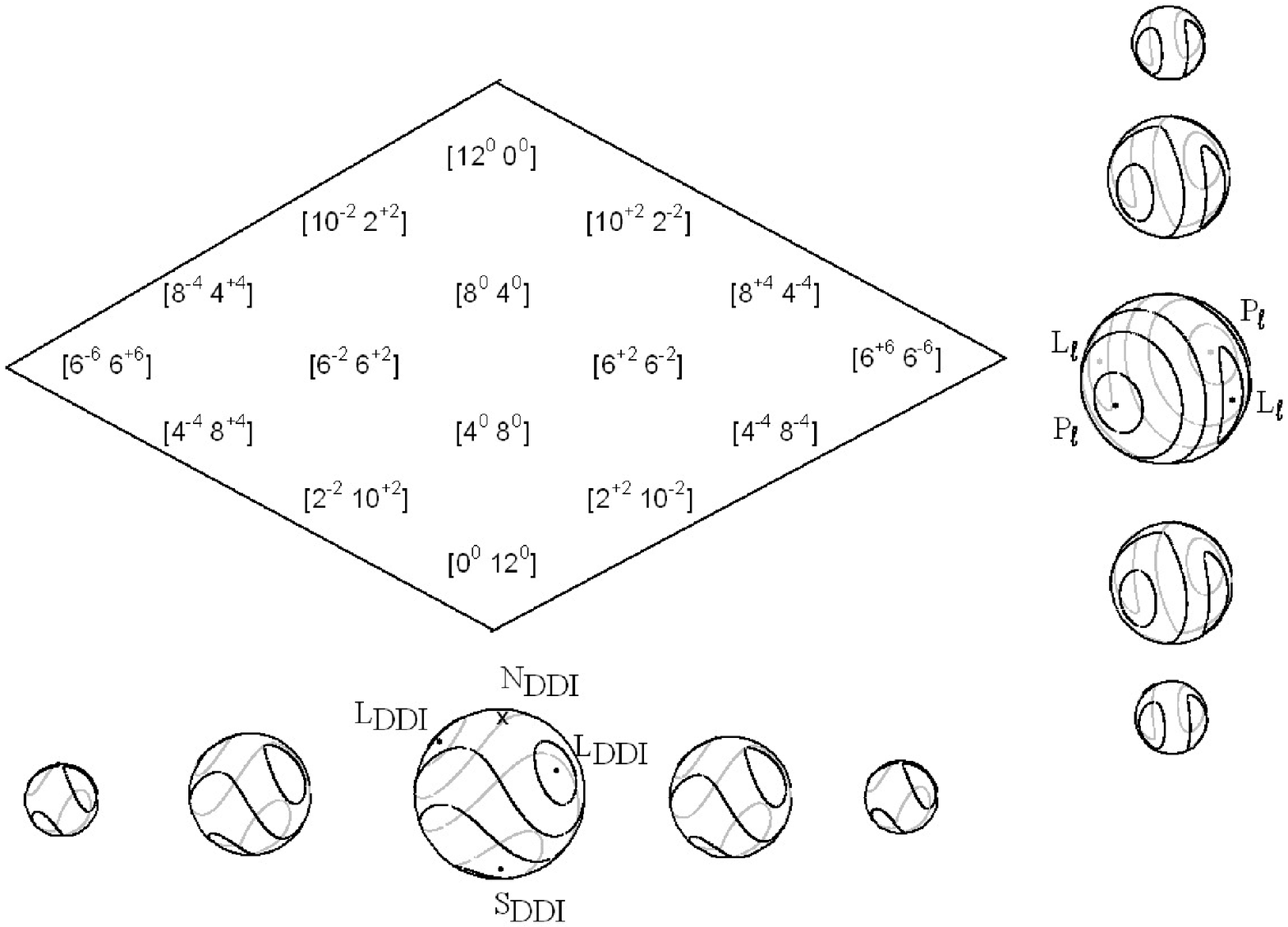}\end{center}
\cpn{DD-I and $\ell$ resonance PPS of $\lbrack 2, 0 \rbrack $ polyad}{DD-I and $\ell$ resonance PPS in $ \lbrack 12, 0 \rbrack $  polyad.  Only selected PPS are shown. \label{diamond12}}   \end{figure}  \end{landscape}

\begin{landscape}  \begin{figure}[hbtp]  
\begin{center}\includegraphics[width=7.55in]{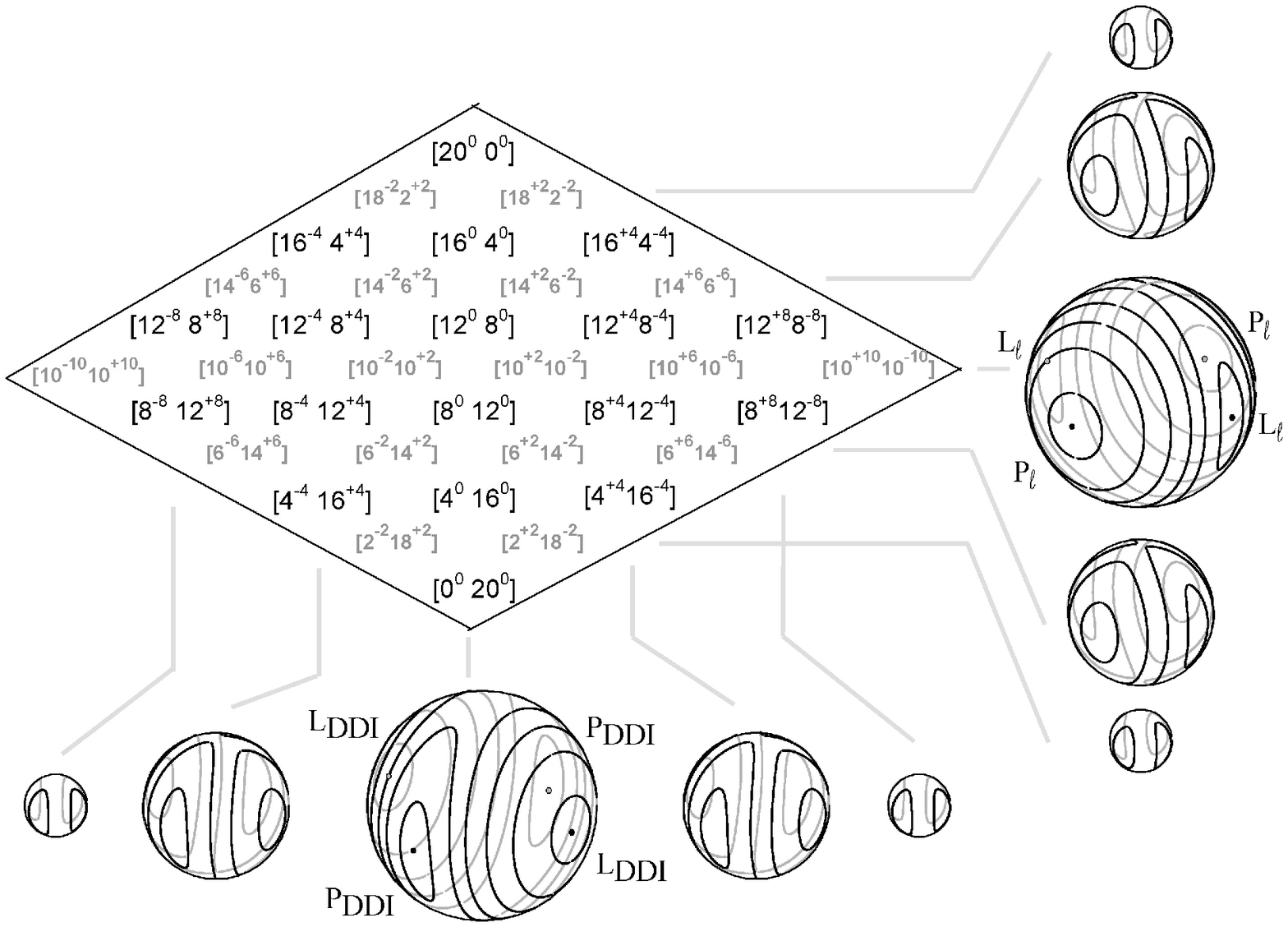}\end{center}
\cpn{DD-I and $\ell$ resonance PPS of $\lbrack 20, 0 \rbrack$ polyad}{DD-I and $\ell$ resonance PPS in $\lbrack 20, 0 \rbrack $ polyad.  Only selected PPS are shown, as indicated by the grey lines. \label{diamond20}}   \end{figure}  \end{landscape}

The DD-I and $\ell$ PPS for polyads $\lbrack 4,0 \rbrack $, $\lbrack 12,0 \rbrack$ and $\lbrack 20,0 \rbrack$ are presented in  Figs.~\ref{diamond4}-\ref{diamond20}.  Note that the two sets of spheres do not coexist within the same Hamiltonian.   The zero-order quantum numbers on these figures are no longer meaningful in labeling trajectories in $H_{DDI}$ and $H_\ell$, since $n_4, n_5$ (or $\ell_4, \ell_5$) are mixed by the resonance term.  As the DD-I and $\ell$ PPS do not qualitatively change across the subpolyads with same $N_b$, in the latter two figures the number of PPS is reduced to make the figures legible.

\newpage  

\newpage   

\newpage 

\newpage  \newpage

\noindent \underline{The $H_{DDI}$ System } \,\,\, This system had been considered previously by Rose and Kellman with slightly different parameters but the same qualitative results \cite{RosePre}.  A comparison across Figs.~\ref{diamond4}-\ref{diamond20} shows that as $N_b$ increases, two new families of critical points emerge in two bifurcations on the DD-I PPS.  Between $N_b=$6-8 (Figs.~\ref{diamond4}, \ref{diamond12}), two {\it local} mode critical points L$_{DDI}$ are born at the north pole of the PPS onto the great circle  ($\psi_a =0, \pi$).  Then between $N_b=$14-6 (Figs.~\ref{diamond12}, \ref{diamond20}) two {\it precessional} mode critical points P$_{DDI}$ emerge at the south pole onto another great circle ($\psi_a =\frac{\pi}{2}, \frac{3\pi}{2}$).  Both L$_{DDI}$ and P$_{DDI}$ migrate towards the equator of the PPS (where $J_a=$0) with increasing $N_b$.

\begin{figure}[hbtp]  
\begin{center} \includegraphics[width=4.25in]{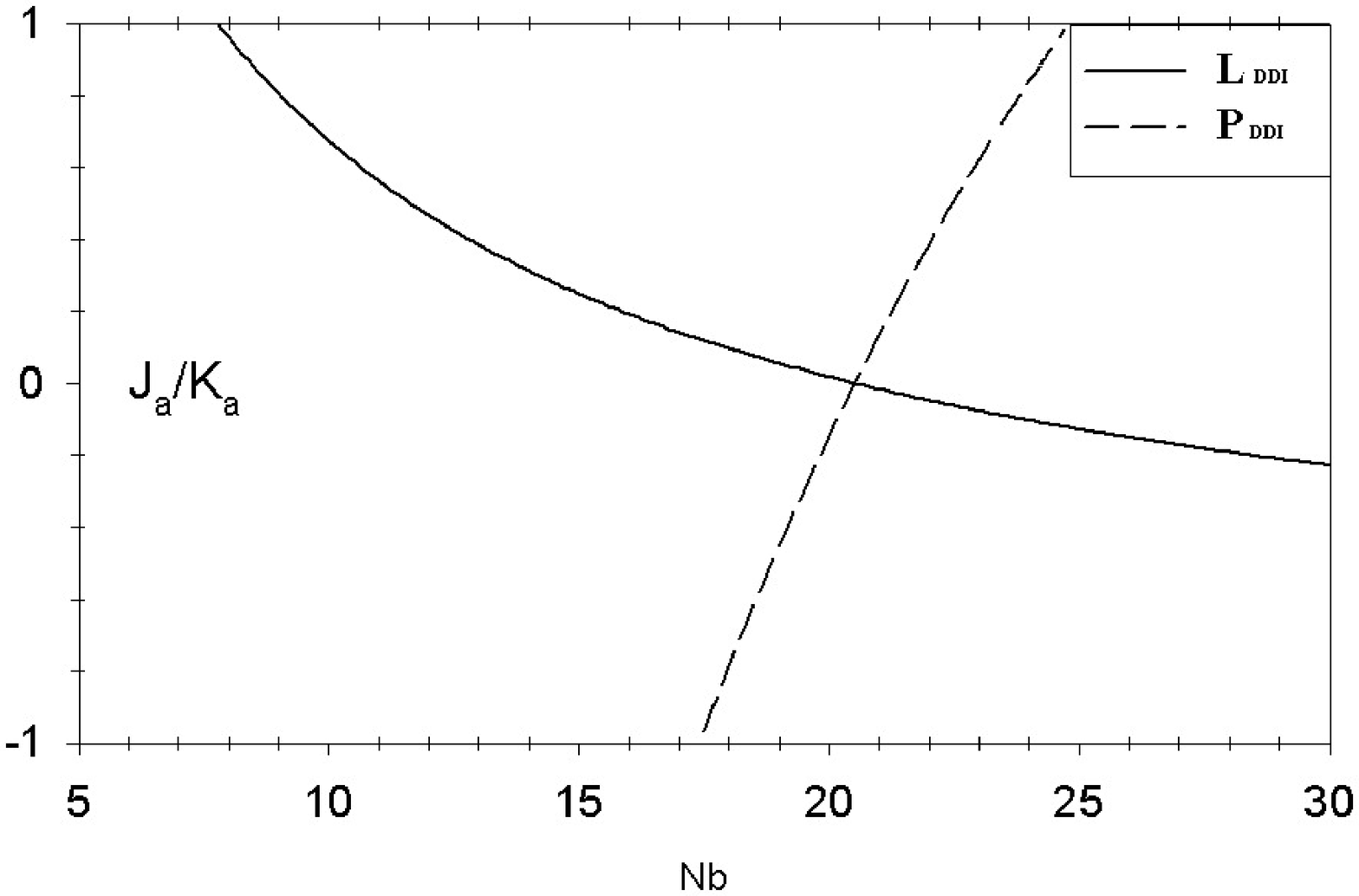}\end{center} 
\cpn{Bifurcation of critical points in $H_{DDI}$}{Bifurcation of critical points in $H_{DDI}$ for the central column subpolyads ($\ell_4=\ell_5=0$).    \label{indddibifs}} 
\end{figure}

The migration of the L$_{DDI}$ and P$_{DDI}$ is depicted in Fig.~\ref{indddibifs}, which plots these critical points in the same manner as Fig.~\ref{fullbifs}.  Similar to Fig.~\ref{fullbifs}, here both critical points in $H_{DDI}$ are born at $J_a/K_a = \pm 1$ and migrate towards the equator, finally reaching $J_a/K_a=0$ at about $N_b=20$.  

\newpage  \newpage 

\begin{figure}[hbtp] 
\begin{center} \includegraphics[width=4.7in]{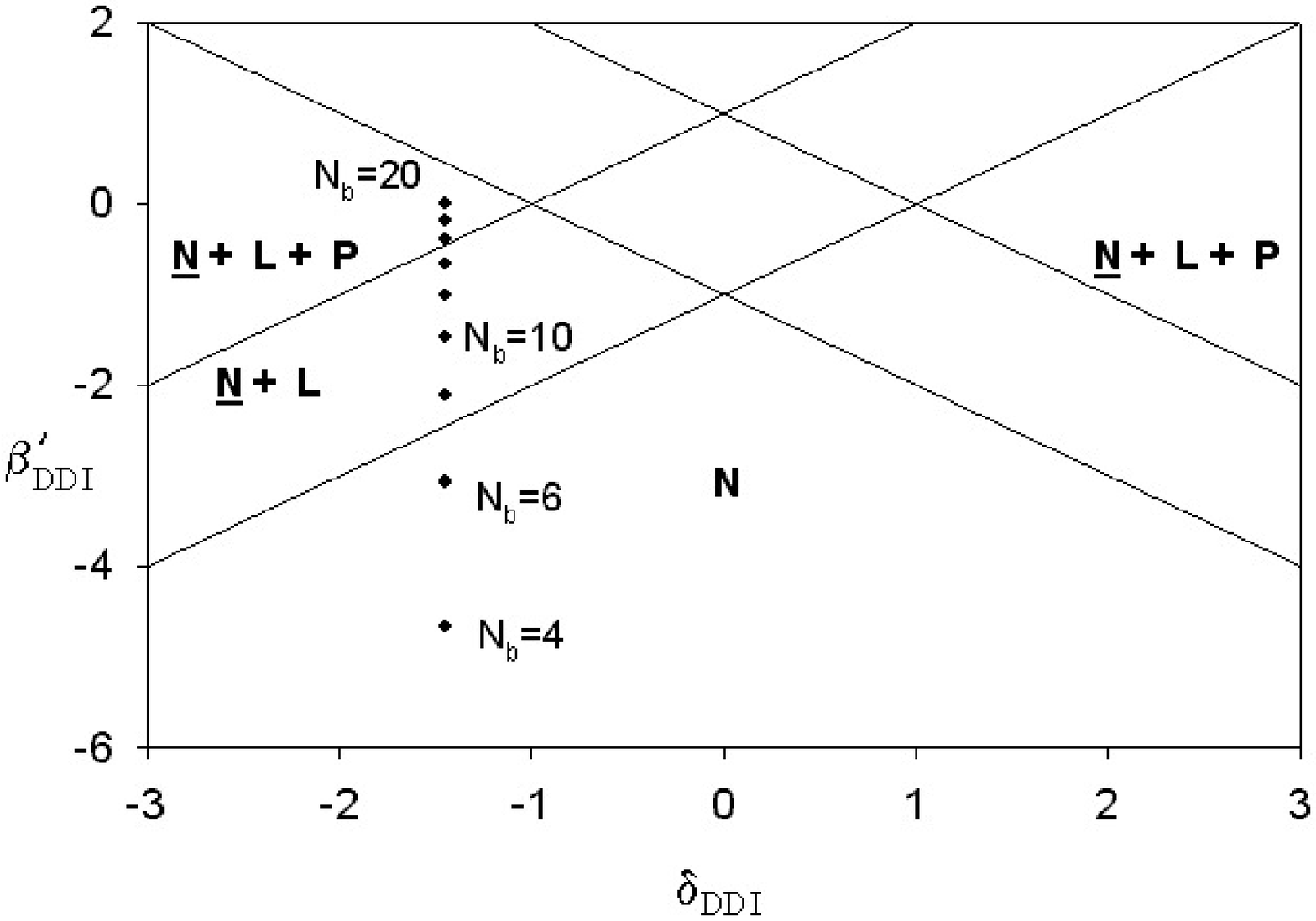}\end{center} 
\cpn{Catastrophe map of $H_{DDI}$}{Catastrophe map of $H_{DDI}$.  The representative points are for the central DD-I PPS with $N_b=$ 4-20.  {\bf N}, {\bf L} and {\bf P} denote the normal, local and precessional mode critical points.  The unstable critical points of a given region are underlined.  \label{DDIcat}}  
\end{figure}

Fig.~\ref{DDIcat} summarizes the structure of the central DD-I PPS ($N_b=$4-20) on the $2:2$ catastrophe map.  The control parameters $\delta_{DDI}, \beta'_{DDI}$ are defined in eqns. (4.1-4.2) of \cite{RosePre}.  $\beta'$ characterizes the strength of the 2:2 resonance, while $\delta$ reflects the detuning between the two coupled frequencies.  Here, the resonance strength $s_{45}$ is a constant.  The trans- and cis- bending frequencies, starting as $\omega_4 < \omega_5$, are tuned towards each other as $x_{55} < 0 < x_{44}$.  The representative points on Fig.~\ref{DDIcat} cross two zone boundaries as $N_b$ increases: first the trans- normal critical point (north pole) is destabilized in a bifurcation, then the cis- normal critical point is destabilized by another bifurcation at higher $N_b$.

Ref.~\cite{RosePre} assumes a planar model of C$_2$H$_2$ bending (see Fig.~1 thereof), and the L$_{DDI}$ and P$_{DDI}$ critical points have the same Cartesian motion as the {\bf L} and {\bf Pre} in Fig.~\ref{animationstills}.  These two modes can be interpreted simply as follows.  The angle $\psi_a$ is defined as the relative phase between the trans- and cis- normal mode oscillators.  When exactly in resonance, these two oscillators have the same frequency.  As shown in Fig.~\ref{Superposition} (a) and (b), superimposing these two oscillators on the same plane with relative phases $0$ or $\pi$ results in only one C-H bender being excited ({\bf L}).  Changing the relative phase to $\frac{\pi}{2}$ or $\frac{3\pi}{2}$ causes the two C-H benders to be out of phase by $\frac{\pi}{2}$ or $\frac{3\pi}{2}$ ({\bf Pre}).   

\newpage    

\newpage \begin{figure}[hbtp] 
\begin{center} \includegraphics[width=4.76in]{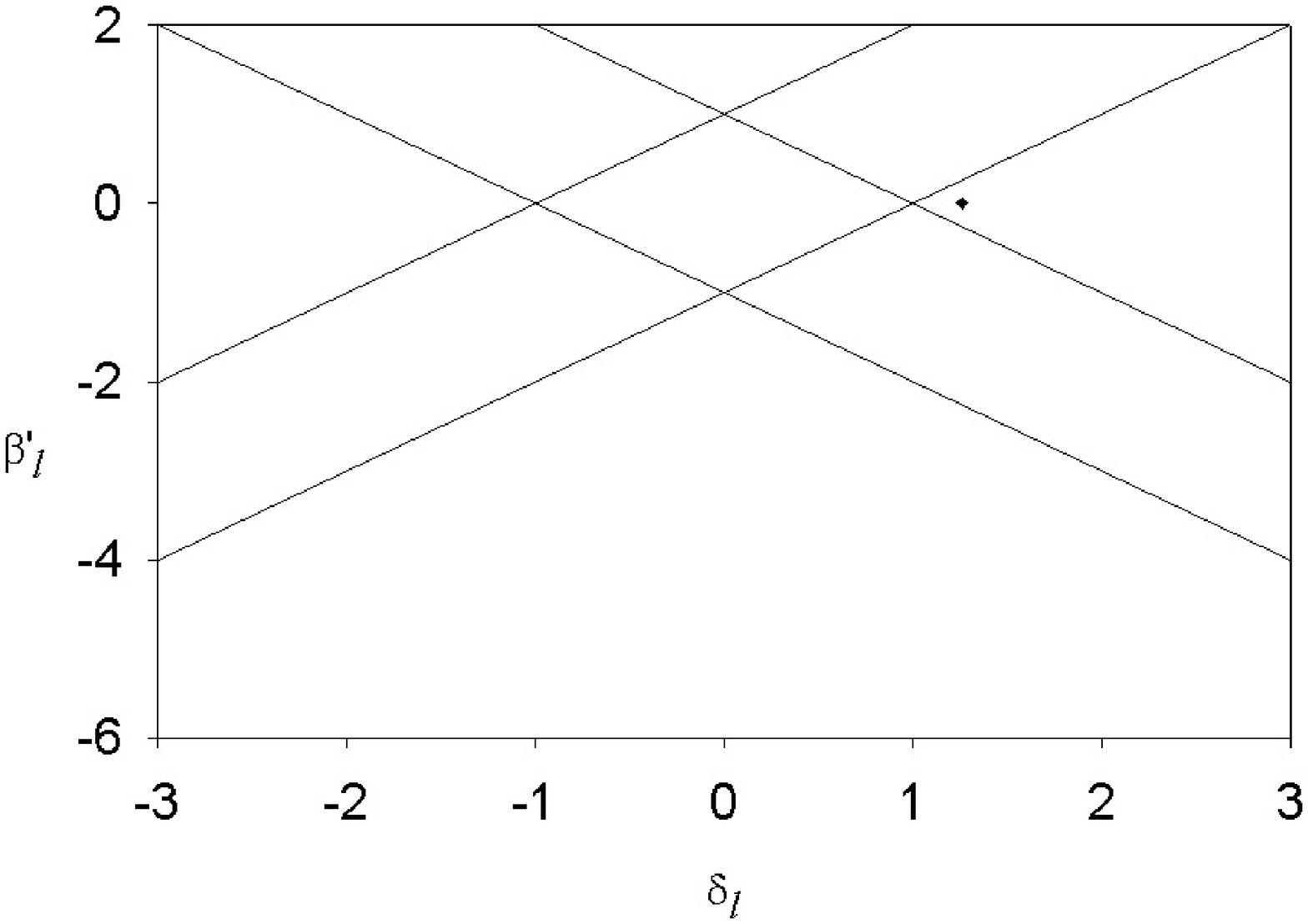} \end{center}
\cpn{Catastrophe map of $H_\ell$}{Catastrophe map of $H_\ell$.   The zone structures are the same as Fig.~\ref{DDIcat}.  At all $N_b$ values, the representative points of the central $\ell$ PPS are located at the same point. \label{ellrescat}} 
\end{figure} 

\newpage \begin{figure}[hbtp] 
\begin{center} \includegraphics[width=5.79in]{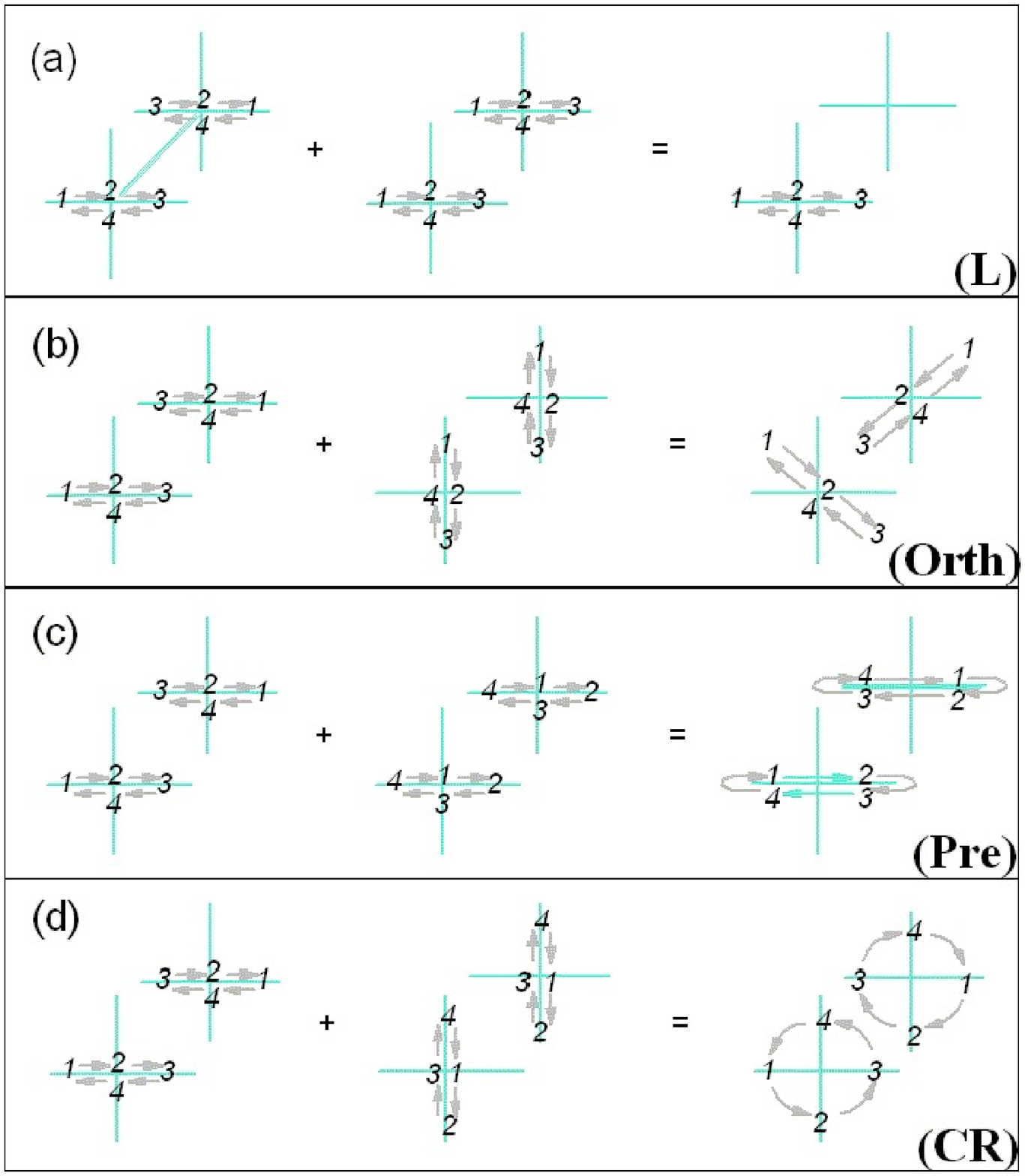}\end{center}
\cpn{New bending modes as superpositions of the normal ones}{New bending modes as superpositions of the normal trans- and cis- modes.  The coordinates are the same as in Fig.~\ref{x1y1}.   For qualitative purposes, the trans- and cis- vibrations are assumed to have the same frequency and amplitude.   In panels (a-d), the relative ({\it phase angle, dihedral angle}) between the normal mode oscillators are in the same order as in eqn. (\ref{CR}).  Superimposing trans- (first column) and cis- (second column) motion result in the four new types of motion found through critical points analysis of the full Hamiltonian $H_{bend}$.  \label{Superposition}}
\end{figure}  \newpage

\noindent \underline{The $H_\ell$ System } \,\,\, Unlike $H_{DDI}$, the Hamiltonian $H_{\ell}$ satisfies a special degeneracy condition.  The quantum ZOS $|n_4^{\ell_4},n_5^{\ell_5} \rangle$ and $|n_4^{-\ell_4},n_5^{-\ell_5} \rangle$ always have the same energy.  This is a consequence of the time reversal symmetry, when the directions of both angular momenta $\ell_4$, $\ell_5$ are reversed \cite{JonasThesis}.  In other words, the two ``frequencies" coupled by $V_\ell$ are always in exact resonance.  Any finite $V_\ell$ term will induce a non-local bifurcation of critical points on the zero-order system \cite{Xiao-Kellman}.  Such a bifurcation is indeed shown when we compare the $\ell$ PPS between Figs.~\ref{spherediamond} and \ref{diamond4}. When $V_\ell$ is turned on, both poles of the $\ell$ sphere become unstable.  The following two new families are stable critical points:
\begin{align}
\mbox{L}_\ell & \mbox{: \,\,\,}& J_b &= 0, \mbox{\,\,\,} \psi_b =0, \pi \label{Lell} \\
\mbox{P}_\ell & \mbox{: \,\,\,}& J_b &= 0, \mbox{\,\,\,} \psi_b =\frac{\pi}{2}, \frac{3\pi}{2} \label{Pell}
\end{align}

Because they are both $2:2$ type resonances, $H_{DDI}$ and $H_\ell$ have the same zone structure on their catastrophe maps.  The latter is shown in Fig.~\ref{ellrescat}.  The coupling strength is taken to be $r_{45}^0$ (which is a constant), and $\beta' \equiv 0$.  On the catastrophe map, the representative points for the central $\ell$ subpolyads in Figs.~\ref{diamond4}-\ref{diamond20} share a single location for all $N_b$ values.    

Formally, the $L_{\ell}$ and $P_{\ell}$ modes are like the local and precessional modes in $H_{DDI}$, although the angles $\psi_a$ and $\psi_b$ have very different meanings.  While $\psi_a$ is the relative phase angle between planar trans- and cis- motions, the physical interpretation of $\psi_b$ is rather complex.   When $\ell_4=\ell_5=0$, $\psi_b$ corresponds to the dihedral angle between the two planar normal C-H benders.  

\noindent \underline{Connection with Critical Points in $H_{bend}$ } \,\,\,  A comparison between the critical points in $H_{DDI}$ and $H_\ell$ and those found in the full $H_{bend}$ of $\S$ 4.4 reveals the following connections:

\begin{enumerate}
\item The 16 conditions of ($\psi_a, \psi_b$) for critical points in $H_{bend}$ (Table~\ref{16cond}) can be produced from combining the great circles on DD-I and $\ell$ PPS.  Table~\ref{16cond} can thus be regarded as a set of generalized great circles in $H_{bend}$.

\item DD-I resonance induces bifurcation of critical points in ($J_a, \psi_a$) coordinates with increasing $N_b$.   The new L$_{DDI}$ and P$_{DDI}$ then migrate towards $J_a=0$.  The $\ell$-resonance, on the other hand, induces L$_{\ell}$ and P$_{\ell}$ critical points, both located at $J_b=0$, as soon as $V_\ell$ is included.  There is no additional bifurcation in $H_\ell$ as $N_b$ increases.   All these observations are consistent with the results in Fig.~\ref{fullbifs}.

\item The four types of non-normal PO in Fig.~\ref{animationstills} can be produced by superimposing equal amounts of trans- and cis- vibrations at the same frequency, with relative phase angle $\psi_a$ and dihedral angle $\psi_b$.   This superposition is illustrated in Fig.~\ref{Superposition}: adding the trans- and cis- bends with (relative phase, dihedral angle) at ($0,0$) and ($0, \frac{\pi}{2}$) result in the {\bf L} and {\bf Orth} type vibrations (panels a and b), while ($\frac{\pi}{2}, 0$) and ($\frac{\pi}{2}, \frac{\pi}{2}$) result in the {\bf Pre} and {\bf CR} type vibrations (panels c and d).  

\item Finally, the critical points in $H_{bend}$ can be regarded as formed from one critical point in $H_{DDI}$ and one in $H_{\ell}$, in the manner listed in Table~\ref{grandstructure}.  The bifurcations along $N_b$ are caused by DD-I, while the nature of the four new modes is determined by both DD-I and $\ell$ resonances.
\end{enumerate}

\begin{table}[hbt]
\cpn{Proposed composition of critical points in $H_{bend}$} {Proposed composition of critical points in $H_{bend}$ from those in $H_{DDI}$ and $H_{\ell}$.  For the notations in columns 1 and 3, refer to Figs.~\ref{diamond4}-\ref{diamond20}.  \label{grandstructure}}
\begin{center} \vspace{0.2in}  
\begin{tabular}{|c|c||c|c|} \hline \hline \, & \, & \, & \,  \\
$H_{DDI}$ and $H_{\ell}$ & $H_{bend}$  & $H_{DDI}$ and $H_{\ell}$ & $H_{bend}$   \\ \, & \, & \, & \, \\  \hline \, & \, & \, & \, \\  
N$_{DDI}, L_{\ell}$ & {\bf Trans} & L$_{DDI} , L_{\ell}$ & {\bf L}   \\  \, & \, & \, & \, \\  
N$_{DDI} , P_{\ell}$ & {\bf Trans} & P$_{DDI} , L_{\ell}$ & {\bf Orth}  \\ \, & \, & \, & \, \\
S$_{DDI} , L_{\ell}$ & {\bf Cis}  & L$_{DDI} , P_{\ell}$ & {\bf Pre}   \\ \, & \, & \, & \, \\ 
S$_{DDI} , P_{\ell}$ & {\bf Cis}   & P$_{DDI} , P_{\ell}$ & {\bf CR}   \\ \, & \, & \, & \, \\ \hline \hline
\end{tabular}  \end{center}  \end{table}

\noindent This composition is only a qualitative one.  Most notably, it fails in predicting the stability of critical points in $H_{bend}$.  Although all new critical points in $H_{DDI}$ and $H_{\ell}$ are stable, those in $H_{bend}$ exhibit three different types of linear stability.  Only two families, the {\bf L} and {\bf CR}, are bi-stable (EE).  Another discrepancy is that while $H_{bend}$ has 4 distinct bifurcations, simply considering the single resonance systems according to Fig.~\ref{indddibifs} would suggest that {\bf L} and {\bf Pre}, {\bf Orth} and {\bf CR} are born in only two bifurcations.  These differences are likely to be caused by the fact that in the full Hamiltonian, the two directions $(J_a, \psi_a)$ and $(J_b, \psi_b)$ are strongly coupled to each other, instead of forming independent subsystems $H_{DDI}+H_{\ell}$.  

\section[Quantum Survival Probabilities]{\underline{Quantum Survival Probabilities}}
\addtocontents{toc}{\protect\vspace*{7pt}}

Due to its resemblance to the transition state of Fig.~\ref{isomerization}, the local bending mode is expected to play an important role in the acetylene-vinylidene isomerization dynamics.   Recently Carter \et performed a Car-Parrinello type calculation on this system \cite{Carter}.  Dozens of classical trajectories are integrated, with the forces at each step obtained from {\it ab initio} calculations.  Surprisingly, many trajectories go back and forth across the barrier many times, before settling in the acetylene potential well.  The authors give a kinematic explanation: The time interval between the hydrogen crossing the barrier and swinging back is not long enough for the energy in the reaction coordinate to dissipate effectively.  This interpretation reconciles an existing discrepancy of vinylidene lifetime -- 0.04-4.6 $ps$ in \cite{Lineberger,C2H2Carrington} and 3.5 $\mu s$ in \cite{Explosion}.  The picosecond timescale is believed to be that of the initial decay of vinylidene, while the microsecond one is the vinylidene lifetime averaged over many recrossings. 

Such a ``recurring state" is necessarily decoupled from the rest of the vibrational manifold.  This is supported by the observation of Levin \et \cite{Explosion}.  In their Coulomb explosion experiment, while the vinylidene molecules have energy well above the reaction barrier, the estimated {\it dilution factor} is only $\approx$ 0.5, indicating that this is coupled to about one other state.   In another independent study, Schork and K\"{o}ppel compared the intrinsic lifetime of vinylidene to the local density of acetylene vibrational states, and concluded that extensive IVR is unlikely, at least for the lowest vibrational state of vinylidene \cite{C2H2Koeppel}.  Srivastava \et also suggested the acetylene-vinylidene isomerization is going to deviate significantly from the RRKM limit, due to the relatively low density of states \cite{LehmannFieldC2H2}.

The survival probability $P(t)$ has been an important tool in characterizing the dynamics of quantum states \cite{FieldLocalMode,TempsHCO1}.  For a quantum state $\vert \Psi \rangle $ written as an expansion in the eigenstate basis
\begin{align}
|\Psi (t) \rangle=\sum c_i \mbox{\,} e^{-iE_i t/ \hbar} |\phi_i \rangle
\end{align}
\noindent $P(t)$ is defined as the overlap between $\vert \Psi(0) \rangle$ and $\vert \Psi(t) \rangle$: 
\begin{align}
P(t) = & {|\langle \Psi(t)|\Psi(0)\rangle|}^2={| \left( \sum_i \langle \phi_i| c_i^* e^{iE_i t/ \hbar} \right) \left( \sum_j c_j |\phi_j \rangle \right) |}^2  \nonumber\\
= & \left( \sum_{i} |c_i|^2 \cos[E_i t / \hbar] \right)^2 + \left( \sum_{j} |c_j|^2 \sin[E_j t /\hbar ] \right)^2 \nonumber\\
=& \sum_i |c_i|^4 + 2 \sum_{i,j; i\ne j} |c_i|^2 |c_j|^2 \cos[(E_i-E_j)t / \hbar ]  \label{PtDef} \end{align}

The survival probability of an eigenstate is trivial: $P(t) \equiv $1.  For other initial states, the initial decay from $P(0)=$1 is dominated by how many states are directly coupled to $|\Psi \rangle$ \cite{WongGrubeleIVR}.  Oscillations (quantum beats) in $P(t)$ at intermediate time scales describe the usually partial recurrences of the initial state.   Finally, the long time average $\sum |c_i|^4 $, also known as the dilution factor \cite{NesbittField}, gives an estimate of the number of states participating in the IVR of the initial state.   

$P(t)$ is the quantum analogue of the classical autocorrelation function \cite{Siebrand}.  When it remains near unity for a sufficiently long time, then $\vert \Psi \rangle $ is strongly localized in some representation (as opposed to spreading over all space).  In the statistical (RRKM) limit, $P(t)$ quickly decays to a value close to the local density of states.

In $\S$ 3.3.3, we hypothesized that semiclassical wavefunctions are likely to localize near the critical points that are also extremum points in the reduced classical phase space.  To test this claim, $P(t)$ of states corresponding to {\bf Trans}, {\bf Cis}, {\bf L} and {\bf CR} overtone states are calculated for polyads $N_b=2$ to $20$.  While the {\bf Trans} and {\bf Cis} states are the normal ZOS $\vert n_4^0, 0^0 \rangle$ and $\vert 0^0, n_5^0 \rangle$, the {\bf L} and {\bf CR} states have to be constructed.  Here we use the method described by Field \et in \cite{FieldLocalMode} (which also contains calculation of $P(t)$ for selected {\bf L} states).   The {\bf L} and {\bf CR} states are defined as \fn{A derivation for the {\bf L} state can also be found in $\S$ 5.3.2. of \cite{DuanThesis}.}:
\begin{align}
|\mbox{L}_n \rangle &= \frac{1}{2^{{n}/{2}}} (\hat{a}_{4d}^\dagger+\hat{a}_{5d}^\dagger)(\hat{a}_{4g}^\dagger+\hat{a}_{5g}^\dagger)^{\frac{n}{2}} |0^0, 0^0 \rangle \\
| \mbox{CR}_n  \rangle &= \frac{1}{2^{{n}/{2}}} (\hat{a}_{4d}^\dagger+\hat{a}_{5d}^\dagger)(\hat{a}_{4d}^\dagger-\hat{a}_{5d}^\dagger)^{\frac{n}{2}} |0^0, 0^0 \rangle 
\end{align}
\noindent These are the ``perfect" {\bf L} and {\bf CR} states in the sense of containing equal amounts of trans- and cis- components.

\begin{landscape} \begin{figure}[hbtp]  
\begin{center}\includegraphics[width=7.53in]{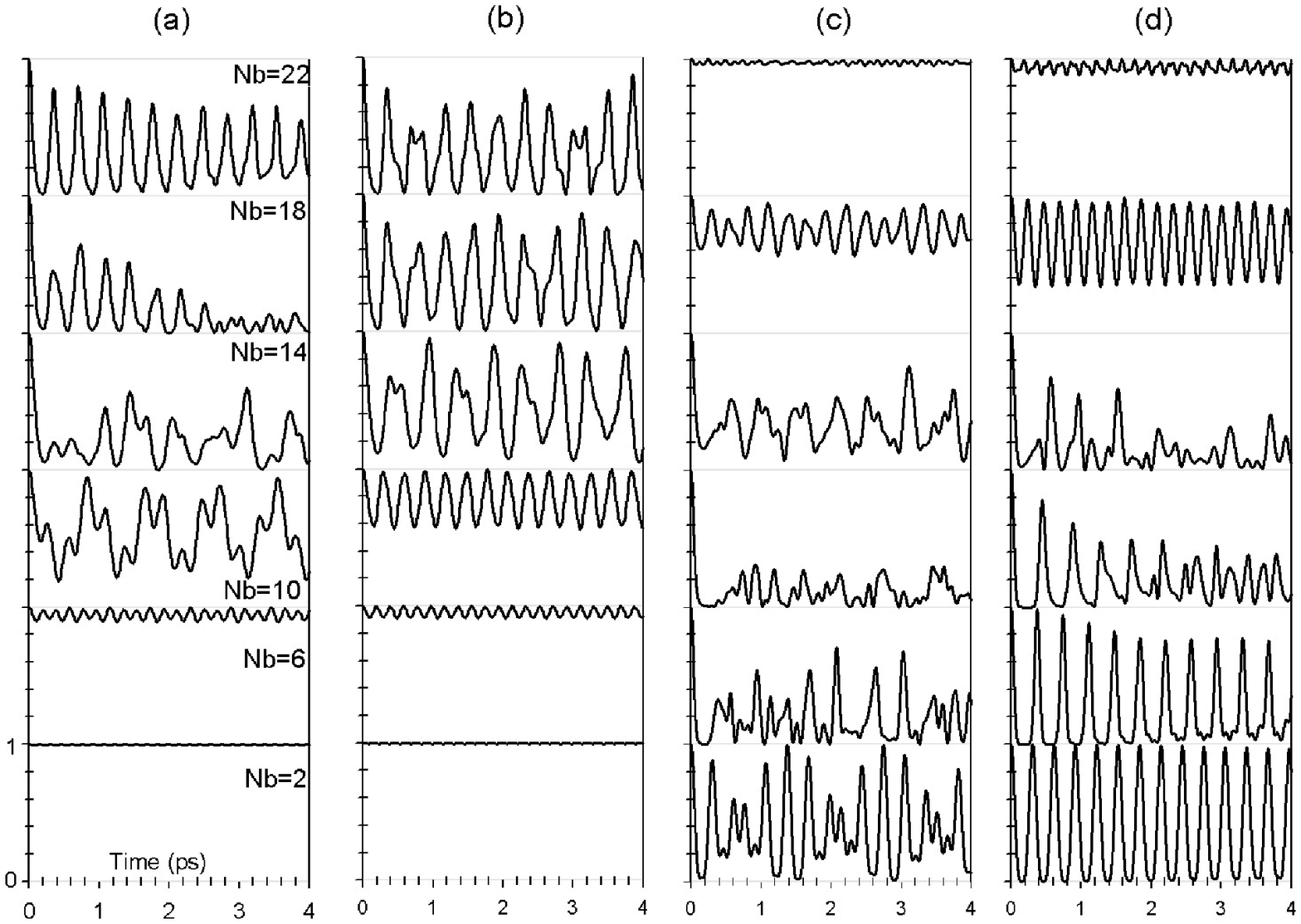}\end{center}
\cpn{Survival probability of selected bending states}{Survival probability of selected bending states.   Panels (a) - (d) display the P(t) of {\bf Trans}, {\bf Cis}, {\bf L} and {\bf CR} states with $N_b=$ 2 - 22, respectively. \label{SurvivalProb}} 
\end{figure}\end{landscape}

Fig.~\ref{SurvivalProb} shows the $P(t)$ values for the first 4 picoseconds.   Similar results had been obtained by Jacobson \et in Fig.~3 in \cite{C2H2FieldHeff}.   In panels (a,b), up to $N_b=$6, the {\bf Trans} and {\bf Cis} states have $P(t) \approx 1$, indicating that these overtone states are well localized near the respective critical points.  At $N_b=$10, the periodic oscillations become stronger, but the recurrences still reach close to 1.  The stability appears to be lost for the {\bf Trans} state at $N_b=$10 and for the {\bf Cis} state at $N_b=$14.  These changes occur soon after the first bifurcations of the {\bf Trans} and {\bf Cis} critical points in Fig.~\ref{fullbifs}, at $N_b=$8 and 10 respectively.  However, even at $N_b=$22 their dilution factors are still 0.4 and 0.3 for $|22^0, 0^0 \rangle$ and $|0^0, 22^0 \rangle$, respectively.  These two states are now coupled to 71 other ZOS states in the polyad.  Had the IVR been purely statistical, the dilution factor would have been $\frac{1}{72}$, an order of  magnitude smaller than the actual dilution factors.  This suggests the vibrational dynamics are far from the RRKM limit, even after then normal modes have been destabilized.

In panels (c) and (d) of Fig.~\ref{SurvivalProb}, before $N_b=14$ the $P(t)$ evolution of {\bf L} and {\bf CR} states remain strongly oscillating.  The almost sinusoidal oscillation between 0 and 1 at $N_b=$2 is due to the fact that each of these states is coupled to another {\bf L} or {\bf CR} states that is degenerate to it.  Between $N_b=$18-22, both {\bf L} and {\bf CR} states have $P(t)$ oscillating slightly under unity.  On the bifurcation diagram of Fig.~\ref{fullbifs}, this corresponds to ($J_a=0$) which is where the (EE)-type critical points {\bf L} and {\bf CR} approach their ``prefect" shape in the Cartesian coordinates.  Hence, quantum wavefunctions localize around these now stable modes.  Similar conclusions were reached by Jacobson \et through a visual match between the classical PO and semiclassical wavefunctions \cite{Jacobson15000,Jacobson10000}.

\newpage \newpage

In conclusion, the states in Fig.~\ref{SurvivalProb} with near unity $P(t)$ are in agreement with the (EE)-type critical points found in the classical analysis.  $P(t)$ of {\bf L} states in $N_b=$22 shows a lack of significant IVR, which is similar to the behavior of the isomerization states discussed in \cite{Explosion}.  The only difference is that the pure bending acetylene states presumably are still below the isomerization barrier.   With additional vibrational energy in the stretching DOF, these states or the superposition of several such states will be capable of going back and forth across the barrier many times.

\section[Summary and Conclusion]{\underline{Summary and Conclusion}}
\addtocontents{toc}{\protect\vspace{2.0in}}
\addtocontents{toc}{\normalfont\normalsize{\noindent Chapter \hfill Page}\protect\vspace*{30pt}}

The critical points analysis is performed on the C$_2$H$_2$ pure bending effective Hamiltonian.  In the $\lbrack N_b, 0 \rbrack $ polyads, 4 new families of critical points ({\bf L}, {\bf Orth}, {\bf Pre} and {\bf CR}) are born out of the normal {\bf Trans} and {\bf Cis} critical points in distinct bifurcations as $N_b$ is increased.  The bifurcation points where the new families are born correspond to qualitative changes in the classical phase space structure.  Computer-generated animations give visual insight into the nature of their motions in Cartesian space.   Similar bifurcation structure is obtained for the $\lbrack N_b, \ell \rbrack$ polyads with $\ell=2, 6$ and $10$.

Three of these new families are consistent with the results of other researchers, who had used more elaborate methods.  The bi-unstable {\bf Pre} family, on the other hand, can only be uncovered through an explicit search of critical points like ours, as presented in this thesis.

Separate consideration of the DD-I or $\ell$ resonances qualitatively accounts for the origin and nature of these four new modes.  Using the method of $\S$ 3.1, the dynamics induced by DD-I or $\ell$ resonance alone is analyzed.   Combining the critical points in the single-resonance Hamiltonians yields the same types of motion as those obtained as critical points in the full bending Hamiltonian.

The calculated quantum survival probability shows the break down of the normal modes description at intermediate polyad, as well as the emergence of {\bf L} and {\bf CR} modes as new stable modes of vibration at $N_b=$22.  These results demonstrate that the classical phase space structure is indeed reflected in the dynamics of the corresponding quantum system.%
\addtocontents{toc}{\protect\vspace*{12pt}}
\chapter[\protect\uppercase{Bifurcation Analysis of C$_2$H$_2$ Stretch-Bend}]{Bifurcation Analysis of C$_2$H$_2$ Stretch-Bend}\label{ch.ch5}%
\addtocontents{toc}{\protect\vspace{0.25in}}%

\section{Introduction}
\addtocontents{toc}{\protect\vspace*{7pt}}

Beyond the pure bending subsystem, the next step is extending the critical point analysis to all 7 vibrational DOF (4 bending and 3 stretching) of C$_2$H$_2$  explicitly.  In such a multidimensional system, an analytic detection method would be superior to both visual inspection and numerical search.

\subsection{5.1.1 The Effective Hamiltonian}
\addtocontents{toc}{\protect\vspace*{5pt}}

Three stretch-bend effective Hamiltonians of C$_2$H$_2$ have been published up to date \cite{Herman1995,Herman1999,Field2001}.  

\begin{figure}[hbtp]  
\begin{center}\includegraphics[width=4.756in]{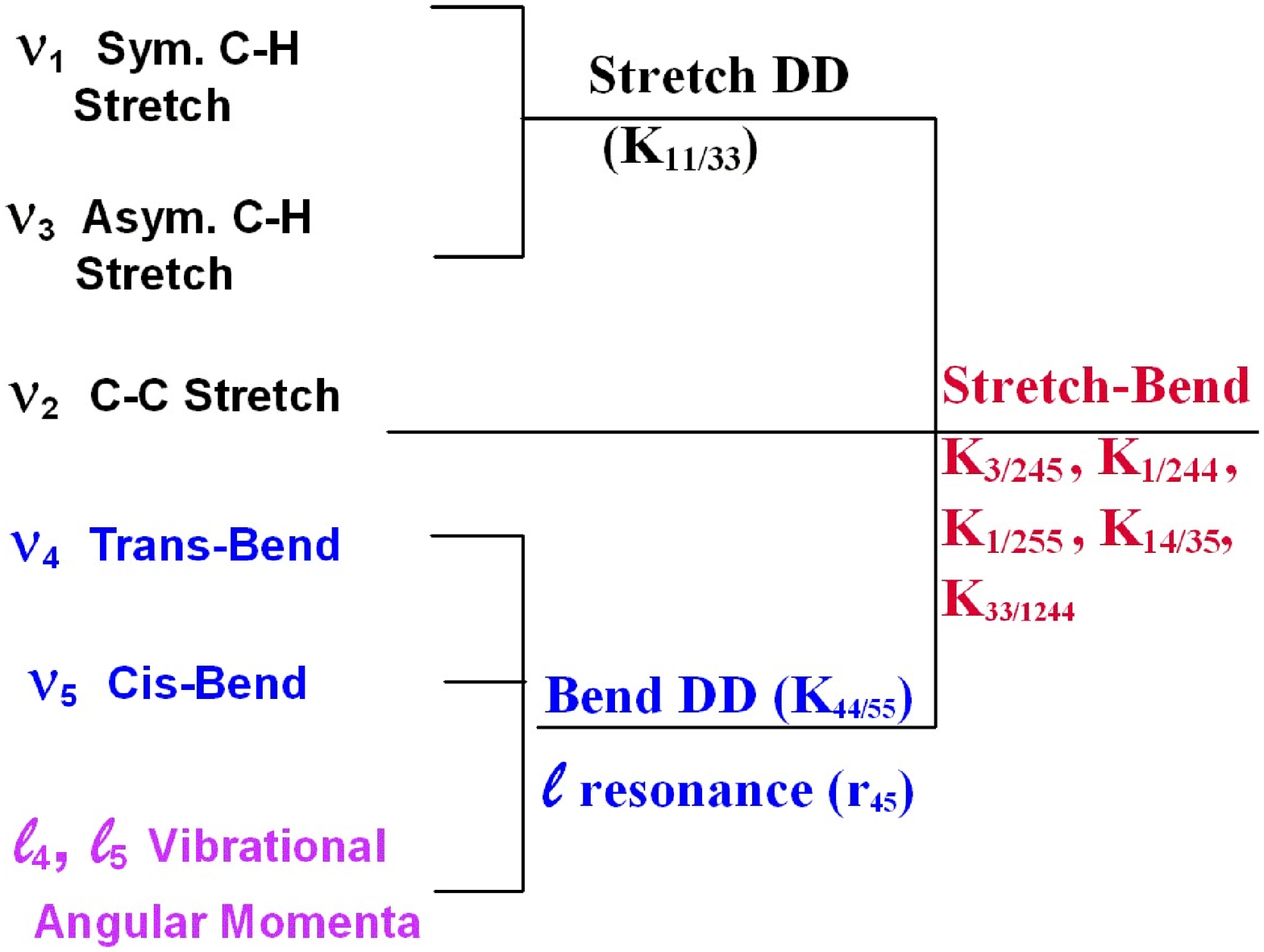}\end{center}
\cpn{Resonance couplings among the stretch-bend normal modes} {Resonance couplings among the stretch-bend normal modes, according to the Hamiltonian of \cite{Herman1995}. On the left are the normal modes of Fig.~\ref{normalmodes}. \label{sbcoupling}}
\end{figure}

In this chapter, we use the first Hamiltonian of Herman \et, because none of the other Hamiltonians has been studied theoretically \cite{Gaspard,DerivState}.  The Hamiltonian in \cite{Herman1995} contains 8 resonances: one Darling-Dennison type $K_{11/33}$ resonance coupling between the two normal C-H stretch modes, the $K_{44/55}$ (known as DD-I in Chapter 4) between the two normal bending modes, the $\ell$ resonance $r_{45}$, and $K_{3/245}$, $K_{1/244}$, $K_{1/255}$, $K_{14/35}$, $K_{33/1244}$ resonances which couple between the stretch and bend DOF.  This coupling structure is illustrated in Fig.~\ref{sbcoupling}.     In a typical stretch-bend polyad, the normal ZOS $\vert n_1, n_2, n_3, n_4^{\ell_4},n_5^{\ell_5} \rangle$ are all coupled by a complex web of resonances except for two special cases.  The pure bending polyads analyzed in Chapter 4 form an isolated subsystem with polyad number $N_s=0$.  The C-C stretch overtones $\vert 0,n_2,0,0^0,0^0 \rangle$ are not coupled by any resonance; therefore they are eigenstates of the Hamiltonian.

\newpage   \newpage   

In producing this Hamiltonians (as well as the other later ones), the fitting is based on only a small part of the states predicted by the polyad model.  A high-lying polyad typically contains dozens of states, out of which only a few have been experimentally detected and included in the fit.  This allows for some uncertainty in the high-order coefficients in the fit.  Especially lacking are states that contain both stretching ($N_s \geq 1$) excitation and high bending excitation ($n_4+n_5 \geq 12$).  

Using Heisenberg's Correspondence Principle in eqn. (\ref{Heisenberg}), the 7 DOF quantum Hamiltonian in \cite{Herman1995} is transformed to a classical one with 14 action-angle variables:  ($\tau_i, \phi_i$) for $i=1-5$ and ($\kappa_j, \chi_j$) for $j=4-5$.  The actions are related to the zero-order quantum numbers by
\begin{align}
\tau_i &=n_i  +\frac{1}{2}  & \mbox { \,\,\,\,\,\,for\,\,} i&=1, 2, 3 \label{eqn5.1} \\
\tau_i &=n_i +1 & \mbox { \,\,\,\,\,\,for\,\,} i&=4, 5  \label{eqn5.2} \\
\kappa_j & = \ell_j & \mbox {\,\,\,\,\,\,for\,\,} j&=4,5 \label{eqn5.3}
\end{align}

To simplify our analysis, the last five high-order parameters in Table~\ref{sbparam},  $y_{244}, K_{33/1244}, k_4, r_{445}, r_{545}$,  are ignored in the rest of this Chapter.   The classical Hamiltonian \cite{Gaspard} then has the following form:
\begin{align}
H_{sb}  = &H_0+H_v \label{sbhamil} 
\end{align}
with
{\small \begin{align}
H_0  = \sum_{i=1}^{5} \omega_i \tau_i+\sum_{i,j=1; i\le j}^5 x_{ij}\tau_i \tau_j + \sum_{i,j=1;i \le j}^5 g_{ij}\kappa_i \kappa_j \label{sbH0}\end{align} }
{\small \begin{align}
H_v= & \frac{K_{1133}}{2} \tau_1 \tau_3 \cos \lbrack 2 (\phi_1- \phi_3) \rbrack \nonumber\\
& +\frac{K_{3/245}}{4}\sqrt{\tau_2\tau_3}\{\sqrt{(\tau_4-\kappa_4)(\tau_5+\kappa_5)}\cos[\phi_2-\phi_3+\phi_4-\chi_4+\phi_5+\chi_5] \nonumber\\
& \quad +\sqrt{(\tau_4+\kappa_4)(\tau_5-\kappa_5)}\cos[\phi_2-\phi_3+\phi_4+\chi_4+\phi_5-\chi_5]\} \nonumber\\
& +\frac{K_{1/244}}{2}\sqrt{\tau_1\tau_2(\tau_4^2-\kappa_4^2)}\cos[\phi_1-\phi_2-2\phi_4] \nonumber\\
& +\frac{K_{1/255}}{2}\sqrt{\tau_1\tau_2(\tau_5^2-\kappa_5^2)}\cos[\phi_1-\phi_2-2\phi_5] \nonumber\\
& +\frac{K_{14/35}}{2}\sqrt{\tau_1\tau_3}\{\sqrt{(\tau_4-\kappa_4)(\tau_5-\kappa_5)}\cos[\phi_1-\phi_3+\phi_4-\chi_4-\phi_5+\chi_5] \nonumber\\
& \quad +\sqrt{(\tau_4+\kappa_4)(\tau_5+\kappa_5)}\cos[\phi_1-\phi_3+\phi_4+\chi_4-\phi_5-\chi_5] \} \nonumber\\
& +\frac{K_{44/55}}{2}\sqrt{(\tau_4^2-\kappa_4^2)(\tau_5^2-\kappa_5^2)}\cos[2(\phi_4-\phi_5)] \nonumber\\
& +\frac{r_{45}}{2}\sqrt{(\tau_4^2-\kappa_4^2)(\tau_5^2-\kappa_5^2)}\cos[2(\chi_4-\chi_5)]   \label{sbclassicalH}
\end{align} }

The values of the parameters in $H_{sb}$ are listed in Table~\ref{sbparam} \fn{It was found later that the $K_{14/35}$ value appears as $29.044$ in \cite{Herman1995} and as $29.944$ in \cite{Gaspard}.  This appears as a minor discrepancy compared to its later revision to $15.66$ in \cite{Herman1999} and then to $16.614$ in \cite{Field2001}.}.   The 3 polyad numbers correspond to 3 classical constants of motion:
\begin{align}
P= & 5 \tau_1 + 3\tau_2 + 5\tau_3 + \tau_4 + \tau_5 = N_t + \frac{15}{2} \label{ClassicalP} \\
R= & \tau_1 + \tau_2+ \tau_3 = N_s+\frac{3}{2} \label{ClassicalR}  \\
L= & \kappa_4+\kappa_5 = \ell  \label{ClassicalL} 
\end{align}

\begin{table}[hbt]
\cpn{C$_2$H$_2$ stretch-bend effective Hamiltonian} {C$_2$H$_2$ stretch-bend effective Hamiltonian from \cite{Herman1995} and published in \cite{Gaspard} for the classical Hamiltonian.   The parameters are in units of cm$^{-1}$. \label{sbparam}}
\begin{center}  \vspace{0.2in}
\begin{tabular}{|c|r|c|r|c|r|c|r|} \hline \hline \, & \, & \, & \, & \, & \, & \, & \, \\
$\omega_1$ & 3501.537 & $x_{15}$ & -10.09  & $x_{45}$ & -2.311  & $K_{14/35}$ & 29.944 \\ \, & \, & \, & \, & \, & \, & \, & \, \\
$\omega_2$ & 2013.425 & $x_{22}$ & -7.802 & $x_{55}$ & -2.492 &  $K_{44/55}$ & -12.909 \\ \, & \, & \, & \, & \, & \, & \, & \, \\
$\omega_3$ & 3417.644 & $x_{23}$ & -5.882 & $g_{44}$ & 0.4181 &   $r_{45}$ & -6.09 \\ \, & \, & \, & \, & \, & \, & \, & \, \\
$\omega_4$ & 621.692 & $x_{24}$ & -12.841 & $g_{45}$ & 6.603 &  --- & --- \\ \, & \, & \, & \, & \, & \, & \, & \, \\
$\omega_5$ & 746.773 & $x_{25}$ & -1.829 & $g_{55}$ & 3.676 & $y_{244}$ & 0.1522 \\ \, & \, & \, & \, & \, & \, & \, & \, \\
$x_{11}$ & -24.758 & $x_{33}$ & -27.483 & $K_{11/33}$ & -102.816  & $K_{33/1244}$ & 6.38 \\ \, & \, & \, & \, & \, & \, & \, & \, \\
$x_{12}$ & -11.199 & $x_{34}$ & -10.617 & $K_{3/245}$ & -16.698  & $k_{4}$ & -1.315   \\ \, & \, & \, & \, & \, & \, & \, & \, \\
$x_{13}$ & -103.386 & $x_{35}$ & -8.676 & $K_{1/244}$ & 6.379 & $r_{445}$ & 0.1255 \\ \, & \, & \, & \, & \, & \, & \, & \, \\
$x_{14}$ & -12.98 & $x_{44}$ & 3.595 &  $K_{1/255}$ & 6.379 & $r_{455}$ & -0.225  \\   \, & \, & \, & \, & \, & \, & \, & \, \\ \hline \hline
\end{tabular}  \end{center}  \end{table}

\subsection{5.1.2 Overview of Existing Studies}
\addtocontents{toc}{\protect\vspace*{5pt}}

To our knowledge, the stretch-bend effective Hamiltonians has never been analyzed with all the resonances.  We are aware of only three relevant studies:

\begin{itemize}
\item Pals and Gaspard \cite{Gaspard} investigated the recurrences of the classical trajectories.   Although the study did include the stretching DOF, the authors were mostly focused on the pure bending subsystem.

\item Hasegawa and Someda \cite{DerivState} analyzed the quantum dynamics using a perturbative method.  The focus was the short-time evolution of 3 types of quantum ZOS: $|0,0,0,n_4^0,0^0 \rangle$, $|1,3,0,6^0,0^0 \rangle$ and $|0,0,n_3,0^0,0^0 \rangle$.

\item Kellman \et investigated the planar system ($\ell_4=\ell_5=0$) with 3 resonances: $K_{11/33}, K_{3/245}$ and $K_{44/55}$ in a diabatic correlation approach \cite{KellmanEPJD}.  A series of states termed the ``primary subpolyad" are identified, which carries most of the intensity in the experimental spectra.  The states are then fit to a single-resonance Hamiltonian in an effort to interpret the spectral patterns. 
\end{itemize}

Due to the dimensionality of the problem, none of these studies explicitly considered all the major resonances in the effective Hamiltonian.  Since the approach presented in this thesis is designed to be dimensionality-independent, we believe it is more suitable for analyzing the stretch-bend acetylene system.

\section{Preliminary Considerations}
\addtocontents{toc}{\protect\vspace*{7pt}}

In Chapter 4 it was shown that the normal bending modes are destabilized at increased internal energy (polyad numbers) by the resonance couplings, and new stable bending modes are born in the bifurcations.  In the stretch-bend system, the PO search by Prosmiti and Farantos also indicates that the highly excited stretch-bend system retains some regularity \cite{C2H2Farantos1}.  Therefore one can expect a similar to the pure bending case to exist in the stretch-bend dynamics, provided there are sufficient excitation and coupling.   

\subsection{5.2.1 The Stretch Overtone Polyads}
\addtocontents{toc}{\protect\vspace*{5pt}}

The critical points analysis starts with solving for all critical points in the 4 DOF (7 DOF - 3 polyad numbers) reduced classical Hamiltonian.  In obtaining the preliminary results, we focus on a single series of polyads of interest, as opposed to varying all three polyad numbers independently.  

Of the five normal modes of C$_2$H$_2$ depicted in Fig.~\ref{normalmodes}, the pure bending subsystem was analyzed in Chapter 4, and the C-C stretch ($\tau_2$) overtones form an isolated subsystem.   This leaves the obvious question:  what could happen to the two C-H normal stretches, as they are excited to higher energy?    Although the C-H stretching dynamics has been actively studied using two-mode models \cite{C2H2HCAO}, their coupling to the other vibrational modes is complex and poorly understood.   This motivates us to investigate the following polyad series, which includes the symmetric and antisymmetric C-H stretch overtones \fn{Here in the purely classical analysis, the quantization requirement that the action in each mode exceed the zero-order energy (see eqns.~\ref{eqn5.1}, \ref{eqn5.2}) will be ignored.}:
\begin{eqnarray}
\{ \tau_1, \tau_2, \tau_3, \tau_4, \tau_5, \kappa_4, \kappa_5 \} = \{ \tau_1, 0, 0, 0, 0, 0, 0 \}, \{0, 0, \tau_3, 0, 0, 0, 0 \} \label{CHovertone}
\end{eqnarray}
\noindent  The polyads containing them are found by substituting eqn. (\ref{CHovertone}) into eqns. (\ref{ClassicalP}-\ref{ClassicalL}) to give 
\begin{align}
P &=5 R, & L &=\ell=0
\end{align}
\noindent These $\lbrack P, R, \ell \rbrack=\lbrack 5 R, R, 0 \rbrack$ polyads are henceforward referred to as the {\it stretch overtone polyads}.   The upper limit of $R$ in this study is set at 8, as the effective Hamiltonian in \cite{Herman1999} includes up to 6 quanta of C-H overtones excitation ($R=7.5$).  

\subsection{5.2.2 Stability of the Normal C-H Stretch Overtones}
\addtocontents{toc}{\protect\vspace*{5pt}}

The first consideration is the stability of the normal C-H stretch overtone states.   For this purpose, the classical trajectories very close to the normal mode overtones in eqn. (\ref{CHovertone}) are integrated.  The overtone is classically stable if the deviation of nearby trajectories remains small.   The onset of large-amplitude oscillations indicates the destabilization of the corresponding overtone trajectory.

\begin{figure}[hbtp]  
\begin{center}\includegraphics[width=4.92in]{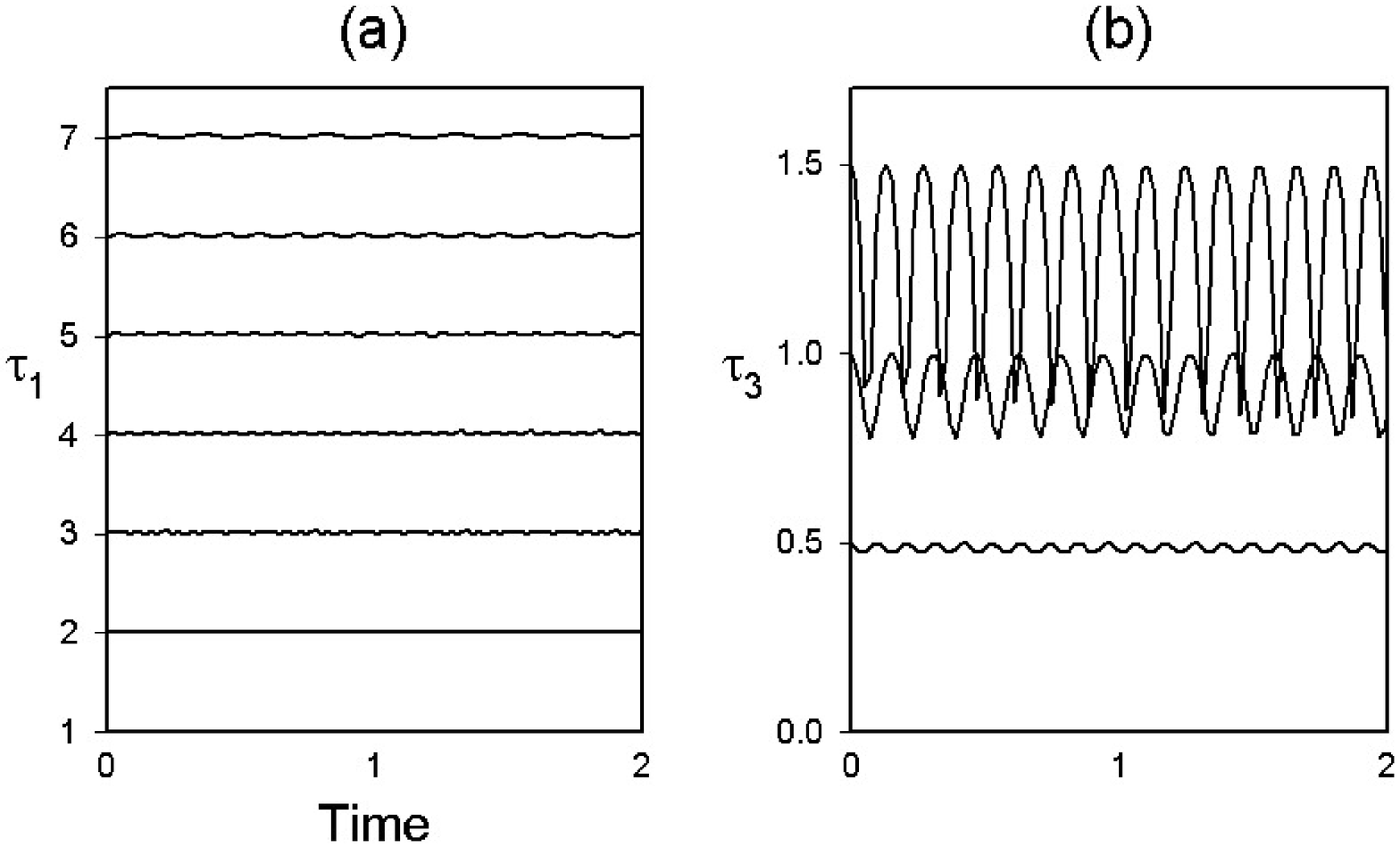}\end{center}
\cpn{Classical stability of C-H normal stretch overtones}{Classical stability of C-H normal stretch overtones.  Panel (a) and (b) are the $\tau_1$ and $\tau_3$ values of trajectories very close to the respective overtone conditions in eqn. (\ref{CHovertone}).  \label{OvertoneStability}}
\end{figure}

The results are displayed in Fig.~\ref{OvertoneStability}.  In panel (a), the symmetric stretch overtone $\tau_1$ remains stable to at least $\tau_1=6.5$ ($n_1=6$).  In contrast, the antisymmetric $\tau_3$ overtone becomes unstable at as low as $\tau_3=1.5$ ($n_3=1$).  Two resonances may be responsible for the destabilization of $\tau_3$ overtones:  $K_{11/33}$ and $K_{3/245}$.

\newpage     \newpage 

\noindent \underline{Single $K_{11/33}$ Resonance } \,\,\,  The single $K_{11/33}$ resonance system 
\begin{align}
H_{11/33}  = H_0+\frac{K_{1133}}{2} \tau_1 \tau_3 \cos \lbrack 2(\phi_1-\phi_3) \rbrack
\end{align}
\noindent is analyzed for the stretch overtone polyads using the formulation of $\S$ 3.1, with $m:n=2:2$.   After the other actions $\tau_2, \tau_4, \tau_5, \kappa_4, \kappa_5$ are set to be zero, ($\tau_1, \tau_3$) form a two-mode system.  The 2 DOF reduced phase space is described be canonical variables $I_z, \psi$:
\begin{align}
I_z & = \frac{\tau_1-\tau_3}{2}, & \psi & = \phi_1- \phi_3  \label{K1133Canonical}
\end{align}

\begin{figure}[hbtp]  
\begin{center}\includegraphics[width=4.5in]{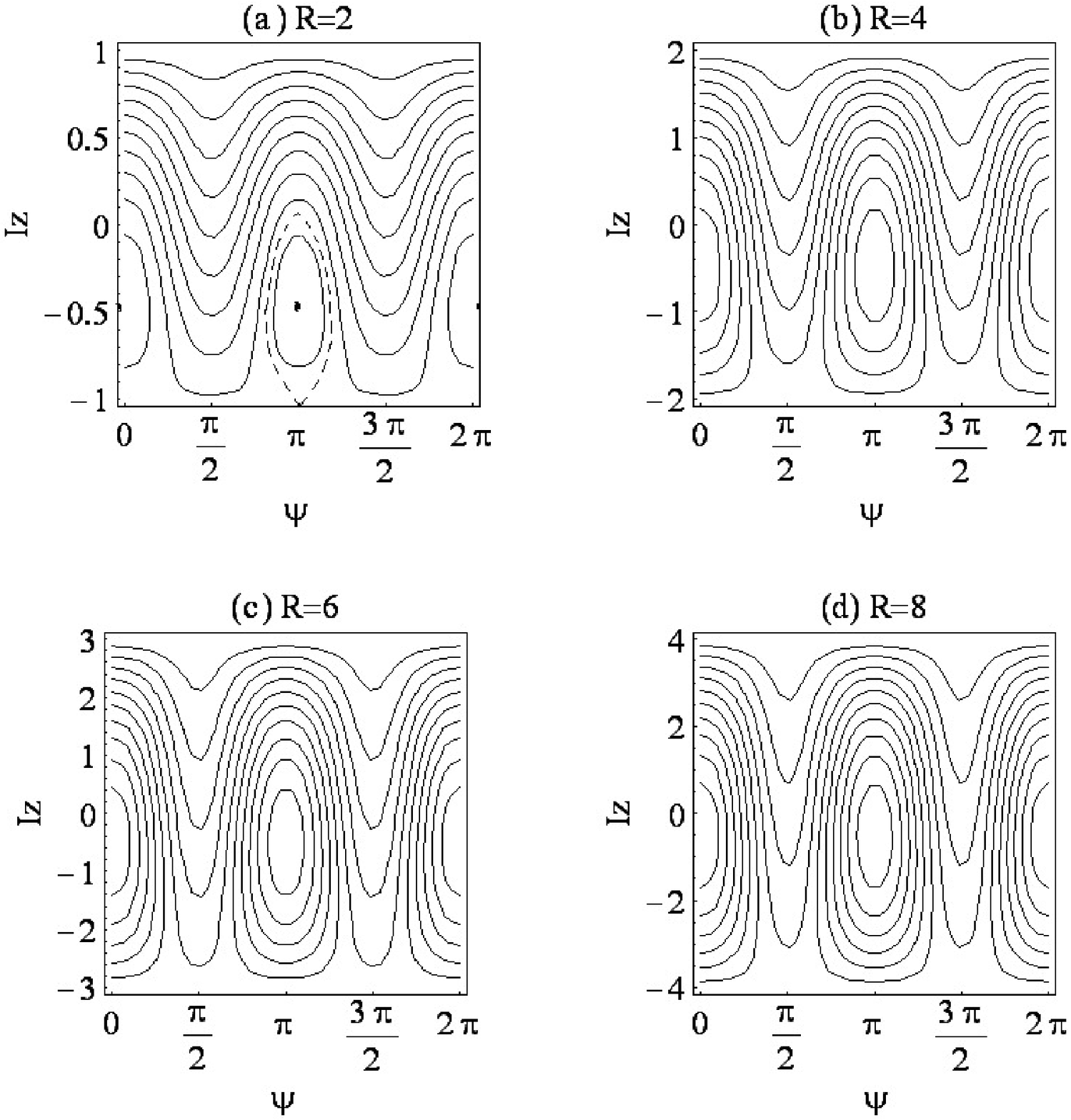}\end{center}
\cpn{Reduced phase space of $H_{11/33}$ Hamiltonian}{Reduced phase space of $H_{11/33}$ Hamiltonian, and the constant energy contours.  In panels (a)-(d) $R=2, 4, 6$ and $8$ respectively.  The dashed line in panel (a) labels the separatrix crossing the unstable $\tau_3$ overtones (the $I_z=-1$ line).  The black dots at $\psi=0, \pi$ and $2\pi$ indicate the stable local mode critical point. \label{PPS1133}}
\end{figure}

The constant energy contours (at arbitrary energy) of $H_{11/33}$ in the ($I_z, \psi$) space are plotted in Fig.~\ref{PPS1133}.  Each of these panels is the Mecartor projection of the corresponding PPS, as in panel (a) of  Fig.~\ref{phasesphere}.  The top and bottom of $I_z$ represent the $\tau_1$ and $\tau_3$ overtones, respectively.  The contours display the same qualitative features as semiclassical trajectories of the eigenstates.  The critical points (dark dots in panel a) at $\psi=0, \pi$ and $2\pi$ are the stable (E) local C-H stretch critical points \fn{The points at $0$ and $2\pi$ are identified with each other.  They are also energetically degenerate with the one at $\pi$, as the two local C-H stretches are classically equivalent.}.  The local model of C-H stretch has been previously used in studying the two coupled C-H stretch oscillators \cite{ChildLawtonC2H2,Klemperer}.

The curves at the top of all panels of Fig.~\ref{PPS1133} that run across $\lbrack 0, 2\pi \rbrack$ indicate the $\tau_1$ overtone is a stable critical point.  The $\tau_3$ overtone, in contrast, becomes {\it unstable} on the PPS from $\tau_3=2$ onward.  These qualitative observations verify the stabilities shown in Fig.~\ref{OvertoneStability}.

\newpage     \newpage 

\noindent \underline{Single $K_{3/245}$ Resonance } \,\,\,  The single $K_{3/245}$ resonance system 
{\small \begin{align}
H_{3/245}  & = H_0+\frac{K_{3/245}}{4}\sqrt{\tau_2\tau_3}\{\sqrt{(\tau_4-\kappa_4)(\tau_5+\kappa_5)}\cos[\phi_2-\phi_3+\phi_4-\chi_4+\phi_5+\chi_5] \nonumber\\
& \quad +\sqrt{(\tau_4+\kappa_4)(\tau_5-\kappa_5)}\cos[\phi_2-\phi_3+\phi_4+\chi_4+\phi_5-\chi_5]\} \label{k3245}
\end{align} }
\noindent was formally analyzed by Rose and Kellman in \cite{Rose2345}.  However, they failed in performing a calculation using actual parameters of C$_2$H$_2$.   Here we perform the calculation for the stretch overtone polyads, with actions $\tau_1=\kappa_4=\kappa_5=0$ \fn{In \cite{Rose2345}, the elimination of $\kappa_i$ from consideration is not rigorous, since the $K_{3/245}$ resonance does couple $\ell_4, \ell_5$, and therefore $\kappa_4, \kappa_5$.}.   The $\tau_3$ overtones are then contained in the action space with
\begin{align}
\tau_2 = \tau_4 = \tau_5 = R-\tau_3
\end{align}
\noindent Because the derivation in \cite{Rose2345} failed to use the proper $d_i$ for the doubly-degenerate bends,  the canonical variables ($J, \varphi$) \fn{The notations are intentionally different from those used in $H_{11/33}$ above to avoid confusion.} of the 1 DOF reduced phase space are redefined as:
\begin{align}
J &= 14 \tau_2 + \tau_3 - 7 \tau_4 - 7 \tau_5, & \varphi &= \phi_3-\phi_2-\phi_4-\phi_5      
\end{align}

\begin{figure}[hbtp]  
\begin{center}\includegraphics[width=4.5in]{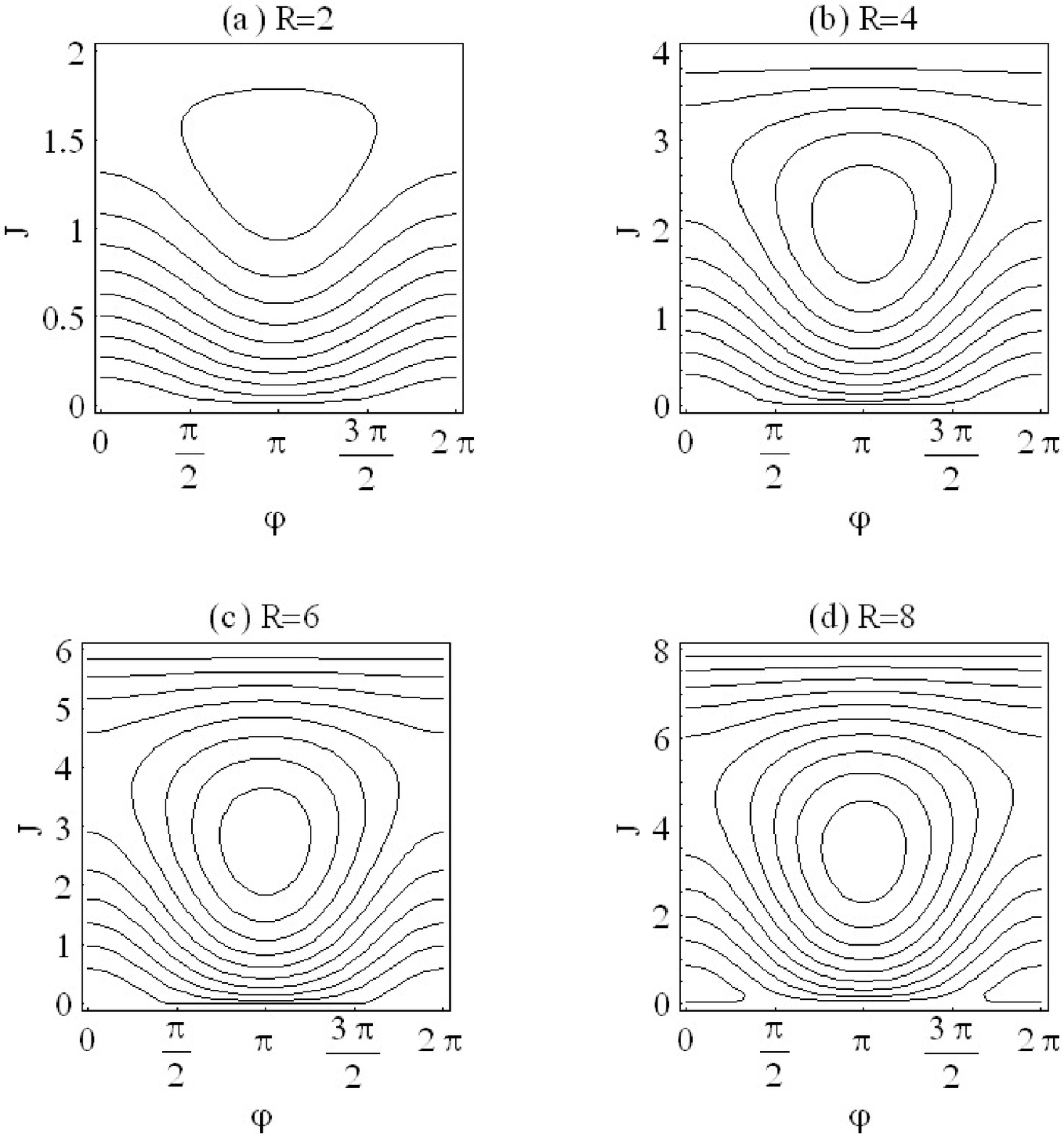}\end{center}
\cpn{Reduced phase space of $H_{3/245}$ Hamiltonian}{Reduced phase space of $H_{3/245}$ Hamiltonian, and the constant energy contours.  In panels (a)-(d) $R=$ 2, 4, 6 and 8.  \label{PPS3245}}  \end{figure}

Fig.~\ref{PPS3245} displays the arbitrary constant energy contours in ($J, \varphi$) space.  The top of each panel (maximum $J$) corresponds to the $\tau_3$ overtones and the bottom to 
\begin{align}
\tau_2 & =\tau_4=\tau_5 = R, & \tau_3 &=0\nonumber\\
\end{align}
\noindent In panel (a), the $K_{3/245}$ resonance induces a bifurcation of the $\tau_3$ overtone. In this process, a stable critical point is created at $\varphi=\pi$.  However, the overtone itself remains stable at higher $R$ (panels b-d).   

\newpage     \newpage 

From the above separate consideration of $H_{11/33}$ and $H_{3/245}$, the $\tau_3$ overtones are first destabilized by the $K_{11/33}$ resonance below $R=2$, and the {\it local} C-H stretch becomes the stable mode.

\subsection{5.2.3 Effect of Stretch-Bend Resonances}
\addtocontents{toc}{\protect\vspace*{7pt}}

After its creation, the local C-H stretch may be further perturbed by the stretch-bend resonances to couple with the bending DOF.   Of the four stretch-bend resonances ($K_{3/245}$, $K_{1/244}$, $K_{1/255}$ and $K_{14/35}$), it is unlikely that they are equally important at all parts of a stretch overtone polyad.  We use a simplified version of Chirikov analysis \cite{Chirikov} in order to estimate of which resonances are important in affecting the local C-H stretch.  When $\kappa_4, \kappa_5$ are left out of consideration, the ``strength" of each of these 4 resonances as well as $K_{11/33}$ can be measured by how far the zero-order frequencies $\omega_i= \partial H_0 / \partial \tau_i$ are tuned towards the integer ratio $m_i$ corresponding to an exact resonance condition.  At the exact integer ratio, $\omega_i$ satisfy  \cite{OxtobyRice}:
\begin{align}
\sum_i m_i \frac{\partial H_0}{\partial \tau_i} = 0 \label{resonanceplane}
\end{align}
\noindent The vector with integer components $\{ m_1, m_2, m_3, m_4, m_5 \}$ corresponds to the resonance vectors mentioned in $\S$ 2.1.  In this case they are:
\begin{align}
K_{11/33}: & \{ 2,0,-2,0,0 \}, & K_{3/245}: & \{ 0,-1,1,-1,-1 \} \nonumber\\
K_{1/244}: & \{ 1,-1,0,-2,0 \}, & K_{1/255}: & \{ 1,-1,0,0,-2 \} \nonumber\\
K_{14/35}: & \{ 1,0,-1,1,-1 \} & \, & \, \nonumber
\end{align}

For each resonance, eqn. (\ref{resonanceplane}) defines a 4-dimensional hypersurface in the $\{ \tau_1, \tau_2, \tau_3, \tau_4, \tau_5 \}$ space.  Because of the conservation of $P$ and $R$, only 3 of the actions can be independently varied within a polyad.  It is therefore possible to represent these hypersurfaces as 2-dimensional surfaces, which are referred to as {\it resonance planes} in a 3-dimensional volume ($J_1$, $J_2$, $J_3$)
\begin{align}
J_1= & \tau_1-\tau_3 & \in & [ -R, R ] \\
J_2= & \tau_2  & \in & [ 0, R ] \\
J_3= & \tau_4-\tau_5 & \in & [ -2 R, 2 R ] 
\end{align}
\noindent This volume is further constrained by the requirement that all the $\tau_i$ should be non-negative.

\begin{figure}[hbtp]  
\begin{center}\includegraphics[width=5.747in]{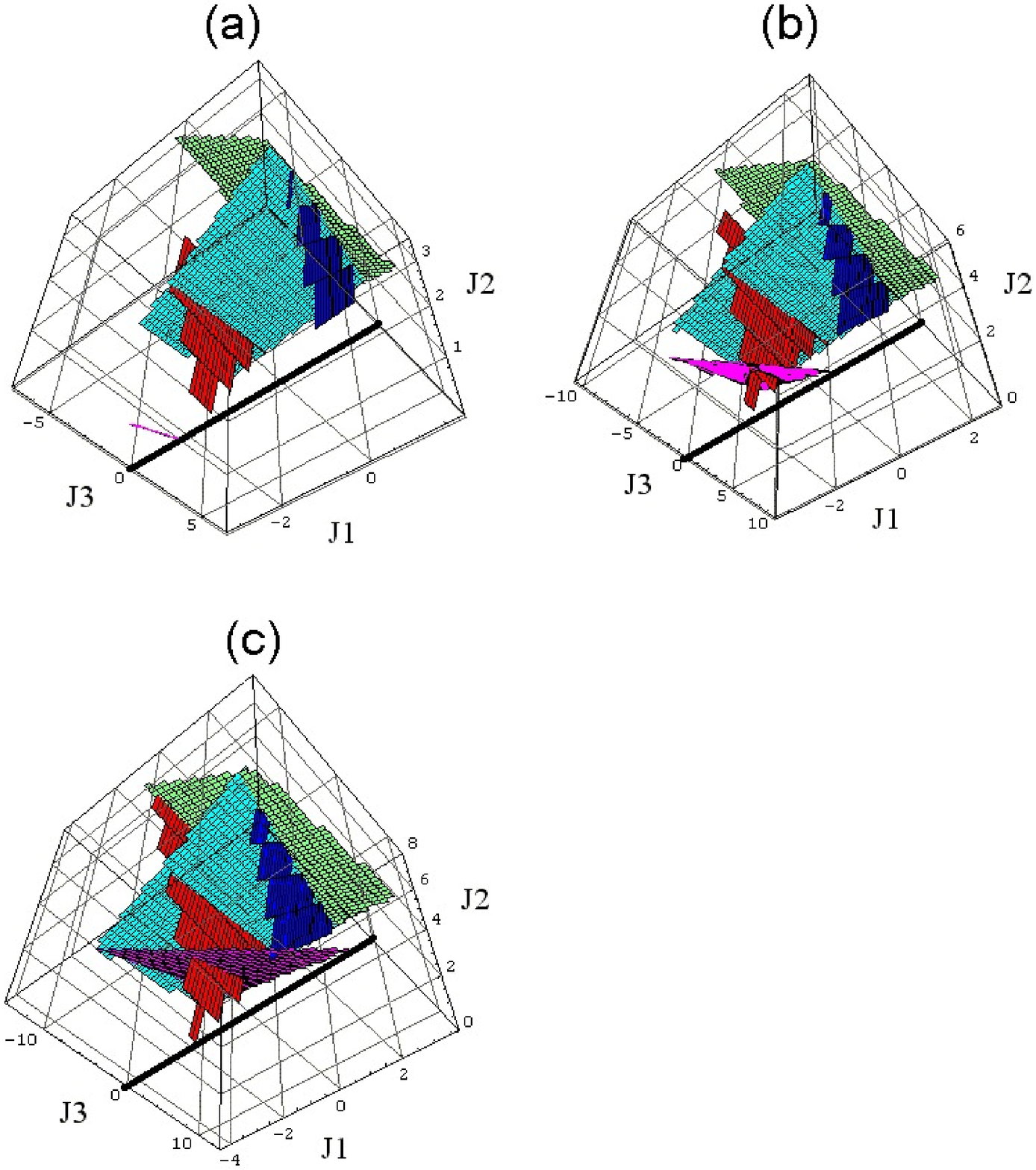}\end{center}
\cpn{Frequency resonance planes in the stretch overtone polyads}{Frequency resonance planes in the stretch overtone polyads.  Panels (a), (b) and (c) display the cases with $R=$ 4, 6 and 8, respectively.  The color coding is: $K_{11/33}$ (red), $K_{3/245}$ (turquoise), $K_{1/244}$ (magenta), $K_{1/255}$ (green), $K_{14/35}$ (navy). The C-H stretch system with $\tau_2=\tau_4=\tau_5=0$ is indicated by the thick black line in each panel. \label{SBChirikov}}
\end{figure}

Fig.~\ref{SBChirikov} depicts the resonance planes in ($J_1, J_2, J_3$).  In panel (a), at $R=4$ the C-H stretch system (thick line) is separately in contact with $K_{11/33}$ (red) and $K_{1/244}$ (magenta) resonance planes at two places.  The $K_{1/244}$ resonance only perturbs but does not destabilize the $\tau_1$ overtones, according to Fig.~\ref{OvertoneStability}.  In panels (b) and (c), the C-H stretch system first interacts with the  $K_{11/33}$ resonance plane, which is next intersected by the $K_{1/244}$ plane.   In all 3 panels the $K_{3/245}$ and $K_{1/255}$ planes are located at another side of the ($J_1, J_2, J_3$) space, and can only interact with the C-H stretch system via $K_{11/33}$ and other resonance planes.  Because the $K_{14/35}$ resonance term vanishes in the absence of bending excitation, the $K_{14/35}$ its plane (navy) also cannot directly interact with the C-H stretch system.

\newpage   \newpage 

\section{Critical Points Analysis}
\addtocontents{toc}{\protect\vspace*{7pt}}

Based on the results of $\S$ 5.2.2-5.2.3, in this subsection we compute critical points in the following three cases:
\begin{enumerate}
\item The ($\tau_1, \tau_3$) subsystem with $K_{11/33}$ resonance;
\item The ($\tau_1, \tau_2, \tau_3, \tau_4$) subsystem with both  $K_{11/33}$ and $K_{1/244}$ resonances;
\item The full $H_{sb}$ of eqn. (\ref{sbclassicalH}) with all 7 resonance couplings.
\end{enumerate}

\subsection{5.3.1 Computational Details}
\addtocontents{toc}{\protect\vspace*{5pt}}

Using the procedure described in Appendix A, a canonical transformation is selected that expresses $H_{sb}$ in the following new action-angle variables: 3 pairs of trivial action-angle variables ($P, R, L, \theta_P, \theta_R, \theta_L$) for the polyad numbers (defined in \ref{ClassicalP}-\ref{ClassicalL}) and their conjugate angles, and 4 pairs of non-trivial ones ($I_1, I_2, I_3, I_4, \Psi_1, \Psi_2, \Psi_3, \Psi_4$) which span the 8-dimensional reduced phase space.  The angles $\Psi_1-\Psi_4$ are chosen to correspond to the $K_{11/33}$, $K_{1/244}$, $K_{44/55}$ and $\ell$ resonances, respectively.  The simplest canonical transformation we have found so far is:
\begin{align}
I_1=& 12 \tau_1 + 8 \tau_2 + 11 \tau_3 + 2 \tau_4 + 2 \tau_5 + 2 \kappa_4 + 2 \kappa_5, & \Psi_1=& \phi_1-\phi_3 \\
I_2=& 6 \tau_1 + 3 \tau_+ 6 \tau_3+ \tau_4+ \tau_5 + \kappa_4 + \kappa_5, & \Psi_2=& \phi_1-\phi_2-2\phi_4 \\
I_3=& 6 \tau_1 + 4 \tau_2 + 6 \tau_3 + \tau_4 + \kappa_4 + \kappa_5, & \Psi_3=& \phi_4-\phi_5 \\
I_4=& -12 \tau_1 - 8 \tau_2 - 12 \tau_3 - 2 \tau_4 - 2 \tau_5 - \kappa_4 - 2 \kappa_5, & \Psi_4=& \chi_4-\chi_5 
\end{align}
\noindent  $\Psi_1$ is the relative phase angle between the symmetric $\tau_1$ and antisymmetric $\tau_3$ oscillators.   $\Psi_3$ and $\Psi_4$ are identical to $\psi_a$ and $\psi_b$ in Chapter 4, respectively.  

Critical points in the reduced phase space are defined by:
\begin{align}
\frac{\partial H_{sb} }{\partial \Psi_1} =\frac{\partial H_{sb}}{\partial \Psi_2 }=\frac{\partial H_{sb}}{\partial \Psi_3}=\frac{\partial H_{sb}}{\partial \Psi_4} & =0 \label{SBFP2} \\
\frac{\partial H_{sb}}{\partial I_1} =\frac{\partial H_{sb}}{\partial I_2}=\frac{\partial H_{sb}}{\partial I_3}=\frac{\partial H_{sb}}{\partial I_4} & =0 \label{SBFP1} 
\end{align}

In eqns. (\ref{SBFP2}), $\Psi_i$ only appear in the form of $\cos$ functions.  Similarly to the treatment of the pure bending subsystem ($\S$ 4.3.1), a sufficient condition for them to be simultaneously satisfied is:

{\footnotesize \begin{align}
& \sin[2\Psi_1]=\sin[\Psi_2]=\sin[2\Psi_3]=\sin[2\Psi_4]=\sin[\Psi_2-2\Psi_3] \nonumber\\
& \quad = \cos[\Psi_1+\Psi_3 \pm \Psi_4]\sin[\Psi_1+\Psi_3 \mp \Psi_4] =\sin[\Psi_1-\Psi_2+\Psi_3 \pm \Psi_4]=0 \label{sbangles}
\end{align}}

One can therefore fix $\Psi_i$ to the discrete values satisfying eqn. (\ref{sbangles}), and only solve the remaining 4 equations in (\ref{SBFP1}).   The latter are transformed by the following substitutions:
\begin{eqnarray}
&u_1 =\sqrt{\tau_1}, u_2 =\sqrt{\tau_2}, u_3 =\sqrt{\tau_3}, \nonumber\\
&u_4 =\sqrt{\tau_4+\kappa_4}, u_5 =\sqrt{\tau_4-\kappa_4}, u_6 =\sqrt{\tau_5+\kappa_5}, u_7 =\sqrt{\tau_5-\kappa_5} \label{sbtransform} 
\end{eqnarray}
\noindent and then multiplied by appropriate factors to remove their denominators.  The result is 7 simultaneous polynomial equations:
\begin{align}
& \frac{\partial H_{sb}(u_i)}{\partial I_1(u_i)} \cdot (u_1 u_3)^{1/2} = 0 \label{3EQ1} \\
& \frac{\partial H_{sb}(u_i)}{\partial I_2(u_i)}  \cdot (u_1 u_2 u_4 u_5)^{1/2} = 0 \label{3EQ2} \\
& \frac{\partial H_{sb}(u_i)}{\partial I_3(u_i)}  \cdot (u_4 u_5 u_6 u_7)^{1/2} = 0 \label{3EQ3} \\
& \frac{\partial H_{sb}(u_i)}{\partial I_4(u_i)}  \cdot (u_4 u_5 u_6 u_7)^{1/2} = 0 \label{3EQ4} \\
& 5 u_1^2 +3 u_2^2+5 u_3^2 + \frac{1}{2}(u_4^2 + u_5^2+ u_6^2 + u_7^2 ) = P \label{sbconstrain1} \\
& u_1^2 +u_2^2 +u_3^2 = R \label{sbconstrain2} \\
& u_4^2 - u_5^2 + u_6^2 - u_7^2 = 0 \label{sbconstrain3}
\end{align}
\noindent The 7 unknown variables $u_1-u_7$ are then solved for in {\it PHCpack} \cite{PHC}.  

\noindent \underline{Treatment of Subsystems } \,\,\, When any of the $u_i$ vanishes, at least one of eqns. (\ref{3EQ1}-\ref{3EQ4}) becomes unphysical.   This happens when there is zero action in any normal mode $\tau_i$, or when $\tau_j = \pm \kappa_j$.  

\noindent \,\,\,
 
Special consideration is required for these cases.  The contributions from the resonance(s) involved with the vanishing action should also vanish from the Hamiltonian.  The equations defining the critical points then must be adjusted accordingly.  As a pedagogical example, in the H$_2$O system the two O-H normal stretch critical point families have no bend action \cite{Zi-MinH2O1}.  In deriving these families from the critical points analysis, the  two stretch-bend Fermi resonances have to be removed.  Critical points in the resulting 2 DOF subsystem then correspond to the normal stretch modes.
  
In order to locate {\it all} the possible critical points in the stretch-bend system, all combinations of $u_i=0$ have to be considered.  In each case, the relevant zero-order and resonance terms in the classical Hamiltonian are removed.  A new canonical transformation (Appendix A) may be necessary so that the new angles correspond to the remaining resonances.  Here we treat the two subsystems outlined in the beginning of $\S$ 5.3 using this kind of special consideration.

\subsection{5.3.2 Results}
\addtocontents{toc}{\protect\vspace*{7pt}}

\begin{figure}[hbtp] 
\begin{center}\includegraphics[width=5.73in]{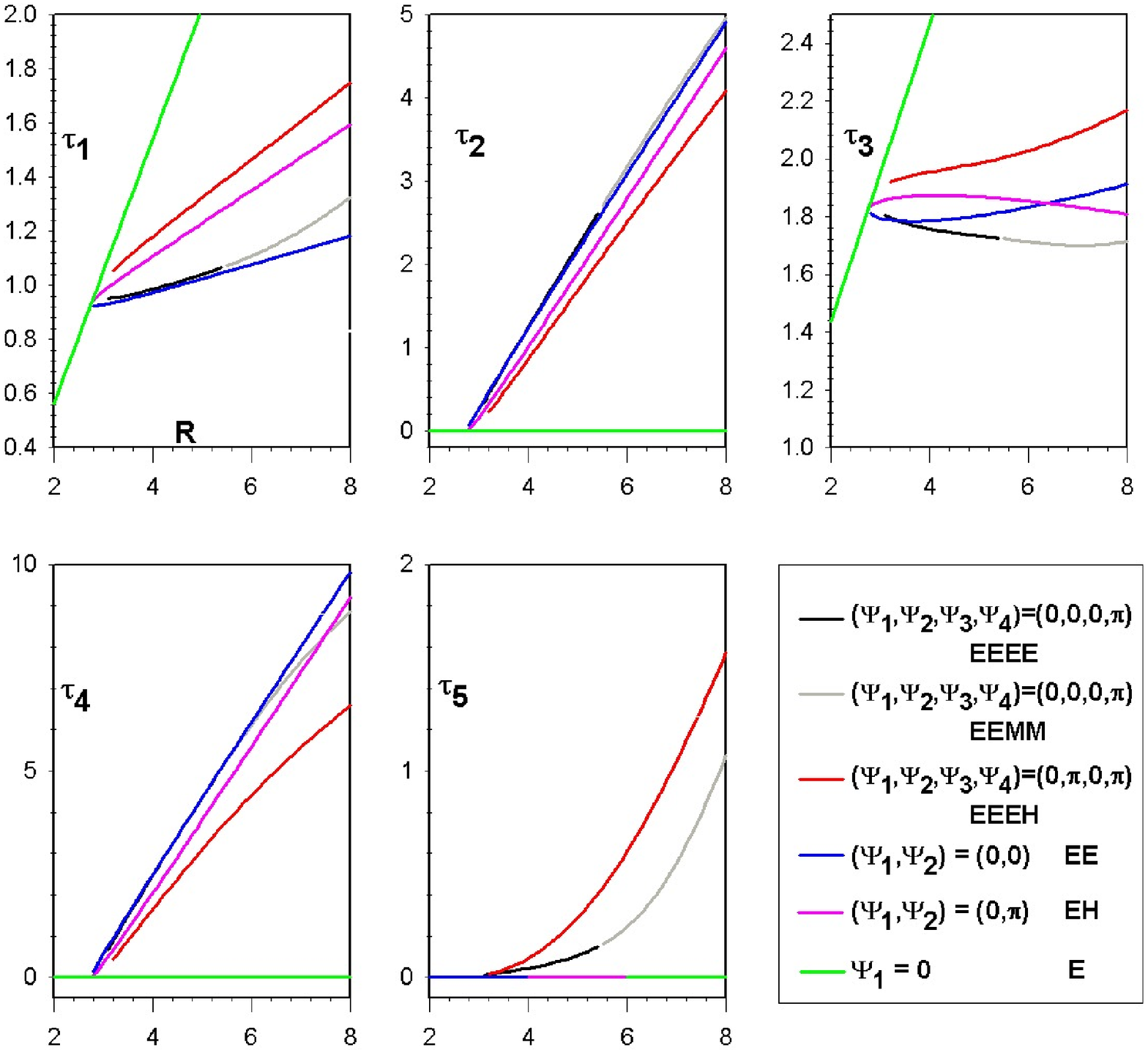}\end{center}
\cpn{Critical points in the stretch overtone polyads} {Critical points in the stretch overtone polyads $\lbrack 5R, R, 0 \rbrack$.  Black, gray and red: critical points in $H_{sb}$.  Navy and magenta: the $K_{11/33}+ K_{1/244}$ system with $\tau_5=0$. Green:  The $K_{11/33}$ system, with $\tau_2=\tau_4=\tau_5=0$.  \label{sbbifurfig}}
\end{figure}

The resulting critical points consist of families of curves in the ($\tau_i, \kappa_j$) 7-dimensional action space, parameterized by the polyad number $R$.   Associated with each family is a discrete set of $\Psi_i$ values.  All these solutions turn out to have $\kappa_4 = \kappa_5=0$.  In Fig.~\ref{sbbifurfig}, we graph the results in 5 separate panels with each $\tau_i$ versus $R$ separately.  In the full system, only those with significantly non-zero $\tau_i$ are displayed here.  In the $K_{11/33}$ subsystem, $\tau_2= \tau_4= \tau_5=\kappa_4=\kappa_5=0$.  In the $K_{11/33}+K_{1/244}$ subsystem, $\tau_5=\kappa_4=\kappa_5=0$.

With $R \in \lbrack 2, 8 \rbrack $, we found one family for the $K_{11/33}$ subsystem corresponding to the local C-H stretch, two families for the  $K_{11/33}+K_{1/244}$ subsystem, and two families for $H_{sb}$ system.

\newpage  \newpage 

When there is only $K_{11/33}$ resonance, one family of stable (E) critical points bifurcates out of the $\tau_3$ overtone.  The new family has 
\begin{align}
\Psi_1 = 0 \label{K1133Family}
\end{align}
\noindent which is the local C-H stretch mode on Fig.~\ref{PPS1133}.  

When both $K_{11/33}+K_{1/244}$ are included and $\tau_5=\kappa_5=0$, the reduced phase space employs four variables ($I_1, I_2, \Psi_1, \Psi_2$).  Two new families of critical points are created at:
\begin{align}
(\Psi_1, \Psi_2) = (0, 0) , (0, \pi) 
\end{align}
\noindent The $(0, 0)$ family has (EE) stability and the $(0, \pi)$ family has (EH) stability.   The C-H stretching motion of these critical points is qualitatively like the local C-H stretch motion, because the $\Psi_1$ value for both families is the same as in eqn. (\ref{K1133Family}).   The $\Psi_2$ angle, as the phase angle of a three-mode resonance $K_{1/244}$, is yet to be given a clear physical meaning.

In the full $H_{sb}$, solving the critical point equations (\ref{3EQ1}-\ref{sbconstrain3}) results in two more families of critical points with
\begin{align}
(\Psi_1,\Psi_2,\Psi_3,\Psi_4) = (0,0,0,\pi), (0,\pi,0,\pi) 
\end{align}
\noindent Although there is no explicit constraint, both of these families have $\kappa_4=\kappa_5=0$.   The $(0, 0, 0, \pi)$ family start with (EEEE) stability and switches to (EEMM) at $R=5.45$. The $(0, \pi, 0, \pi)$ family has (EEEH) stability for up to $R=$ 8.  These two families also have $\Psi_1=0$,  indicating a local-type stretch in the C-H DOF.  The values of two other angles $\Psi_3=0, \Psi_4=\pi$, according to Table~\ref{16cond}, suggest that the bending motion is of the local ({\bf L}) type.

A striking feature in Fig.~\ref{sbbifurfig} is that the critical points in both the $K_{11/33}+K_{1/244}$ and full Hamiltonian appear to have bifurcated out of the local C-H stretch family (green line).  The latter then bifurcates out of the $\tau_3$ overtones.  We believe that the new families of critical points in Fig.~\ref{sbbifurfig} correspond to novel types of dynamics in the respective stretch overtone polyads.

These critical points in the stretch-bend Hamiltonian do not correspond to PO in the full phase space when there are up to 3 frequencies in the cyclic angles:
\begin{align}
\dot{\theta}_P =\frac{\partial H_{sb}}{\partial P}, \mbox{\,\,\,\,\,} \dot{\theta}_R =\frac{\partial H_{sb}}{\partial R}, \mbox{\,\,\,\,\,} \dot{\theta}_\ell =\frac{\partial H_{sb}}{\partial \ell} 
\end{align}
\noindent The last frequency, $\dot{\theta}_\ell$, turns out to be zero for all the critical points graphed in Fig.~\ref{sbbifurfig}, in the same manner as $\dot{\theta}_b$ does in the $\lbrack N_b, 0 \rbrack$ pure bending polyads when $\ell=0$ (see Chapter 4).  The first two frequencies, $\dot{\theta}_P$ and $\dot{\theta}_R$, are not uniquely defined.  Instead, an alternative choice of the canonical transformation can change the definitions of $\theta_P$ and $\theta_R$ as well as their frequencies.  

\section{Summary}
\addtocontents{toc}{\protect\vspace*{7pt}}

The critical points analysis is performed on the C$_2$H$_2$ stretch-bend effective Hamiltonian, for the purpose of clarifying the fate of normal C-H stretch mode under increasing excitation.  Preliminary results indicate that the normal antisymmetric C-H stretch overtone (critical point) is first destabilized by $K_{11/33}$ resonance to produce the local C-H stretch.  Then the local stretch bifurcates into at least 4 families of stretch-bend critical points at higher excitation after the inclusion of stretch-bend resonances.  Such an observation suggests that the stretch-bend dynamics could be influences by a chain of bifurcations, each induced by perhaps the addition of one resonance.

Currently we are working to clarify the physical meaning of these critical points, as well as their role in the classical phase space.  A more complete classification, including all the subsystems (combinations of $\tau_i=0$) is also in progress.%
\addtocontents{toc}{\protect\vspace*{12pt}}%
\chapter[\protect\uppercase{Conclusions and future directions}]{Conclusions and future directions}\label{ch.ch6}%
\addtocontents{toc}{\protect\vspace{0.25in}}

\section[Conclusions]{\underline{Conclusions}}
\addtocontents{toc}{\protect\vspace*{7pt}}

A generalized method of critical points analysis is proposed for studying the dynamics of vibrationally excited molecules.  The classical form of an effective Hamiltonian with polyad number(s) is canonically transformed to reduce the dimensionality.  In the reduced phase space, the critical points are systematically found as roots of analytic equations.  Their number and linear stability are followed as the polyad number(s) is varied.  Critical points that are linearly stable in all directions are expected to indicate regions of regular (quasiperiodic) motion, which correspond to modes of vibration.  These critical points constitute the most important invariant structure in phase space.   A change in their number and/or stability, called a bifurcation, indicates qualitative changes in both the dynamics and spectra of the system.

The analysis is carried out in the C$_2$H$_2$ pure bending system.  With increasing polyad number $N_b$, four new families of critical points are born in successive bifurcations of the normal mode families.  After their first bifurcation, the normal modes families become unstable.  The two new all-stable families {\bf L} and {\bf CR} correspond to the new modes dominating the bottom and top (respectively) of the high-lying polyads. The four new families can be qualitatively interpreted as superpositions of resonant modes caused when DD-I or $\ell$ resonances act separately on the zero-order system.  

The same analysis is extended to the C$_2$H$_2$ stretch-bend system.  Preliminary results are presented and discussed for polyads $\lbrack 5 R, R, 0 \rbrack$, which contain the stretch overtones.  As the stretch polyad number $R$ is increased,  first the antisymmetric C-H normal stretch overtone is first substituted by the local C-H stretch as the stable critical point, in a bifurcation induced by $K_{11/33}$ resonance.  Then $K_{1/244}$ and other resonance couplings set in sequentially, creating new families of critical points.  These critical points are expected to correspond to novel modes of stretch-bend vibration.

\section[Summary of Contributions]{\underline{Summary of Contributions}}
\addtocontents{toc}{\protect\vspace*{7pt}}

In this thesis we have
\begin{itemize}
\item Introduced a generalized critical points analysis method formulated for arbitrary DOF and multiple polyad numbers.  It locates new modes of vibration by following the critical points in the reduced phase space.  The method scales well with additional dimensionality of the problem.
\item Found four new modes of vibration in the acetylene pure bending system, and provided a qualitative explanation of their origin and nature in terms of single $DD-I$ or $\ell$ resonances.
\item Studied for the first time the acetylene stretch-bend system with all the resonances.  Preliminary results suggest a series of resonances that act sequentially to couple the C-H stretching overtones to the remaining DOF.
\end{itemize}

\section[Future Work]{\underline{Future Work}}
\addtocontents{toc}{\protect\vspace*{7pt}}

Chapter 5 only focused on the dynamics of stretch overtone polyads of acetylene.  A comprehensive critical points analysis of the full phase space remains to be carried out.  The resulting critical points could be used to assign the many strongly perturbed eigenstates in polyads $\lbrack 4, 20, 0 \rbrack ^{u+}$ and $\lbrack 5, 25, 0 \rbrack ^{u+}$.  These eigenstates appear as neither normal mode ZOS, nor attributable to perturbations from a single resonance source \cite{Herman1999}.  

The method formulated in this thesis has opened the door to understanding the dynamics of other high DOF systems with polyad structure.  These systems include the other C$_2$H$_2$ isotopomers, formaldehyde (H$_2$CO) \cite{Formaldehyde}, methane (CH$_4$) \cite{methane} and even myoglobin \cite{Myoglobin}.  The last case indicates that even in large biomolecules, under favorable conditions a few strongly coupled modes may remain dynamically isolated for a (relatively) prolonged time.  Analyze these new systems will greatly expand the application of the polyad Hamiltonian model.  

\,\,\,

New theoretical inquiries also arise from this thesis, especially with regard to the mathematical theories of relative equilibria.  Most existing studies are concerned with rigorously conserved symmetries, such as the angular momentum of an isolated body ($\S$ 3.3.2).  In the case of molecules, however, both the effective Hamiltonian and polyad numbers are approximate.  In equating the dynamics of the effective Hamiltonian and that of the true molecular Hamiltonian, the critical points analysis needs to be structurally stable with regard to small polyad-breaking terms.  These terms become increasingly important at high vibrational excitation, especially near the threshold of an isomerization barrier \cite{JacobsonChildHCP1,JacobsonChildHCP2}.  A deeper insight into the effect of these terms would be of great practical importance to our method.

\addtocontents{toc}{\protect\vspace*{12pt}}%
\appendix
\addtocontents{toc}{\noindent\normalfont{APPENDIX}\hfill}
\addtocontents{toc}{\protect\vspace{0.25in}}

\chapter[\protect\uppercase{Canonical Transformation}]{Canonical Transformation}\label{app.1}
\addtocontents{toc}{\protect\vspace*{10pt}}

This appendix explicitly derives the canonical transformation that results in a reduction of the classical Hamiltonian with polyad number(s).   The application of eqns. (\ref{Heisenberg}) to an $N$-mode quantum effective Hamiltonian (\ref{generalHeff}) results in a classical Hamiltonian with $N$ pairs of action-angle variables ($\tau_i,\phi_i$)
\begin{align}
H(\tau_i,\phi_i)=H_0(\tau_i)+ H_v (\tau_i,\phi_i) \label{oldgeneralH}
\end{align}

Each resonance coupling in $H_v$ can be expressed as an $N$-vector. Let there be $M$ ($M \leq N$) resonance vectors that are linearly independent of each other: 
\begin{eqnarray}
\vec{V}_i =\{ N_{i1}, N_{i2}, \ldots, N_{iN} \} \mbox{\,\,\, for } i=1, \ldots, M \label{trans1}
\end{eqnarray}
\noindent There exist a total number of ($N-M$) polyad numbers \cite{KellmanVector,FriedEzra}
\begin{eqnarray}
\vec{P}_j =\{ P_{j1}, P_{j2}, \ldots, P_{jN} \} \mbox{\,\,\, for } j=1, \ldots, (N-M) \label{trans2}    
\end{eqnarray}
\noindent which correspond to vectors perpendicular to all $\vec{V_i}$:
\begin{eqnarray}
\vec{P}_j \cdot \vec{V}_i =0
\end{eqnarray}

The reduction of eqn. (\ref{oldgeneralH}) consists of finding a canonical transformation which (1) linearly combines $\tau_i$ into $M$ new actions $I_i$ and $N-M$ polyad numbers $P_j$; (2) linearly combines $\phi_i$ into $M$ angles $\Psi_i$ conjugate to $I_i$ and $N-M$ cyclic angle $\theta_j$.  After this transformation, the reduced Hamiltonian will be spanned by only $2N-2M$ new variables ($I_i, \Psi_i$):
\begin{eqnarray}
H(I_i,\Psi_i)=H_0(I_i,P_j)+ H_v (I_i,\Psi_i,P_j) \label{reducedgeneralH}
\end{eqnarray}

This transformation can be expressed in matrix notation as
\begin{eqnarray}
\left( \begin{array}{c} \Psi_1 \\ \cdots \\ \Psi_M \\ \theta_1 \\ \cdots \\ \theta_{N-M}  \\ I_1 \\ \cdots \\ I_M \\ P_1 \\ \cdots \\ P_{N-M} \end{array} \right)=
\left( \begin{array}{cc}A&0\\0&B \end{array} \right) \left( 
\begin{array}{c} \phi_1 \\ \cdots \\ \cdots \\ \cdots \\ \cdots \\ \phi_N \\ \tau_1 \\ \cdots \\ \cdots \\ \cdots \\ \cdots \\ \tau_N \end{array} \right) \label{generalcanonical}
\end{eqnarray}

Both $A$ and $B$ are $N \times N$ matrices.  The first $M$ rows of $A$ are the $\vec{V}_i$ vectors in eqn. (\ref{trans1}), while the last $N-M$ rows of $B$ are the $\vec{P}_j$ vectors in eqn. (\ref{trans2}).   Using the symplectic formulation of canonical transformations (Chapter 9.3 of \cite{Goldstein}), we require:
\begin{eqnarray}
\left( \begin{array}{cc}A&0\\0&B \end{array} \right) \left( \begin{array}{cc}0&-E_N\\E_N&0 \end{array} \right) {\left( \begin{array}{cc}A&0\\0&B \end{array} \right)}^T = \left( \begin{array}{cc}0&-E_N\\E_N&0 \end{array} \right)  \label{symtrans} 
\end{eqnarray}
\noindent  This is equivalent to
\begin{eqnarray}
A B^T = B A^T = E_N 
\end{eqnarray}

In solving for the unknown elements in matrices $A$ and $B$, there are usually more unknowns than the number of independent equations.  The transformation then is not uniquely determined.  

Nevertheless, the classical dynamics (and physical properties in general) should not be dependent on the choice of a special coordinate system.   Consider a critical point in coordinates ($I_i, \Psi_i$) where
\begin{eqnarray}
\frac{\partial H}{\partial \Psi_i} = \frac{\partial H}{\partial I_i}  = 0 \label{GeneralFP}
\end{eqnarray}
\noindent In an alternative coordinate system ($J_i, \Phi_i$), the critical point equations can be derived using the chain rule from calculus:
\begin{align}
\left( \begin{array}{c} \frac{\partial H}{\partial {J_1}} \\ \cdots \\ \frac{\partial H}{\partial {J_{M}}} \end{array} \right) = &\left( \begin{array}{ccc} \frac{\partial {I_1}}{\partial {J_1}}& \cdots & \frac{\partial {I_M}}{\partial {J_1}} \\ \cdots & \cdots & \cdots \\ \frac{\partial {I_1}}{\partial {J_M}}& \cdots & \frac{\partial {I_M}}{\partial {J_M}} \end{array} \right) \left( \begin{array}{c} \frac{\partial H}{\partial {I_1}} \\ \cdots \\ \frac{\partial H}{\partial {I_M}} \end{array} \right) =0 \label{generalaction} \\
\left( \begin{array}{c} \frac{\partial H}{\partial {\Phi_1}} \\ \cdots \\ \frac{\partial H}{\partial {\Phi_{M}}} \end{array} \right) = &\left( \begin{array}{ccc} \frac{\partial {\Psi_1}}{\partial {\Phi_1}}& \cdots & \frac{\partial {\Psi_M}}{\partial {\Phi_1}} \\ \cdots & \cdots & \cdots \\ \frac{\partial {\Psi_1}}{\partial {\Phi_M}}& \cdots & \frac{\partial{\Psi_M}}{\partial {\Phi_M}} \end{array} \right) \left( \begin{array}{c} \frac{\partial H}{\partial {\Psi_1}} \\ \cdots \\ \frac{\partial H}{\partial {\Psi_M}} \end{array} \right) =0 \label{generalangle}
\end{align}
\noindent  The two square matrices are determined by the transformation between ($I_i, \Psi_i$) and ($J_i, \Phi_i$).  From eqn. (\ref{GeneralFP}), the two column vectors  $\left( \frac{\partial H}{\partial I_i}\right)$ and $\left( \frac{\partial H}{\partial \Psi_i}\right)$ vanish at the critical points.  The two column vectors on the left, $\left( \frac{\partial H}{\partial J_i}\right)$ and $\left( \frac{\partial H}{\partial \Phi_i}\right)$ then must also vanish.  Therefore the critical points are indeed invariant under different choices of the canonical transformation.

Apart from the uncertainty in the canonical transformation, there is also uncertainty in choosing the polyad numbers $P_j$ as well as the cyclic angles $\theta_j$.  The main consequence is that the $N-M$ frequencies $\dot{\theta}_i$ associated with a critical point are also arbitrary.  A preferred definition is not evident from the general consideration \fn{The only exceptions are some trivial cases.  For example, if a zero-order mode $i$ is {\it uncoupled}, then using $\phi_i$ as a cyclic angle is more intuitive.}.  Additional system-specific constraints, such as relating $\dot{\theta}$ variables to the vibrational frequencies in the Cartesian coordinates, may be necessary to address the choice of a non-unique set of action-angle variables.   

In summary, when reducing the classical Hamiltonian using the polyad number(s), there is freedom in choosing both the polyad number(s) and the canonical transformation defining the reduced phase space.  However these different choices should lead to the same critical points and physical behavior.

\chapter[\protect{TOPOLOGY OF $\lbrack N_b, 0 \rbrack$ PURE BENDING PHASE SPACE}]{Topology of $\lbrack  N_{\lowercase{b}}, 0 \rbrack$ Pure Bending Phase Space}\label{app.2}
\addtocontents{toc}{\protect\vspace*{10pt}}

\begin{center}\underline{B.1. The Poincar\'{e}-Hopf Index Theorem} \end{center}

The {\it Poincar\'{e}-Hopf Index Theorem} \cite{IndexTheorem}, initially proposed by Poincar\'{e} and later extended by Hopf,  provides a constraint on the possible combination of the critical points in the reduced classical phase space.  

The theorem states that {\it ``The index of a vector field with finitely many zeros on a compact, oriented manifold is the same as the Euler characteristic of the manifold"}  \cite{Weissen}.   In our cases, the manifold is the reduced phase space while the vector field is the flow generated by Hamilton's equations of motion.  An index $g_i$ is assigned to each critical point (i.e. a zero of the vector field) based on its linear stability.  The theorem indicates that {\it the sum of all indices $g_i$ is equal to the Euler characteristics $\chi$}.  $\chi$ is also known as the {\it topological index} since it is entirely determined by the topology of the manifold.

A critical point is non-degenerate if its stability matrix $A$ in eqn. (\ref{linearEOM}) has no zero eigenvalues.  At such a point, $g_i = \pm 1$ according to one of the following two equivalent criteria: (1) $(-1)^n$, with $n$ the number of eigenvalues with positive real part; (2) the sign of the determinant of $A$.  Examples of $g_i$ in 1-3 DOF Hamiltonian systems are listed in Table~\ref{taba.1}.

\begin{center}  \begin{table}[hbt]
\cpn{Indices of critical points in Hamiltonian systems} {Indices of critical points in Hamiltonian systems with 1-3 DOF. \label{taba.1}}
\vspace{0.2in}
\begin{center} \begin{tabular}{|c|c||c|c|} \hline\hline
\, & \, & \, & \, \\
Linear Stability & $g_i$ & Linear Stability &  $g_i$ \\   \hline \, & \, & \, & \, \\
E & +1 & EEE & +1 \\  \, & \, & \, & \, \\
H & -1 & EEH & -1 \\  \, & \, & \, & \, \\
EE & +1 & EHH & +1 \\ \, & \, & \, & \, \\
EH & -1  &HHH & -1 \\ \, & \, & \, & \, \\
HH & +1 & EMM & +1 \\ \, & \, & \, & \, \\
MM & +1 &  HMM & -1 \\  \, & \, & \, & \, \\  \hline \hline
\end{tabular}  \end{center} \end{table}  \end{center}

\begin{figure}[hbtp] 
\begin{center}  \includegraphics[width=5.08in]{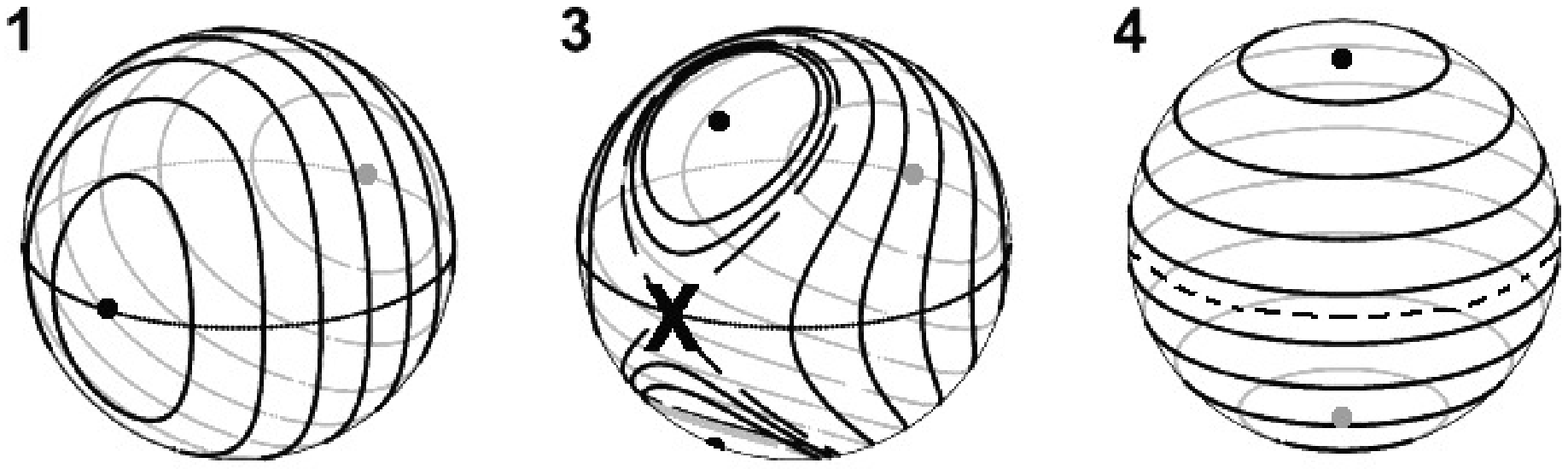}\end{center}  
\cpn{Conservation of topological index on the PPS}{Conservation of topological index on the PPS.  A large dot indicates a stable (E) critical point and the ``X" in sphere 3 indicates an unstable (H) critical point.   \label{IndexSphere}} 
\end{figure}

On a 2-dimensional manifold, $\chi$ is related to the number of ``holes" (genus) $h$ on the manifold:
$$ \chi = 2-2 h $$
\noindent Therefore a torus (one ``hole") has $\chi=0$ while a sphere (no ``hole") has $\chi=2$.  The second case has been extensively used by Kellman \et to verify the critical points on a PPS.  It proves to be especially useful in determining the stability of critical points located at the poles \cite{SvitakMNRes}.  Fig.~\ref{IndexSphere} demonstrates the Poincar\'{e}-Hopf index theorem, for selected PPS in Fig.~\ref{catmap}.  Spheres ``1" and ``4" each has a total of two stable critical points, while sphere ``3" has 3 stable and 1 unstable critical points.  In all cases, the sum of the stability indices by Table~\ref{taba.1} equals 2, as expected for a sphere.

\newpage   \newpage

The application of the Poincar\'{e}-Hopf Theorem to higher dimensional manifolds is limited, mostly due a lack of means of direct visualization.  A few known cases are summarized in Table~\ref{taba.2}.

\begin{center}  \begin{table}[hbt]
\cpn{Topological indices of 2- and 4-dimensional manifolds} {Topological indices of 2- and 4-dimensional manifolds, from \cite{ZhilinskiiIndex}. \label{taba.2}}
\vspace{0.2in}
\begin{center}
\begin{tabular}{|c|c|c|} \hline\hline \, & \,  & \,\\  
Symbol & Manifold & $\chi$ \\  \, & \, & \, \\  \hline \, & \,  & \,\\  
$S^2$ & 2-sphere & 2 \\ \, & \,  & \,\\  
$T^2$ & 2-torus  & 0 \\ \, & \,  & \,\\  
$S^4$ & 4-sphere & 2 \\ \, & \,  & \,\\  
$T^4$ & 4-torus  & 4 \\ \, & \,  & \,\\  \hline \hline
\end{tabular}  \end{center}   \end{table}  \end{center}

\begin{center}\underline{B.2. Topology of [$N_b, 0$] Bending Phase Space} \end{center}

This subsection proposes a topological description of the critical points found in the C$_2$H$_2$ pure bending polyads $\lbrack N_b, 0 \rbrack$.  

\begin{figure}[hbtp]
\begin{center}  \includegraphics[width=3.028in]{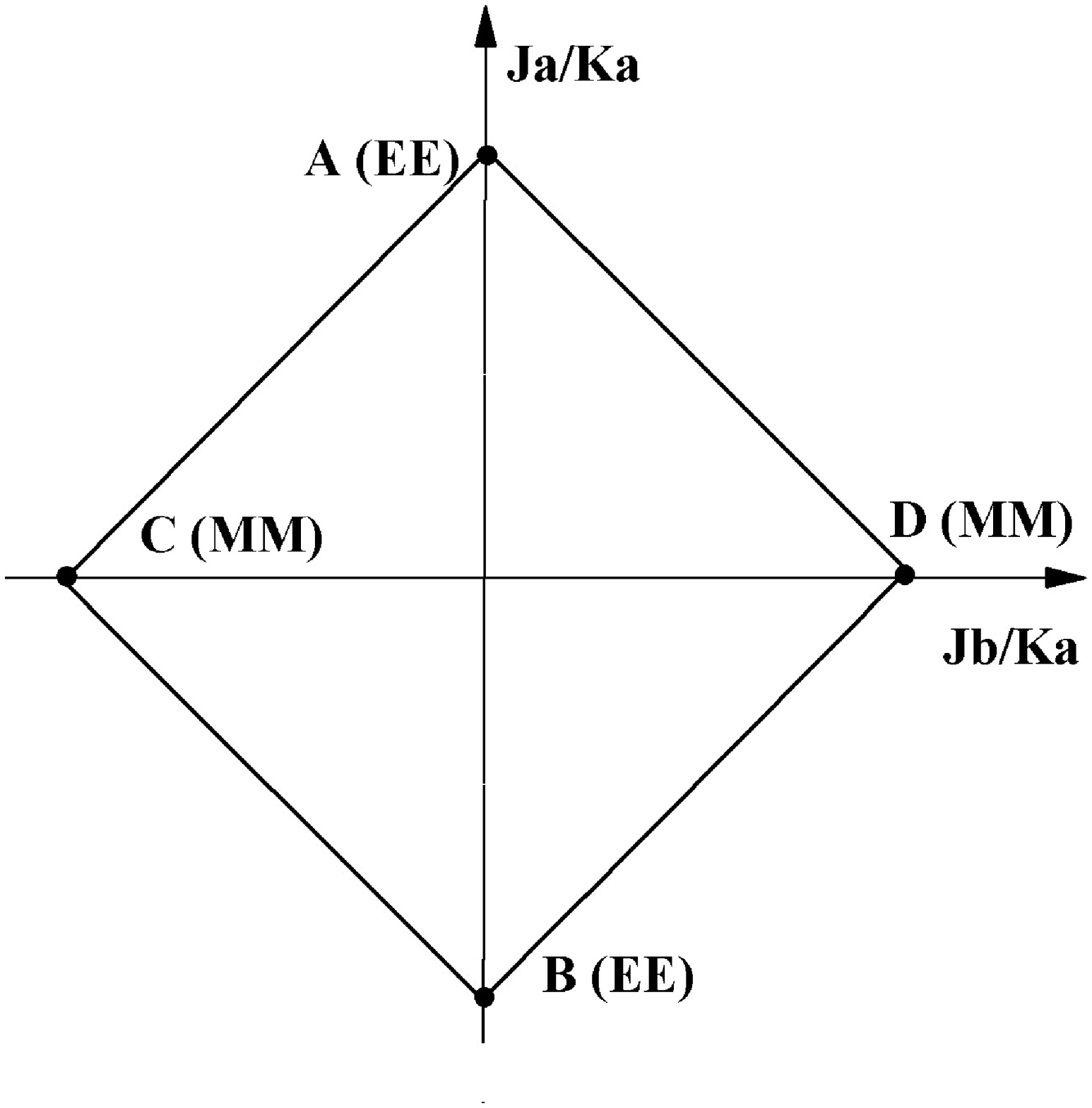}\end{center}  
\cpn{Critical points in $\lbrack N_b, 0 \rbrack$ polyads at low $N_b$} {Critical points in $\lbrack N_b, 0 \rbrack$ polyads at low $N_b$. \label{figa.1}}
\end{figure}

Currently, there is no sign (such as the display of monodromy \cite{ZhilinskiiMonodromy}) indicating that this classical phase space changes its topology with the $N_b$.   We first consider the low polyad end with $N_b \leq 7$, before any bifurcation takes place.  In this case there are no critical points except possibly where $|J_a|+|J_b|=K_a$.   In Figs.~\ref{spherediamond}-\ref{diamond20} of Chapter 4, these locations appear as the boundary of the diamond-shaped space.  Here this space is schematically illustrated in Fig.~\ref{figa.1}.  Because the coordinate system ($J_a, \psi_a, J_b, \psi_b$) is singular at these locations,  the Hamiltonian is transformed to a local-mode representation using the {\it x-K relationship}, which is described in Chapter 7.6 of \cite{JacobsonThesis}.   The result is shown in Fig.~\ref{figa.1}.  It was found that only the four vertices in Fig.~\ref{figa.1} are critical points: points A and B ($J_a=\pm K_a$, $J_b=0$) with (EE) stability and points C and D ($J_a=0$, $J_b=\pm K_a$) with (MM) stability.   
 
\newpage  \newpage

The two $J_b = \pm K_a$ families of critical points correspond to states with maximum $\ell_i$ in each of the normal mode oscillators \fn{Since $\ell_i \leq n_i$, there cannot be pure ``$\ell_i$ -overtones" in the form of e.g. $n_4=0, \ell_4 \ne 0$.}: 
$$\vert {(N_b/2)}^{\pm (N_b/2)}, {(N_b/2)}^{\mp (N_b/2)} \rangle$$   
\noindent These two families do not participate in the subsequent bifurcations.    

When $N_b >7$, at each bifurcation point in Fig.~\ref{fullbifs}, the sum of indices will be conserved if (1) each of the classical normal {\bf Trans} and {\bf Cis} motion, being a 2-dimensional oscillator, is doubly-degenerate; and (2) each point on the four new families of critical points corresponds to a quadruply-degenerate set of PO \fn{In classical mechanics, the two carbon and two hydrogen atoms are assumed distinguishable.   The quadruply-degenerate PO are related to each other a 90$^o$ rotation around the C-C bond and a mirror plane perpendicular to it.}.  Then the Poincar\'e-Hopf index theorem can be satisfied in the following manner.  Substituting the stability index of each family and its degeneracy into eqns. (\ref{4bifurcations}), before and after each bifurcation we have:
\begin{align}
\mbox{{\bf L}, {\bf CR}:\,\,\,\,\,} & \mbox{(EE)} \rightarrow \mbox{4 (EE) + 2(EH) ;\,\,\,} & (+2) & = 4 \times (+1) + 2 \times (-1) \\
\mbox{{\bf Orth}:\,\,\,\,\,} & \mbox{(EH)} \rightarrow \mbox{4 (EH) + 2(HH) ;\,\,\,} & (-2) & = 4 \times (-1) + 2 \times (+1) \\
\mbox{{\bf Pre}:\,\,\,\,\,} & \mbox{(HH)} \rightarrow \mbox{4 (HH) + 2(EH) ;\,\,\,} & (+2) &= 4 \times (+1) + 2 \times (-1) 
\end{align}
\noindent As points C and D do not participate in these bifurcations, only these critical points need to be considered.   In summary, during each of the four bifurcations of the {\bf Trans} and {\bf Cis}, the sum of stability indices remains unchanged.%
\addtocontents{toc}{\protect\vspace*{12pt}}%
\setlength{\baselineskip}{14pt plus1pt minus0pt}
\bibliography{refer} \label{bibl}
\end{document}